\documentclass[12pt,a4paper]{article}
\pdfoutput=1
 \usepackage{jheppub}
 \usepackage{amsmath}
 \usepackage{amsfonts}
 \usepackage{mathtools}
 \usepackage{amssymb}
 \usepackage{caption}
 \usepackage{subcaption}

\title{Anomalous transport  at weak coupling}

\author{Subham Dutta Chowdhury, Justin R. David}
\affiliation{Centre for High Energy Physics, Indian Institute of Science,\\
C. V. Raman Avenue, Bangalore 560012, India.}
\emailAdd{subham, justin@cts.iisc.ernet.in}

\abstract{
We evaluate the contribution of  chiral fermions in $d=2, 4, 6$, chiral bosons, a chiral 
gravitino like theory   in $d=2$  and chiral gravitinos in $d=6$ 
to all the leading parity odd transport coefficients at one loop. 
This is done by using finite temperature field theory 
to evaluate the relevant Kubo formulae.  For chiral fermions and chiral bosons  the 
relation between the parity odd transport coefficient and the microscopic anomalies
including gravitational anomalies  agree 
with that found by using the general methods of hydrodynamics and the argument involving 
the  consistency of the Euclidean vacuum. 
For the gravitino like theory  in $d=2$ and  chiral gravitinos in $d=6$, we show that 
relation between the  pure gravitational anomaly and parity odd  transport
breaks down.  
From  the  perturbative 
calculation we clearly identify the terms that contribute to the anomaly polynomial,
but not to the transport coefficient for gravitinos. 
We also develop a  simple  method  for evaluating  the angular integrals in the one loop diagrams 
involved in the Kubo formulae. 
 Finally we show that charge  diffusion mode of an ideal  2 dimensional  Weyl gas in the presence of a finite chemical potential 
  acquires  a speed, which is equal to half  the speed of light.}

\begin{document}
 \maketitle

\section{Introduction}

The modifications of the macroscopic equations of hydrodynamics due to the presence of 
quantum anomalies of the underlying theory has been the focus of recent interest. 
It began with the observation of parity odd terms in the constitutive relation of the charge 
current  in the holographic dual of ${\cal N}=4$ super Yang Mills at finite temperature and 
 chemical potential \cite{Bhattacharyya:2007vs,Erdmenger:2008rm,Banerjee:2008th}. 
This was then understood from  general considerations using  the equations of hydrodynamics,  
anomalous conservation laws in presence of background fields and 
the second 
law of thermodynamics \cite{Son:2009tf}.  
These parity odd transport coefficients are non-dissipative .  This fact  was used to 
determine the relation between microscopic anomalies and  the macroscopic 
parity odd transport coefficient using a equilibrium partition function \cite{Banerjee:2012cr,Jensen:2012jh}.

Using  general considerations from anomalies and the second law of thermodynamics
or using the  equilibrium partition function method
it is not possible to determine the precise numerical constant that determine the relation between the 
parity odd transport coefficients and the gravitational anomalies or mixed gauge-gravitational
anomalies of the microscopic theory.  One  reason for this is that both these methods 
rely on performing derivative expansions in velocities. 
Any relation between the gravitational anomalies or mixed gauge-gravitational 
anomalies would involve a jump in the derivative expansion. 
 For example consider hydrodynamics in  $d=2$, the gravitational anomaly occurs at 
the  2nd derivative in the  conservation law, however it affects the constitutive 
relations  at the zeroth order in the  derivative expansion \cite{Jensen:2012kj}.  
There are three methods in  literature which relates these anomalies to the 
parity odd transport coefficients. 
The first is the direct  perturbative method of evaluating the transport coefficient of 
interest  using the corresponding Kubo formula. 
This is usually done for free chiral fermions and then arguing that 
it is not renormalized, either using holography or in perturbation theory 
\cite{Landsteiner:2011cp,Landsteiner:2011iq,Landsteiner:2012kd,Golkar:2012kb}.  
The second method is using the consistency of the Euclidean vacuum,  
developed by \cite{Jensen:2012kj,Jensen:2013rga}.
The third method relies on performing   one loop integration  to obtain the theory 
on the spatial slice to relate the the anomalies to the Chern-Simons couplings  and then 
arguing that these are one loop exact\cite{DiPietro:2014bca,Golkar:2012kb}. The Chern-Simons couplings are in turn 
related to the transport coefficients \footnote{We thank Kristan Jensen and Zohar Komargodski for informing us 
about this approach.}. 
To make this  discussion concrete  let us consider $d=2$ and the 
the   parity odd term in the constitutive relation for the stress tensor \footnote{We have 
written down the stress tensor in the anomaly frame.} given by 
\begin{equation}
 T^{\mu \nu} = (\epsilon + P ) u^\mu u^\nu  -P \eta^{\mu\nu} +
 \lambda^{(2)} ( u^\mu \epsilon^{\nu\rho} u_\rho +  u ^\nu \epsilon^{\mu \rho} u_\rho) . 
\end{equation}
Here $u^\mu$  is the velocity profile of the fluid.  From general considerations of 
the equilibrium partition function and conservation laws 
\cite{Valle:2012em,Jain:2012rh} \footnote{See \cite{Banerjee:2013fqa} for an alternative approach.}
it can be shown that transport coefficient $\lambda^{(2)}$  is of the form 
\begin{equation}
 \lambda^{(2)} = \tilde c_{2d} T^2  -c_s \mu_-^2, 
\end{equation}
where $T$ is the temperature and $\mu_-$ is the chiral chemical potential. 
The Euclidean partition method shows that 
$c_s$ is  the coefficient of the charge current anomaly defined by 
the conservation equation 
\begin{equation}
 \partial_\mu j^\mu = c_s \epsilon^{\mu\nu} F_{\mu\nu}, 
\end{equation}
$F_{\mu\nu} $ is the background field strength. 
However the relation between
 $\tilde c_2d$ and the  pure gravitational anomaly $c_g$  defined by  
 \begin{equation}\label{2dgravan}
  \partial_\nu T^{\mu\nu} = F^\mu_\nu j^\nu + c_g \epsilon^{\mu\nu} \nabla_\nu R, 
 \end{equation}
where $R$ is the background curvature cannot be shown using methods which involve expansions 
in derivatives. This is because the coefficient $\tilde c_{2d}$ occurs at the zeroth order in derivative in the 
stress tensor. From (\ref{2dgravan}) it is seen that pure gravitational anomaly is effective at the 2nd order in the derivative 
expansion. 
It was argued using an argument which involves the  consistency of the Euclidean vacuum that 
\begin{equation}\label{gravanr}
 \tilde c_{2d} = - 8\pi^2 c_g. 
\end{equation}
A similar relation was obtained between  the mixed gravitational anomaly and  the parity 
odd transport coefficient in $d=4$ determined by the two point function of the charge current and the 
stress tensor. 
In $d=4$ the relation of the type in (\ref{gravanr}) has been 
 verified by  a direct perturbative calculation 
done in   \cite{Landsteiner:2011cp} as well as using  the holographic dual of ${\cal N}=4$ Yang-Mills 
\cite{Landsteiner:2011iq}. It was also shown within perturbation theory the relation in 
(\ref{gravanr}) 
is not renormalized for theories of  chiral fermions which interact via  Yukawa couplings. 
Similar equations relating parity odd transport and anomalies including gravitational anomalies were
obtained for arbitrary even dimensions \cite{Loganayagam:2011mu,Loganayagam:2012pz,
Jensen:2012kj,Jensen:2013rga}
and summarized by the `replacement rule'.
However the relation  between gravitational anomalies and transport 
of the type given in equation (\ref{gravanr}) for dimensions other that of $d=4$  has not yet been verified by a direct 
perturbative calculation. 
Further more it has been suspected that higher spin chiral fermions like gravitinos 
violate the relations of the type (\ref{gravanr}) \cite{Loganayagam:2012zg,Eling:2013bj,DiPietro:2014bca} .

Motivated by these questions we study the relationship between 
anomalies and parity odd transport coefficients using the direct approach of evaluating the 
respective Kubo formulae.  We consider theories of free chiral fermions in $d=2, 4, 6$ and chiral 
bosons in $d=2$ 
 at finite temperature and  chemical potential. We evaluate all the leading 
parity odd transport coefficients.  This is done using the method 
developed by \cite{Kharzeev:2009pj}  which involves evaluating the Feynman diagrams
in finite temperature field theory and then performing the Matsubara sums. 
This approach is  different  form that is used in \cite{Yee:2014dxa}  to study transport properties of 
free chiral fermions in arbitrary even dimensions that uses  Schwinger-Keldysh propagators
We keep track of contributions that arise from gauge anomalies  as well as pure  gravitational anomalies and  mixed gauge-gravitational
anomalies.  To address the issue of whether higher spin fermions obey the `replacement rule' we study 
the contributions of a chiral gravitino like theory in $d=2$ and chiral gravitinos in $d=6$ to parity odd transport. 

We now briefly summarize the results of this work. 
The Feynman diagrams which contribute to the transport coefficients in  $d=2$ are 
considerably different from that 
in higher dimensions. This is because the spin connection is not present in the action of fermions
in $d=2$.   
From the perturbative calculation in $d=2$ we show that chiral fermions as well as 
chiral bosons obey the relation in (\ref{gravanr}) predicted by the 
argument involving the consistency of  the Euclidean vacuum. The transport coefficients of the 
chiral bosons is identical to that of chiral fermions which is expected from Bose-Fermi duality in 
$d=2$.  For a higher spin fermion, that is a gravitino like theory,  we show that the 
the relation in (\ref{gravanr})  does not hold by an explicit perturbative analysis. 
In $d=4$ we reproduce the results of \cite{Kharzeev:2009pj,Landsteiner:2011cp,Yee:2014dxa}  and show the perturbative results 
for the transport coefficients agree with the `replacement rule'. 
Since gravitinos are not  charged, they do not contribute to the mixed anomaly in $d=4$ and  
 they do not contribute to any of the parity odd transport coefficients as well. 
In $d=6$, for  chiral fermions we reproduce  the results of \cite{Yee:2014dxa} for the contributions 
from the gauge anomaly to parity odd transport.  We also keep track of contributions from 
gravitational anomalies. 
We show that
contributions of chiral fermions to parity odd transport from the 
pure gravitational 
and the mixed anomaly agree with that predicted 
by the argument involving the consistency of the Euclidean 
vacuum.  Next we consider gravitinos in $d=6$ and show that its contribution to 
parity odd transport is not related to the pure gravitational anomaly and therefore 
does not agree with  the prediction obtained  from the replacement rule

From out study of higher spin chiral fermions in $d=2$ and $d=6$ we can draw the following general lesson for their contribution to parity odd transport. 
These transport coefficients arise in  correlators involving only  stress energy tensor. 
We see that the reason that gravitinos  do not obey the `replacement rule' 
is because of the presence of the following term in their  stress energy tensor. 
\begin{equation}\label{extra}
T_{\mu\nu}^{(3/2) extra}  = - \frac{i}{2} \left[ 
\partial_\sigma \left( \bar \psi^\sigma \gamma_\mu \psi_\nu  - \bar\psi_\mu \gamma_\nu \psi^\sigma
\right) + ( \mu\leftrightarrow \nu)  \right]. 
\end{equation}
This arises from the linearization of the Christoffel connection 
in the kinetic term for the gravitinos which is not present for fermions. 
Such a term gives rise to the interaction between the  magnetic  moment of the gravitino with 
the gravitational field \cite{AlvarezGaume:1983ig}.  It is this term that 
modifies the anomaly polynomial  for gravitinos from that
of a chiral spinor. 
However on evaluating Kubo formulae, at the zero frequency and zero momentum 
limit the contribution of the extra term in (\ref{extra})   vanishes. 
Thus the contribution of the gravitino to the transport coefficient is proportional to the chiral spinor. 
In fact it is just $d -1$ times that of of the chiral spinor.   
The $-1$ accounts for the ghosts. 
We observe this phenomenon for the gravitino like theory in $d=2$ and gravitinos in $d=6$
while evaluating the Feynman diagrams involved in the Kubo formula, 
and note that this property is  independent of dimensions. 

Apart form understanding the `replacement rule' our explicit calculations also point 
to a technical simplification in evaluating the integrals involved in the one loop 
diagrams. We see that  it is possible to take the zero frequency and zero momentum 
limit to obtain the transport coefficient before performing the angular integrations
involved in the one loop diagrams.  We develop a prescription for evaluating these integrals. 
This simplifies the calculation considerably 
compared to the ones in \cite{Kharzeev:2009pj,Landsteiner:2011cp,Landsteiner:2012kd}. 
In fact in \cite{Yee:2014dxa}  a guess is made for the 
angular integrals which was verified case by case. 
We will see that the angular integrals are elementary after taking the zero frequency and 
zero momentum limit.

Finally since parity odd transport coefficients appear 
at the zeroth order in the derivative expansion in $d=2$ they can affect linearized hydrodynamic
modes at the leading order. 
We show that the charge diffusion mode develops a velocity 
in presence of the parity odd transport coefficients.   For the Weyl gas we see that this velocity is $1/2$ times the  speed of light.

 The organization of the paper is as follows. 
 In section \ref{an2d} 
 we study parity odd transport in $d=2$, we first obtain the relevant Kubo formula for the 
 transport coefficient and then evaluate the contribution of chiral fermions, chiral bosons and a chiral  gravitino like theory to 
 the two leading 
 transport coefficients.  From the explicit calculation we will see how the `replacement rule' breaks 
down for the case of gravitino like theory in $d=2$. 
In section \ref{an4d} we look at $d=4$ and repeat the same analysis for chiral fermions. 
 In section \ref{an6d} 
  we turn to $d=6$ and study parity odd transport of both chiral fermions and chiral gravitinos. 
 Again we see that chiral gravitinos do not obey the replacement rule. 
In section \ref{dispersion}
 we study linearized hydrodynamic modes which are modified by parity odd transport 
coefficients in $d=2$.  
Appendix \ref{moments} contains details of moments of statistical distributions involved in obtaining the 
transport coefficients. Appendix \ref{an6da}
 contains the details of the calculations of all the leading 
parity odd transport in $d=6$.

\section{Anomalies and transport in  \texorpdfstring{$d=2$}{Lg} } \label{an2d}

In this section we first evaluate the 2 leading  parity odd transport coefficients. They occur 
at the zeroth order in the derivative expansion in the velocities.  
To obtain the Kubo formulae for these transport coefficients,  let us first write down the constitutive 
relations to the zeroth order in derivatives. The stress tensor and the current are given by 
\footnote{We work in the mostly negative signature through out this paper.}
\begin{eqnarray} \label{constrel}
T^{\mu \nu }&=&\left(\epsilon+P\right)u^{\mu }u^{\nu }-P \eta^{\mu \nu }
+\lambda^{(2)}  (u^\mu \epsilon^{\nu \rho }u_\rho +u^\nu \epsilon^{\mu \rho }u_\rho),\nonumber\\
j^{\mu }&=&nu^{\mu }
+\zeta^{(2)} \epsilon^{\mu \nu }u_{\nu }.               
\end{eqnarray}
Here $\epsilon, P, n$ refer to energy density, pressure and the charge density. 
$\lambda^{(2)}$ and $\zeta^{(2)}$ are the two parity odd transport coefficients which are of interest. 
The superscripts refer to the fact that these transport are for $d=2$. 
Note that we have written the constitutive relations in the anomaly frame. 
We work in the metric with the signature $(1, -1)$. 
To obtain the Kubo formula for these, consider the fluid  perturbed  from its rest frame $u^\mu = ( 1,  v^x) $ with $|v^x| <<1$. 
We also perturb the metric and the gauge fields as follows
\begin{eqnarray} \label{backexp}
& &  g_{\mu\nu} = \eta_{\mu\nu} + h_{\mu\nu} , \qquad  h_{tt}, h_{tx} \neq 0,  h_{xx} = 0,  \\ \nonumber
 & & A_t = a_t ,    \qquad A_x= a_x.
\end{eqnarray}
Substituting this expansion in the constitutive relations in (\ref{constrel}), we obtain the following expressions 
to the linear order 
\begin{eqnarray} \label{fcont}
T^{tx}=\epsilon v^x+Pv^x-Ph_{tx},\nonumber\\
 j^t=-\zeta^{(2)}  v^x + \zeta_v h_{tx}.
\end{eqnarray}
Here we have equated  the  first order terms in the equations for  the current and called it $j^t$ for convenience in notation. 
We have also used the fact that $v_x = -v^x + h_{tx}$, to linear order. 
Solving for $v^x$ using the first equation  and substituting in the  constitutive relation for the current we obtain
\begin{eqnarray}
j^t=n-\zeta^{(2)}  \left(\frac{T^{tx}+P h_{tx}}{\epsilon+P} \right)+\zeta^{(2)} h_{tx}.\nonumber\\
\end{eqnarray}
This can be thought of as an Ward  identity   obeyed by the theory. We Fourier transform the equation and then differentiate
with respect to $h_{tx}$ and finally set the perturbations to zero. 
Using the definition
\begin{equation}
\frac{1}{\sqrt{g}}\frac{\delta}{\delta g_{\mu\nu}} \cdot  = -\frac{T^{\mu\nu}} {2}, 
\end{equation}
we  obtain
\begin{equation} \label{w1}
 \langle  j^t(\omega, p ) T^{tx} (-\omega,  -p )  \rangle =   -\zeta^{(2)} 
 \left(\frac{\langle T^{tx}(\omega, p )  T^{tx}(-\omega, -p)\rangle  -P}{\epsilon+P} \right)
 -\zeta^{(2)}.
\end{equation}
Now we  use the relation 
\begin{equation}
 \lim_{p \rightarrow 0, \omega\rightarrow 0}   \langle T^{tx}(\omega, p )  T^{tx}(-\omega, p )\rangle =P, 
\end{equation}
which can be obtained by differentiating the first equation in (\ref{fcont}) with respect to $h_{tx}$  in (\ref{w1}). This leads us   to 
the following Kubo formula 
\begin{eqnarray}\label{2dk1}
- \zeta^{(2)} &=&  \lim_{p \rightarrow 0, \omega \rightarrow 0 }  \langle  j^t(\omega, p ) T^{tx} (-\omega,  -p )  \rangle_R. 
 \end{eqnarray}
 Similarly we consider the first order change in the stress tensor $T^{tt}$  from (\ref{constrel}), which is given by 
\begin{equation}
 T^{tt} =  2 \lambda^{(2)} ( -v^x + h_{tx}). 
\end{equation}
Again eliminating $v^x$ and differentiating the above Ward identity in Fourier space  with respect to $h_{tx}$  we obtain
\begin{eqnarray} \label{2dk2}
- \lambda^{(2)} &=& \frac{1}{2} \lim_{p \rightarrow 0, \omega \rightarrow 0 }\langle  T^{tt}(\omega, p ) T^{tx} (-\omega,  -p )  \rangle_R.
\end{eqnarray}

Before proceeding we make a  few observations related to  the Kubo formulae  for the two leading order transport 
coefficients in (\ref{2dk1}) and (\ref{2dk2}).  
Both  the two point functions are obtained as  the Fourier transform of the 
 real time retarded correlator and the expectation value is taken over states held at finite temperature
 and chemical potential of the theory.  These statements will hold true 
 for all the correlators  involving the various Kubo formulae  studied in this paper.
Also  note that they do not involve any division by momenta, since
these transport coefficients occur at the zeroth order in derivatives. 
We need to take the zero frequency limit first and then the zero momentum limit
 in (\ref{2dk1}) and (\ref{2dk2}). 
 This is similar to  the situation in higher dimensions where 
the static, zero frequency limit needs to be first taken and then the zero momentum limit \cite{Kharzeev:2009pj}. 
Note that these transport coefficients  in $d=2$ can also be obtained by considering one point functions of the stress tensor 
and the current. However we wish to illustrate the role of contact terms and also have a discussion similar to that seen in higher
dimension and therefore we choose to examine two point functions. 
The  transport coefficients of interest occur at the zeroth order in derivatives, they  can in principle  affect 
the hydrodynamic charge and sound modes \footnote{For relativistic fluids in  $1+1$ dimensions, these hydrodynamic 
modes were studied in \cite{David:2010qc}.}
at the zeroth order in the momentum expansions.  We will study the  consequences of these transport coefficients 
in section \ref{dispersion}.

\subsection{Chiral fermions} \label{chifermi}

Our goal in this section is to evaluate 
 transport coefficients  $\zeta^{(2)}$ and $\lambda^{(2)}$ given in 
(\ref{2dk1}) and (\ref{2dk2}) for a theory of free chiral fermions held at  finite
temperature  and finite chemical potential. 
As we have discussed earlier, it is the  Fourier transform of the real time 
retarded correlators which determine the transport coefficients. 
We obtain these correlators by first evaluating them in the Euclidean theory 
and then performing the necessary analytic continuation. 
Consider the Euclidean partition function  ${\cal S}_E$ of free chiral fermions,  coupled to background metric
and gauge field. The subscript $E$ will always refer  to Euclidean. 
It admits the following expansion  around the unperturbed background given in (\ref{backexp}) 
\begin{eqnarray} \label{actexp}
S_{E} &=& S_{E}^{(0)} + \frac{1}{\sqrt{g}}\frac{\delta S_E}{\delta A_\mu} A_\mu + 
\frac{1}{\sqrt{g}}\frac{ \delta S_{E} } {\delta g_{\mu \nu}}  g_{\mu \nu}  +  
\frac{1}{2 \sqrt{g}}\frac{\delta}{ \delta g_{\rho\sigma}}(\frac{1}{\sqrt{g}}\frac{\delta S_{E} }{ \delta g_{\mu\nu} } )
g_{\mu\nu} g_{\rho\sigma}\nonumber\\
&&+ \frac{1}{2\sqrt{g}} \frac{\delta}{ \delta g_{\rho\sigma}}(\frac{1}{\sqrt{g}}\frac{\delta S_{E} }{ \delta A_{\mu} } )A_{\mu} g_{\rho\sigma}
+ \cdots 
\end{eqnarray}
Note that due to minimal coupling, the gauge field $A_\mu$ occurs at most with 
a single power in the above expansion.  
The partition function of the theory is given by 
\begin{equation}\label{part}
 Z_{E}  = \int {\cal D}{\psi^\dagger}{ \cal D}  \psi\exp ( S_{E} ). 
 \end{equation} 
 Here $\psi$ refers to the  Fermion. 
Then the expectation value of  the current and the stress tensor are defined by 
\begin{eqnarray}
\langle j^\mu \rangle_E  &=&  
-\left. \frac{1}{\sqrt{g}}\frac{\delta \ln {\cal Z}_E}{\delta A_\mu}\right|_{a_\mu, h_{\mu\nu} = 0} 
= -\frac{1}{ {\cal Z }_E^{(0)} } \int {\cal D}{\psi^\dagger}{ \cal D} \psi e^{  S_{E}^{(0)} }
\left. \frac{1}{\sqrt{g}}\frac{\delta S_E}{\delta A_\mu}\right |_{a_\mu, h_{\mu\nu}  =0}, 
\\ \nonumber
\langle T^{\mu\nu} \rangle_E &=&  - 2 
\left. \frac{1}{\sqrt{g}}\frac{\delta \ln{\cal Z} _E}{\delta g_{\mu\nu}} \right |_{a_\mu, h_{\mu\nu} = 0}
= -\frac{2}{ {\cal Z}_E^{(0)}  } \int {\cal D}{\psi^\dagger}{ \cal D} \psi e^{ S_{E}^{(0)} } 
\left. \frac{1}{\sqrt{g}}\frac{ \delta S_{E} } {\delta g_{\mu \nu} }  g_{\mu \nu}\right |_{a_\mu, h_{\mu\nu}  =0},
\end{eqnarray}
where  ${\cal Z }^{(0)}$ refers to the partition  function in the absence of any perturbation.
To un-clutter notations,  from now all derivatives in the fields are understood to be evaluated 
at  $ a_\mu, h_{\mu\nu} = 0$ .
The  relevant two point functions  in the Euclidean theory are  given by 
\begin{eqnarray}
\langle j^\mu T^{\rho\sigma} \rangle_E &=&  
2 \frac{1}{\sqrt{g}}\frac{\delta}{\delta g_{\rho \sigma}}(\frac{1}{\sqrt{g}}\frac{\delta  \ln{\cal Z}_E }{  \delta A_{\mu}}).
\end{eqnarray}
Substituting the expansion of the action given in (\ref{actexp}) into the definition of the
partition function we obtain
\begin{eqnarray} \label{jt1}
\langle j^\mu T^{\rho\sigma} \rangle_E 
&=& - \langle j^\mu \rangle_E \langle T^{\rho \sigma} \rangle_E 
+  \frac{2}{{ \cal Z}_E^{(0)} } \int {\cal D}{\psi^\dagger}{ \cal D} \psi e^{  S_{E}^{(0)} }
\frac{1}{\sqrt{g}}\frac{\delta S_E}{\delta A_\mu} \frac{1}{\sqrt{g}}\frac{ \delta S_{E} } {\delta g_{\rho \sigma}}  \\ \nonumber
&&+  \frac{2}{ { \cal Z}_E^{(0)} }  \int {\cal D}{\psi^\dagger}{ \cal D} \psi e^{ S_{E}^{(0)} } 
\frac{1}{\sqrt{g}}\frac{\delta}{\delta g_{\rho \sigma}}(\frac{1}{\sqrt{g}}\frac{\delta  S_E }{  \delta A_{\mu}}). 
\end{eqnarray}
Note the disconnected  contribution of the  first term cancels the disconnected 
contributions from the second term, in the first line of the above equation. 
The non-zero contribution of the correlator arises from the connected contribution 
of the second term and the from the last term. Since $S_E$ is a bilinear in the 
fermions, the  third term is a contact term.
We will see that such terms contribute crucially to the correlator and it is only after 
adding these terms, the  `replacement rule' for  chiral  fermions can be demonstrated. 
The structure of these terms in $d=2$ is different from that in higher dimensions. 
On performing the same manipulations for correlator corresponding 
to $\lambda^{(2)}$, we obtain 
\begin{eqnarray} \label{tt1}
\langle T^{\mu\nu} T^{\rho\sigma} \rangle_E   &= & 
 \frac{1}{\sqrt{g}}\frac{\delta}{\delta g_{\rho \sigma}}(\frac{1}{\sqrt{g}}\frac{\delta  \ln{\cal Z}_E }{  \delta g_{\mu \nu}}), \\ \nonumber
&=& - \langle T^{\mu\nu} \rangle_E \langle T^{\rho \sigma} \rangle_E 
+  \frac{4}{{ \cal Z}_E^{(0)} } \int {\cal D}{\psi^\dagger}{ \cal D} \psi e^{  S_{E}^{(0)} }
\frac{1}{\sqrt{g}}\frac{\delta S_E}{\delta g_{\mu\nu} } \frac{1}{\sqrt{g}}\frac{ \delta S_{E} } {\delta g_{\rho \sigma}}  \\ \nonumber
&&+ \frac{4}{ { \cal Z}_E^{(0)} }  \int {\cal D}{\psi^\dagger}{ \cal D} \psi e^{ S_{E}^{(0)} } 
\frac{1}{\sqrt{g}}\frac{\delta}{\delta g_{\rho \sigma}}(\frac{1}{\sqrt{g}}\frac{\delta  S_E }{  \delta g_{\mu \nu}}).
\end{eqnarray}
Here again the disconnected terms  cancel  and we need to examine only the connected
diagrams to evaluate the above correlator. 

After evaluating the Euclidean two point functions, 
 we perform the following analytic continuation to  obtain the retarded correlator
\begin{eqnarray}\label{merel}
\langle j^\tau (\omega_n, p)   T^{\tau x} (-\omega_n, -p)   \rangle_E =-\langle j^t (\omega, p)   T^{tx} (-\omega, -p)   \rangle_R |_{i\omega_n \rightarrow p_0 + i\epsilon}. 
\end{eqnarray}
Here $\omega_n  = 2\pi n T$, where $T$ is the temperature  and $n \in Z$ refer to the Matsubara frequencies. Note that all external Matsubara frequencies are even multiples of $\pi T$ and 
will be labelled by $n$. 
The overall negative sign results from converting the tensor indices containing 
the Euclidean time $\tau $ to Minkowski time $t$ using the replacement $\tau  \rightarrow i t$. 
Similarly we have 
\begin{eqnarray} \label{merel2}
\langle T^{\tau\tau} (\omega_n, p)   
T^{\tau x} (-\omega_n, -p)   \rangle_E =-i\langle T^{tt} (\omega, p)   T^{tx} (-\omega, -p)   \rangle_R |_{i\omega_n \rightarrow p_0 + i\epsilon} 
\end{eqnarray}
Finally to obtain the transport coefficients, we use the Kubo formula in (\ref{2dk1}) and (\ref{2dk2}). 

Let us now proceed to implement the above procedure in detail. 
The first step is to write down the expansion of the Euclidean action of chiral  
fermions in  background  gauge field and metric. 
The action is given by 
\begin{equation}\label{fcact}
S_E = \int d\tau dx  \sqrt{g} e^{\mu}_a \psi^\dagger \gamma^a  D_\mu  P_{-}  \psi .
\end{equation}
The gamma matrices  convention, we choose to work with, 
are given by 
\begin{equation}
\gamma^\tau = i  \sigma^1, \qquad \gamma^x = i \sigma^2, 
\qquad \gamma_c  = - i \gamma^\tau \gamma^x = - \sigma^3.
\end{equation}
The chiral  projection operators are  defined by 
\begin{equation}
P_{\mp} = \frac{1}{2} ( 1\mp \gamma_c).
\end{equation}
Note that the Euclidean  gamma matrices are anti-Hermitian, $\gamma_c$ is Hermitian and 
$P_-$ projects on to the upper component of spinor, while $P_+$ retains the 
lower component of the spinor.
The gamma matrices obey the algebra
\begin{equation}
\{ \gamma^a, \gamma^b\} = 2 \eta^{ab}_E, \qquad \eta^{ab}_E = - \delta^{ab}, 
\end{equation}
$e^\mu_a$ is the vierbien and  
the covariant derivative $D_\mu$  is defined  by
\begin{equation}
D_\mu =   \partial_\mu   +\frac{1}{2} \omega_{\mu c d} \sigma^{cd} + i A_\mu  , \qquad
\sigma^{cd} = \frac{i}{4} [\gamma^c \gamma^d]   
\end{equation}
In two dimensions, the term involving the 
 spin connection $\omega_{\mu c d}$ vanishes and we do not consider it for the rest of this 
 section.  We will see in the subsequent sections  that the  contact terms in higher dimensions arise from 
 linearizing the spin connection. In $d=2$  the source of the contact terms 
 arise from  the linearization 
 of the metric and the vierbein.

 We proceed to expand the action  (\ref{fcact}) in terms of the following  perturbations in the 
 metric and the gauge field
 \begin{eqnarray}
 g_{\tau \tau } =  -1 + h_{\tau\tau}, \qquad g_{\tau x} = g_{x\tau} = h_{\tau x}, \qquad g_{xx} = -1, 
 \\ \nonumber
 A_\tau =   A_\tau ^{(0)} + a_\tau, \qquad A_x =0.
 \end{eqnarray}
 Here $A_\tau^{(0)}$ refers to the  constant
  background  chemical potential, which will be turned on subsequently. 
 The inverse metric, to the second order in the perturbations is given by
 \begin{eqnarray}\label{mexp}
g^{\mu \nu } =  \left( 
\begin{array}{ll} 
-1-h_{\tau\tau}-h_{\tau\tau}^2-h_{\tau x}^2 & \;\;\;\;-h_{\tau x}-h_{\tau\tau}h_{\tau x}\\
-h_{\tau x}-h_{\tau\tau}h_{\tau x} & \;\;\;-\;1-h_{\tau x}^2 
\end{array}  \right) + O(h^3). 
 \end{eqnarray}
 Using the gauge in \cite{AlvarezGaume:1983ig} the vierbein is given by 
 \begin{eqnarray} \label{vexp1}
e_{a\mu } = 
\begin{pmatrix}
-1+\frac{h_{\tau\tau}}{2}+\frac{1}{8}(h_{\tau x}^2+h_{\tau\tau}^2) && \frac{h_{\tau x}}{2}+\frac{h_{\tau\tau}h_{\tau x}}{8}\\
\\
\frac{h_{\tau x}}{2}+\frac{h_{\tau\tau}h_{\tau x}}{8} && -1+\frac{h_{\tau x}^2}{8} \\
\end{pmatrix} + O(h^3). 
\end{eqnarray}
Finally the inverse vierbein is given by 
\begin{eqnarray}\label{vexp2}
e^{\tau}_{\hat \tau} &=& 1+\frac{h_{\tau\tau}}{2}+\frac{3}{8}(h_{\tau x}^2+h_{\tau\tau}^2) +O(h^3) , \qquad
e^{\tau}_{\hat x} =  \frac{h_{\tau x}}{2}+\frac{3}{8}(h_{\tau x}h_{\tau \tau}) + O(h^3) \nonumber\\
e^{x}_{\hat \tau} &=& \frac{h_{\tau x}}{2}+\frac{3}{8}(h_{\tau x}h_{\tau\tau}) +O(h^3), \qquad 
e^{x}_{\hat x }= 1+\frac{3}{8}h_{\tau x}^2 +O(h^3) .
\end{eqnarray}
where the hatted variables refer to the flat space co-ordinates. 
We now substitute these expansions in (\ref{fcact}) and obtain the following action, 
to the leading orders in the perturbations.
\begin{eqnarray}  
S_E &=& S_E^{(1)} + S_E^{(2)} , \nonumber \\  \label{lexpansion1}
S_{E}^{(1)} &=& \int d^2x \sqrt{g}\left[ \psi^\dagger(x)\gamma^\tau D_\tau^{(0)} P_- \psi(x)+\psi^\dagger(x)\gamma^x\partial_x P_-\psi(x) \right]  \\
&&-\frac{h_{\tau\tau}}{2} \psi^\dagger(x)\gamma^\tau D^{\tau(0)} P_-\psi(x)-
\frac{h_{\tau x}}{2}\psi^\dagger(x)(\gamma^x D^{\tau (0)} + \gamma^\tau \partial^x) P_-\psi(x)\nonumber\\
&& \left[ -\frac{3 h_{\tau\tau}h_{\tau x}}{8}(\psi^\dagger(x) \gamma^x D^{\tau(0)}  P_-  \psi(x)+
\psi^\dagger(x)\gamma^\tau \partial^x  P_-\psi(x)) \right] + \cdots \nonumber\\
\label{lexpansion2}
S_{E}^{(2)} &=& ie\int d^2x \sqrt{g}\left[ ( a_\tau  \psi^\dagger(x)\gamma^\tau P_- \psi(x)
+\frac{h_{\tau x}}{2} a_\tau \psi^\dagger(x)\gamma^x P_- \psi(x) \right] + \cdots 
\end{eqnarray}
where
\begin{equation}
D_\tau^{(0)} = \partial_\tau +i  e A_\tau^{(0)},
\end{equation}
is the covariant derivative in the background chemical potential. 
In (\ref{lexpansion1}) and (\ref{lexpansion1}),  we  have retained only the terms relevant to obtain the correlators of interest. 
Note also, we have used the on-shell condition to simplify  the coefficient of $h_{\tau\tau}$. 
The components of the 
 flat space stress tensor of interest  is then given by 
\begin{equation}\label{stress}
T^{\tau\tau}_{fl}(x)=  \psi^\dagger \gamma^\tau D^{\tau(0)} P_-\psi, \qquad
T^{\tau x}_{fl}( x) =  \frac{1}{2} \psi^\dagger(  \gamma^\tau \partial^x  + \gamma^x ( D^{\tau(0)}  ) P_-\psi .
\end{equation}
Similarly the time component of the  flat space charge current   is given by 
\begin{equation}
j^\tau_{fl} =  - i e \psi^\dagger \gamma^\tau P_- \psi .
\end{equation}
Note that the quadratic terms involving the perturbations in 
(\ref{lexpansion1}) and (\ref{lexpansion2}) are responsible
for contact terms in the two point functions of interest.
These terms contribute in the correlators for $d=2$ because the one needs only $2$ gamma 
matrices to obtain a non-zero  trace along with $\gamma_c$.  
 As  a simple cross  check on these  terms  we 
  considered the stress tensor in  curved space and expanded it  to first order in metric perturbations. 
  The  first derivative of the curved space stress tensor agrees with the second derivative
  of the action with respective to the metric obtained from  \ref{lexpansion1} and \ref{lexpansion2}.

The last ingredient we require to perform the computation of the two point function, is the 
propagator for the  free chiral fermions. From the  first term in the action 
given in (\ref{lexpansion1}),  the Euclidean propagator  at finite 
temperature and finite chemical potential  in momentum space  is given by 
\begin{eqnarray}
\langle  \psi (\omega_m, p ) \psi^\dagger(\omega_{m'} , p') \rangle &=&
S(\omega_m, p) \delta_{m, m'}\delta(p-p') 2\pi ,
\end{eqnarray}
and 
\begin{eqnarray}\label{propagatorf}
S(\omega_m , p )&=&-i{\left[ \gamma^\tau \left(\omega_m -ei\mu -ei\mu _5 \gamma_c\right)+\gamma^x p\right]^{-1} },\nonumber\\
&=&  \left( \begin{array}{cc}
0 & \frac{-i}{i\omega_m+e(\mu -\mu _c)-p}\\
\frac{-i}{i\omega_m+e(\mu +\mu _c)+p } & 0 
\end{array} \right).
\end{eqnarray}
Here $\omega_m = ( 2m + 1) \pi T$ is the Fermionic Matsubara frequency at temperature $T$. 
The label $m$ will refer to Matsubara frequencies which occur as odd multiples of $\pi T$. 
We have chosen the background chemical potential to be 
\begin{equation}
A_\tau^{(0)} = i\mu + i\mu_c \gamma_c,
\end{equation}
where
$\mu$ is the chemical potential corresponding to the vector $U(1)$ and $\mu_c$ is the
chemical potential corresponding to the axial $U(1)$ current \footnote{The  signs for the 
chemical potentials are fixed 
using the  thermodynamic definition $\frac{\partial\ln{\cal Z} }{\partial \mu} =  j^0$. This allows us to 
identify  the Euclidean gauge field to be $A_\tau^E = i \mu$. }. 
 Note that the transformation to the 
momentum space is given by 
\begin{eqnarray}
\psi(x) =  \frac{1}{\beta}\sum_{\omega_m} \int \frac{dp}{2\pi}  \psi(p) e^{ - i (\omega_m \tau +  p x)} , 
\\ \nonumber
\psi^\dagger(x) = \frac{1}{\beta}\sum_{\omega_m} \int \frac{dp}{2\pi} 
 \psi^\dagger(p) e^{  i (\omega_m \tau +  p x)}.
\end{eqnarray}
This explains the relative sign between the frequency and the chemical potential in
the propagator.

\subsection*{$\zeta^{(2)}$ from chiral fermions}
\begin{figure}[!htb]\label{jt2diag}
  \centering
   \hspace*{-2cm}\includegraphics[scale=0.2]{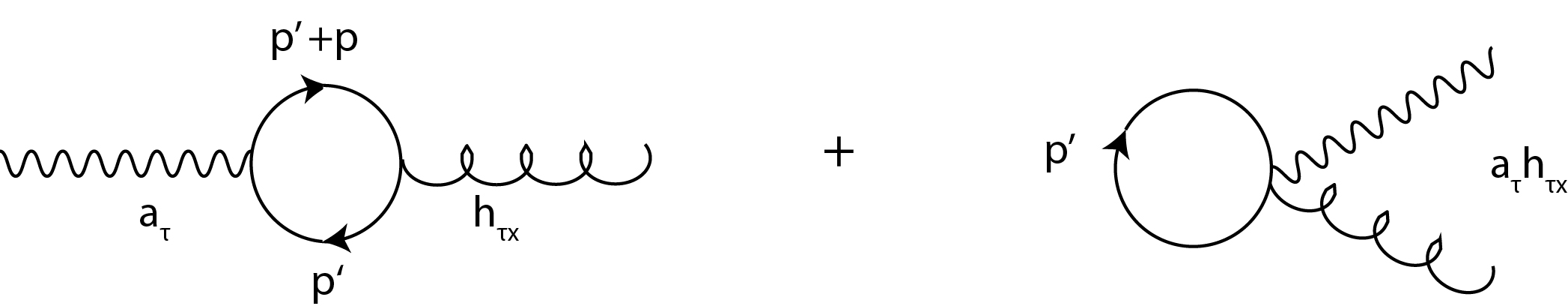} 
   \caption{Diagrams contributing to  $\zeta^{(2)}$ }
\end{figure}   

Let us now evaluate the Euclidean correlator $\langle j^\tau T^{\tau x} \rangle$. 
From (\ref{jt1})  and (\ref{lexpansion2}), we see that there are two contributions. They are given by 
\begin{eqnarray} \label{jt2}
\langle j^\tau (p) T^{\tau x}(-p) \rangle & = & A + B , \\ \nonumber
A =  \langle j^{\tau}_{fl} (p) T^{\tau x}_{fl} (-p) \rangle_c , &\qquad& 
B =   \frac{ie}{2} \int \frac{d^2 p'}{(2\pi)^2}
  \langle \psi^\dagger ( p') \gamma^x P_-  \psi(p')  \rangle.
\end{eqnarray}
Here,  for convenience of notation, we have used $p$ to refer to $(p_0, p_x)  = ( \omega_m,  p_x) $
where $\omega_m$ is the corresponding Matsubara frequency.  The integral in the 
second term  $B$  stands for
\begin{equation}
\int\frac{d^2 p}{(2\pi)^2}   \rightarrow \frac{1}{\beta} \sum_{n} \int \frac{d p_x}{2\pi} 
\end{equation}
Note that the term $B$ arises from the  second contribution in (\ref{jt1}).  
The Fourier transform of the stress tensor and the current, in flat space, are given by 
\begin{eqnarray} \label{momstress}
j^\tau_{fl}(p) &=& - i e \int \frac{d^2 p'}{(2\pi)^2} \psi^\dagger(p' - p ) \gamma^\tau P_- \psi(p') , 
\\ \nonumber
T^{\tau x}_{fl} ( - p )  &=& \frac{1}{2}  \int\frac{d^2 p'}{(2\pi)^2} 
\psi^\dagger( p'+p) \left[ \gamma^\tau i p_x'  + \gamma^x( i p_0 '+  e( \mu + \mu_c \gamma_c) ) \right] P_-  \psi( p'). 
 \end{eqnarray}
 Substituting these expressions in (\ref{jt2}) and performing the necessary Wick contractions, 
 we obtain for term $A$
 \begin{eqnarray}\label{a1}
 A &=&   \frac{ie}{2} \int \frac{d^2 p'}{(2\pi)^2} {\rm Tr} \left[
 \gamma^\tau S(p' + p ) \left( \gamma^\tau i p_x'  + \gamma^x( i p_0 '+  e( \mu + \mu_c \gamma_c) ) \right)S(p')P_+ \right], \\ \nonumber
 &=&-  \sum_{m} \frac{e}{2\beta} \int \frac{dp_x'}{2\pi} \left( 
 \frac{ p'_x + i\omega_m + e \mu_-}{ (  i\omega_m  + i \omega_n + e\mu_-  - ( p_x'+p_x) ) ( i\omega_m + e \mu_- - p_x')  } \right),
 \end{eqnarray}
 where 
 \begin{equation}
  \mu_- = \mu - \mu_c.
 \end{equation}
 Note that there is an over all  $2\pi \beta \delta(0) $ 
 resulting from momentum conservation, which we have factored out. 
We perform the sum over Matsubara  Fermionic frequencies by the usual trick of converting the sum to a contour integral. 
Here are the results for the sum
\begin{eqnarray}
& & \sum_{m} \frac{1}{\beta}  
\left[ {( i \omega_m + e\mu_-  -p_x') ) ( i \omega_m + e\mu_- + i \omega_n - p_x' - p_x) } \right]^{-1} 
\\ \nonumber
 & & =  \frac{  f ( p_x'- e\mu_-)  - f( p_x' +p_x  - \mu_- - i \omega_n )  }
 { i \omega_n - p_x  } , \\ \nonumber
 & & \sum_{m} \frac{1}{\beta} \frac{ i \omega_m + e\mu_-}
 {( i \omega_m + e\mu_-  -p_x' ) ( i \omega_m + e\mu_- + i \omega_n - p_x' - p_x) }  \\ \nonumber
 & & = 
   \frac{  p_x'   f ( p_x'- e\mu_-) - ( p_x + p_x' - i \omega_n) f(  p_x + p_x' - i \omega_n) }{  i \omega_n - p_x}  
 - \frac{1}{2},
\end{eqnarray}
and $f$ is the Fermi-Dirac distribution given by 
\begin{equation}
 f( x) = \frac{1}{ 1 + e ^{\beta x} }.
\end{equation}
Note that the temperature, chemical potential independent constant $1/2$ in the second sum can be ignored, since
it just results in infinite constant. 
Substituting the results for these sums into the expression for $A$ in (\ref{a1}), we obtain
\begin{eqnarray}
 A =  \frac{e}{2} \int_{-\infty}^{\infty} \frac{dp_x'}{2\pi} \left( \frac{ 2p_x'  f( p_x' - e\mu_-) }{ p_x - i \omega_n } 
 - \frac{  (p_x + 2 p_x' - i \omega_n) f(  p_x + p_x' - i \omega_n)}{ p_x - i\omega_n } \right) .
\end{eqnarray}
We can now change variables in the integration for  the 2nd term and combine both the terms as \footnote{Though there is 
a divergence in the integrals when $p_x' \rightarrow -\infty$ it can be shown that this infinite constant
is independent of temperature or chemical potentials. The change of variables can be justified 
on careful treatment of the integrals when $p_x'<0$. }
\begin{equation} \label{a2}
 A =  \frac{e}{2}\frac{ p_x + i \omega_n}{ p_x - i \omega_n}  \int \frac{dp_x'}{2\pi}  f ( p_x' - e\mu_-). 
\end{equation}
Let us now examine the term $B$, again performing the required Wick contraction we obtain 
\begin{eqnarray}
 B &=& - i  \frac{e}{2}\int \frac{d^2p'}{(2\pi)^2 }  {\rm Tr} ( \gamma^x S(p'  ) P_+ ), \\ \nonumber   
&=& \frac{e}{2\beta} \sum_{m}  \int \frac{dp_x'}{2\pi} \frac{1}{ i\omega_m   + e\mu_- - p_x' } .
 \end{eqnarray}
 After performing the Matsubara sum we obtain 
 \begin{equation} \label{b1}
  B = \frac{e}{2} \int_{-\infty}^\infty \frac{dp_x'}{2\pi} f( p_x' - e\mu_-). 
 \end{equation}
Adding the contributions in (\ref{a2}) and (\ref{b1}) the $\langle j^\tau T^{\tau x} \rangle$ correlator is 
given by 
\begin{eqnarray} \label{jtchifi}
\langle j^\tau (p) T^{\tau x}(-p) \rangle &=&  e \frac{ p_x}{ p_x - i \omega_n}  \int_{-\infty}^\infty
 dp_x' f (p_x' - e\mu_-), \\ \nonumber
 &=&  e \frac{ p_x}{ p_x - i \omega_n}  \int_{0}^\infty \frac{dp_x'}{2\pi} ( f( p_x' - e\mu_-)   - f( p_x'  + e\mu_-) ), \\ \nonumber
 &=&   \frac{ p_x}{ p_x - i \omega_n}  \frac{e^2\mu_-}{2\pi}.
 \end{eqnarray}
 To obtain the  second line of the above equation, we have again ignored an infinite constant,
 which is independent of temperature and chemical potential. 
 The last line  is obtained using 
  the results of appendix A,  equation (\ref{int1}),   to perform the integral over the Fermi-Dirac 
 distribution. 
 We can obtain the retarded correlator using  (\ref{merel})
 \begin{equation}\label{jt3}
 \langle j^t (p_0, p_x) T^{t x}(-p_0, -p_x) \rangle_R = -  \frac{ p_x}{ p_x - p_0   }\frac{e^2\mu_-}{2\pi}.
  \end{equation}
  Here $(p_0, p_x)$ refer to the frequency and momentum in Minkowski space. 
  We have ignored the $i\epsilon$ that results from using (\ref{merel}), which is required when 
  one needs to transform this correlator in Fourier space to a real time correlator. 
  Finally using (\ref{2dk1}), the result for the transport coefficient is obtained by first taking the 
  zero frequency limit $p_0 \rightarrow 0$ and then the zero momentum limit $p_x\rightarrow 0$. 
  This results in 
  \begin{eqnarray}\label{valz2}
  \zeta^{(2)} &=&  - \lim_{p_x\rightarrow 0, p_0 \rightarrow 0} \langle j^t (p_0, p_x) T^{t x}(-p_0, -p_x) \rangle_R, \\ \nonumber
  &=& \frac{e^2\mu_-}{2\pi}.
  \end{eqnarray}
  Examining (\ref{jt3}), we see that it was crucial to take the zero frequency limit first and then 
  the zero momentum limit.   Note that both term $A$ and  the `contact' term 
  $B$ in (\ref{jt2}) contribute and they 
  contribute equally in the zero frequency and zero momentum limit.  Also note that in 
  $d=2$ term $B$ contributed since one needed only $2$ gamma matrices to saturate the trace. 
  If once considers the contribution to this transport coefficient from the anti-chiral Fermions
  with chemical potential, then  through a similar analysis one obtains
  \begin{eqnarray}\label{valz22}
  \zeta^{(2)}_{anti-chiral}  &=&   - \frac{e^2\mu_+}{2\pi},
  \end{eqnarray}
  where
  \begin{eqnarray}
  \mu_+ = \mu+\mu_c.
  \end{eqnarray}
  From (\ref{valz2}) and (\ref{valz22}), we see that the transport coefficient vanishes for 
  a Dirac fermion in the absence of the chiral chemical potential $\mu_c$. 
  
  Let us now compare  the transport coefficient for chiral fermions in (\ref{valz2}), 
  with the coefficient of the $U(1)$ anomaly. 
  The anomaly equation satisfied by the chiral current 
  $ j^\mu = e\bar \psi \gamma^\mu P_- \psi$, is given by 
  \begin{equation} \label{2dano}
  \partial_\mu j^\mu =   c_s \epsilon^{\mu\nu} F_{\mu\nu}, \qquad c_s = - \frac{e^2}{4\pi}.
  \end{equation}
  The value of $c_s$ for a Weyl Fermion  in 2 dimensions can be read out at many places in the literature, 
  see for instance equation (2.27) of  \cite{Jensen:2013rga}. 
  We now compare the coefficient of the anomaly and the transport coefficient to 
  see that they are related by 
  \begin{equation}
  \zeta^{(2)} = - 2c_s \mu_-.
  \end{equation}
  which is the `replacement rule' obtained in \cite{Valle:2012em,Jain:2012rh,Jensen:2012kj}. 
 Observe that we have arrived at this relation between the transport coefficient and the microscopic anomaly 
  from explicitly evaluating the diagrams that contribute to the Kubo formula. 
  Note that the `contact' terms gave rise to an equal contribution.
 
  For later purpose it is useful to parametrize 
   the anomaly coefficients with the help of the  anomaly polynomial. 
The anomalies of a theory in $d$ dimensions, are encoded in a 
$d+2$ form constructed out of gauge field strengths $\hat{F}$ and curvature tensor $\hat{R}_{ab}$. 
We define the   curvature two-form $\hat{R}_{ab}$ and the field strength two-form as
\begin{eqnarray}
\hat{R}_{ab} = \frac{1}{2} R_{abcd} dx^c \wedge dx^d, \qquad \hat{F}=\frac{1}{2} F_{ab} dx^a \wedge dx^b.
\end{eqnarray}
Furthermore, let us define a $2k$-form which is a k-th degree polynomial in the curvature 2-forms
\begin{eqnarray}
\hat{R}_k \equiv \frac{1}{2} \hat{R}^{a_1}_{a_2} \wedge \hat{R}^{a_2}_{a_3} \wedge \cdots \hat{R}^{a_k}_{a_1}. 
\end{eqnarray}
The anomaly polynomial 
for a single Weyl fermion in  $d=2$  takes the form \citep{Jensen:2012kj}
\begin{eqnarray}\label{anompoly2d}
{\cal P}_{d=1+1}(\hat{F},\hat{R}) &=& c_s F\wedge F +c_g {\rm tr}(\hat{R} \wedge \hat{R}),
\end{eqnarray}
where,
\begin{eqnarray}
c_s = -\frac{e^2}{4\pi}, \qquad c_g = -\frac{1}{96\pi}.  
\end{eqnarray}

\subsection*{$\lambda^{(2)}$ from chiral fermions}
\begin{figure}[!htb]\label{tt2diag}
  \centering
   \hspace*{-2cm}\includegraphics[scale=0.2]{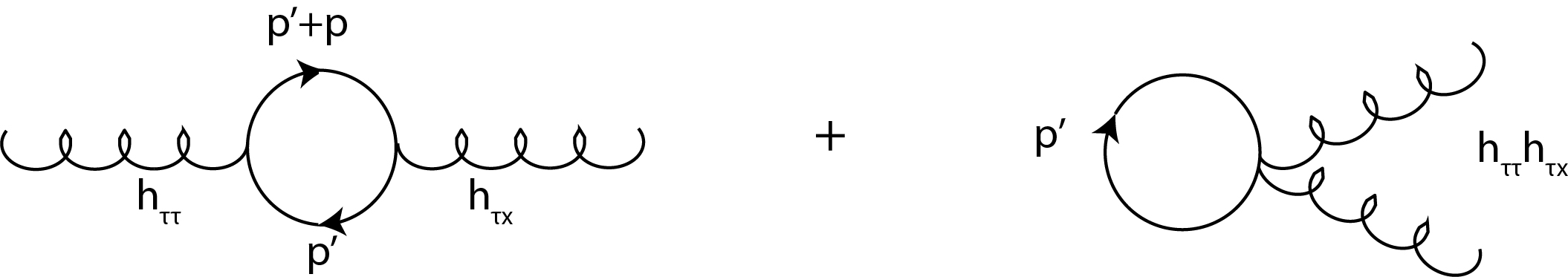} 
   \caption{Diagrams contributing to  $\lambda^{(2)}$ }
\end{figure}   

We evaluate   the correlator $\langle T^{\tau \tau} T^{\tau x}\rangle$. 
Again from (\ref{tt1}) and (\ref{lexpansion1}), we see that  this correlator receives contributions from   two terms which are 
given  by 
\begin{eqnarray}
& &  \langle T^{\tau \tau} (p) T^{\tau x} (-p) \rangle =  A + B, \\ \nonumber
 & & A= 
 \langle T^{\tau \tau}_{fl} (p)  T^{\tau x}_{fl} (-p) \rangle_c  , \qquad 
 B = - \frac{6}{8} \langle  \int \frac{d^2 p}{(2\pi)^2 } \psi^{\dagger} \left[
 \gamma^x ( i p_\tau  + ( \mu + \mu_c \gamma_c)  )  + \gamma^\tau i p_x   \right] P_{-} \psi  \rangle.
\end{eqnarray}
To evaluate the term $A$ we need  
\begin{equation}
 T^{\tau\tau}_{fl} (p) = \int \frac{d^2 p' }{(2\pi)^2} \psi^{\dagger} ( p' - p) \gamma^\tau ( i p_\tau + e (\mu + \mu_c \gamma_c) ) 
 P_{-} \psi(p) .
\end{equation}
Inserting this and the expression for $T^{\tau x}(-p)$ from (\ref{momstress}), and  performing the Wick 
contractions using the propagator in (\ref{propagatorf}), we obtain 
\begin{eqnarray}
 A (\omega_n, p_x)  &=& - \frac{1}{2 \beta } \sum_{m}  \int \frac{dp'}{2\pi} {\rm Tr}  \left( 
 \gamma^\tau [ i (\omega_n  + \omega_m )  + e ( \mu + \mu_c \gamma_c)  ]  \right.  \\ \nonumber
 & & \qquad\left. \times  S( \omega_n  + \omega_m,  p_x + p_x' ) 
 [ i  \gamma^\tau p_x'  + \gamma^x ( i \omega_m + e ( \mu + \mu_c \gamma_c) ] 
 S(\omega_m, p_x' ) P_+  \right) . 
\end{eqnarray}
Substituting  the expression for the propagator,  performing the trace  and the 
Matsubara sum results in 
\begin{eqnarray}
\lim_{p_x \rightarrow 0} A(0, p_x)   &=&  \frac{i}{2}  \int_0^\infty 
\frac{dp_x'}{2\pi}  p_x'   (  f ( p_x' - e\mu_-)  + f( p_x' + e\mu_-) ) , \\ \nonumber
&=&   \frac{i}{8\pi} 
\left( e^2 \mu_-^2 + \frac{\pi^2 T^2}{3} \right) .
\end{eqnarray}
In the last line we have used the identity  (\ref{int3}) to perform the integral. 
Evaluating $B$ using the  propagator given in (\ref{propagatorf}), we obtain
\begin{eqnarray}
 B &=&  \frac{6}{8 \beta } \sum_n \int \frac{d p_x }{ 2\pi} {\rm Tr} 
 \left[ \gamma^x ( i \omega_n  +( \mu + \mu_c \gamma_c)  )  + \gamma^\tau i p_x S(\omega_n ,  p_x )
 \right] , \\ \nonumber
 & =&  \frac{3 i }{2 } \int_0^\infty \frac{dp_x}{2\pi}  p_x\left(  f(  p_x - e\mu_-)  + 
 f ( p_x + e\mu_-) \right) , \\ \nonumber
 & =& \frac{3i }{8 \pi  } \left( e^2 \mu_-^2 + \frac{\pi^2 T^2}{3} \right) .
\end{eqnarray}
In the second line we have performed the resulting Matsubara sum,
 after which we have used (\ref{int3}) to perform the resulting integral. 
 Finally adding the contributions from terms $A$ and $B$  results in 
 \begin{equation} \label{euclid2dl}
  \lim_{p_x \rightarrow 0} \langle T^{\tau \tau} (0, p_x ) T^{\tau x} (0, -p_x ) \rangle 
  =  
  \frac{i }{2 \pi  } \left( e^2 \mu_-^2 + \frac{\pi^2 T^2}{3} \right). 
 \end{equation}
 Using  (\ref{merel2}),  the retarded correlator in Minkowski space is given by 
 \begin{equation}
 \lim_{p_x \rightarrow 0} \langle T^{tt} (0 , p_x) T^{tx} (0 , - p_x ) \rangle 
 = -   \frac{1 }{2 \pi  } \left( e^2 \mu_-^2 + \frac{\pi^2 T^2}{3} \right). 
 \end{equation}
 Substituting this result in the expression (\ref{2dk2}), the transport coefficient of interest is
 given by 
 \begin{equation}\label{flambda2}
 \lambda^{(2)}  =  \frac{1}{4\pi} \left( e^2 \mu_-^2 + \frac{\pi^2 T^2}{3} \right). 
 \end{equation}

 We now compare this transport coefficient with the coefficient of the gravitational anomaly
 of the Weyl Fermion, defined by
 the conservation law 
 \begin{equation}
  \nabla_\mu T^{\mu\nu}  = F^\mu_\nu J^\nu  + c_g \epsilon^{\mu\nu} \nabla_\nu R, 
  \qquad  c_g =  -  \frac{1}{96\pi}.
 \end{equation}
The coefficient $c_g$ for a Weyl  Fermion  is obtained from \eqref{anompoly2d}. 
In \cite{Jensen:2013rga}, the transport coefficient $\lambda^{(2)}$ was parametrized as 
\begin{equation}
  \lambda^{(2)} = \tilde c_{2d} T^2 - c_s \mu_-^2. 
\end{equation}
Comparing with the result from  evaluation of the Kubo formula  given in (\ref{flambda2}), we see 
\begin{equation}
 c_s = - \frac{e^2}{4\pi}, \qquad \tilde c_{2d} = \frac{\pi}{12}.
\end{equation}
The value for $c_s$ agrees with the chiral anomaly of a Weyl fermion in 2 dimensions as in (\ref{2dano}). 
The value for $\tilde c_{2d}$ obeys the relation 
\begin{equation}
 \tilde c_{2d}=  - 8\pi^2 c_g. 
\end{equation}
as argued in \cite{Jensen:2013rga} using the consistency of the Euclidean vacuum. 
We have arrived at these relations by explicitly evaluating 
the diagrams which contribute to the Kubo Formula.  Note that for both transport  coefficients, there was 
a  contribution from the `contact'  term which was important in obtaining these relations. 

\subsection{Chiral bosons}

In 2 dimensions a massless chiral Weyl fermion is dual to a chiral boson. Therefore we expect the contributions 
of a chiral boson to the transport coefficient $\zeta^{(2)}$ and $\lambda^{(2)}$, to be identical to that of the 
chiral fermions studied in section \ref{chifermi}. 
In this section we will show that it is indeed the case. Though the result is expected on general grounds,  performing the 
calculation in detail  will serve as a cross  check of the analysis in 
 section \ref{chifermi}. Also  as far as we are aware transport coefficients have never been 
evaluated directly for chiral bosons. With these motivations we proceed with the analysis of transport 
in this system. 

Chiral bosons in the Euclidean theory obey the constraint \footnote{ 
$\epsilon^{\tau x} = \epsilon_{\tau x} = 1$ and is anti symmetric in its indices.   }
\begin{equation} \label{chicons}
 \partial^\mu \phi = i \frac{\epsilon^{\mu\nu} }{\sqrt{g} } \partial_\nu \phi.
\end{equation}
Note that this is the constraint which in flat space reduces to 
$\partial_{\bar z } \phi  =0$ on using  the change of 
variables to holomorphic
and anti-holomorphic coordinates given in (\ref{holanti}). 
The constraint ensures that the bosons depend only  $z$, this property is 
is  identical to  the 
Weyl fermions considered in section \ref{chifermi}. 
To write down the stress tensor and the current of the chiral boson theory, we impose the constraint
of (\ref{chicons}) in the stress tensor of the boson and the current following \cite{AlvarezGaume:1983ig}.  
They are given by 
\begin{eqnarray}\label{chbocu}
 T^{\mu\nu} &=&  \frac{-1}{8}\left( \partial^\mu  + i  \frac{\epsilon^{\mu\rho}}{\sqrt{g} }\partial_\rho\phi \right) 
 \left( \partial^\nu + i \frac{ \epsilon^{\nu\sigma}}{\sqrt{g}}  \partial_\sigma\phi \right) , \\ \nonumber
 j^\mu &=& i e_b \frac{1}{2} \left( \partial^\mu + i \frac{\epsilon^{\mu\rho} }{\sqrt{g}} \partial_\rho \phi \right),
\end{eqnarray}
where $e_b$ is the charge of the boson. The normalization of stress tensor is consistent with
that used in  conformal field theory in two dimensions.
We have kept the 
metric in these expressions
for the stress tensor  arbitrary, to enable us to evaluate the relevant Kubo formula.
The partition function of the theory is identical to that of chiral Weyl fermions.  The frequencies for the bosons 
in the Euclidean theory are even  multiples of $\pi T$ and are  given by 
\begin{equation} \label{freq}
  \omega_n    =   2n \pi T , \qquad\qquad  n\in Z.
\end{equation}
 The  thermal  two point function for the bosons is given by 
\begin{eqnarray}\label{propagatorcb}
S_B( \omega_n, p) = \langle \phi(\omega_n, p ) \phi( - \omega_{n'},  - p' ) \rangle &=& 
\frac{2}{(i\omega_n )^2-p^2 } 2\pi \beta \delta_{n, n'} \delta( p-p'). 
\end{eqnarray}   
An  important fact to be remembered while performing Wick contractions is that 
the one point function of the charge density in the thermal state is non zero
in the presence of the chiral chemical potential  and is given by 
\begin{equation} \label{charge}
 \langle j^t (t, x)  \rangle = e^2\frac{\mu_-}{2\pi}.
\end{equation}
This can be derived from the partition function of the fermion theory which coincides with 
that of chiral bosons \footnote{We review the thermodynamics 
of the Weyl gas section \ref{dispersion}.  }.  One can show that the expectation value of the 
charge density is given by (\ref{charge}) also directly in the bosonic theory by 
adding the term $\mu_-j^0$ to the Hamiltonian and evaluating the partition function.  
Converting the expectation value of the charge to the Euclidean theory we obtain
\begin{equation}
\langle j^\tau( \tau, x ) \rangle =    ie^2\frac{\mu_-}{2\pi}.
\end{equation}
Let us convert this expectation value to a rule for determining one point functions of the 
operator $\phi$ in momentum space. 
Substituting the expression for $j^\tau$ from (\ref{chbocu}) 
and performing the Fourier transform  to momentum space we obtain 
\begin{equation}\label{curexp}
\langle ( i \omega_n + p_x) \langle \phi(\omega_n, p) \rangle = 2\pi \beta \delta_{n, o} \delta(p)
\frac{e^2}{e_b} \frac{ \mu_-}{\pi}. 
\end{equation}
We will  now determine the relation  of the charge $e_b$  to the charge 
of the chiral fermion $e$ by demanding that the 
 thermal expectation value of the stress-tensor $T^{\tau\tau}$ agrees with that of the 
 Weyl fermions.  The stress tensor in momentum space is given by
 \begin{eqnarray}
T^{\tau\tau}_{fl} ( \omega_n , p ) &=& \frac{1}{8\beta} \sum_{n'} \int \frac{dp'}{2\pi}
[ i \omega_{n'}  - i \omega_n  )  + (p'- p) ][ i  \omega_{n'} + p']   \\ \nonumber
& & \qquad\qquad\qquad \times 
\phi(\omega_{n'}, p') \phi( \ \omega_n - \omega_{n'}, p-p') ,
\end{eqnarray}
 
The thermal expectation value can be written as,
\begin{eqnarray}
\langle T^{\tau\tau}   \rangle &=& A + B . \nonumber\\  
\end{eqnarray}
\begin{eqnarray}\label{abchib}
A &=&  \frac{1}{8\beta} \sum_{n'} \int \frac{dp'}{2\pi}
[ i \omega_{n'}  - i \omega_n  )  + (p'- p) ][ i  \omega_{n'} + p']   \\ \nonumber
& & \qquad\qquad\qquad\qquad\qquad \times 
\langle \phi(\omega_{n'}, p') \phi( \ \omega_n - \omega_{n'}, p-p') \rangle , \\ \nonumber
B &=& \frac{1}{8\beta} \sum_{n'} \int \frac{dp'}{2\pi}
[ i \omega_{n'}  - i \omega_n  )  + (p'- p) ][ i  \omega_{n'} + p']   \\ \nonumber
& & \qquad\qquad\qquad\qquad\qquad \times 
\langle \phi(\omega_{n'}, p') \rangle \langle  \phi( \ \omega_n - \omega_{n'}, p-p')\rangle . 
\end{eqnarray}
We now  evaluate term $A$ by substituting the propagator from (\ref{propagatorcb}), this yields
\begin{eqnarray}
A &=&\frac{1}{4\beta} \sum_{n'} \int \frac{dp'}{2\pi}
\frac{[ i\omega_{n'}  + p']^2}{(i\omega_{n'} )^2 - (p')^2} \nonumber\\
&=&-\frac{1}{4\pi} \int_{-\infty}^{\infty} dp'\,  p'\,  b(p')\nonumber
\end{eqnarray}
Here we have factored out the $2\pi \delta_{n, o} \delta(p)$ which results from momentum conservation.  
 $b(p_x)$ denotes the 
 Bose-Einstein distribution. To obtain the second equality 
 we have used the following  Matsubara sum for even integer frequencies 
\begin{eqnarray}
 \sum_{\omega_{n'} } \frac{1}{\beta} \frac{ i \omega_{n'}  }
 {( i \omega_{n'}   -p_x' ) }  
 = -p_x'   b ( p_x') \nonumber\\
 b(p_x') = \frac{1}{e^{\beta p_x'}-1}. 
\end{eqnarray}
On neglecting an infinite constant independent of temperature we obtain 
\begin{equation}
 A = - \frac{1}{2\pi} \int_0^\infty   dp'\,  p'\,  b(p'). 
\end{equation}
Using the first moment of the 
Bose-Einstein distribution given in ( \eqref{int5}) we obtain 
\begin{equation}
A  = -\frac{\pi T^2}{12}.
\end{equation}
Let us now examine the term $B$ in (\ref{abchib}) 
Substituting  (\ref{curexp}) for the expectation values of the currents we obtain  
\begin{eqnarray}
B &=& - \frac{1}{8} \left(\frac{e^2}{e_g} \frac{\mu_-}{\pi}\right)^2. 
\end{eqnarray}
Here again we have factored out the delta function due to momentum conservation. 
Then putting terms $A$ and $B$ together we obtain 
\begin{equation}
\langle T^{\tau\tau} \rangle = -\frac{1}{4\pi} (\frac{\pi^2 T^2}{3} + \frac{e^4 \mu_-^2}{e_b^2 2\pi}).  
\end{equation}
Comparing this to the energy density of chiral fermions given in (\ref{thermweyl}) \footnote{Note that the  overall negative 
sign in $\langle T^{\tau\tau} \rangle$ is due to the fact we are in Euclidean space. } 
we see that for Bose-Fermi duality we must define the charge of the 
chiral boson to be  given by 
\begin{equation}\label{eberel}
e_b = \frac{e}{\sqrt{2\pi}}.
\end{equation}
Now that we have defined the theory we can proceed to evaluate the transport coefficients.

\subsection*{$\zeta^{(2)}$ from chiral bosons}

\begin{figure}[!htb]
\centering
\hspace*{-2cm}\includegraphics[scale=0.2]{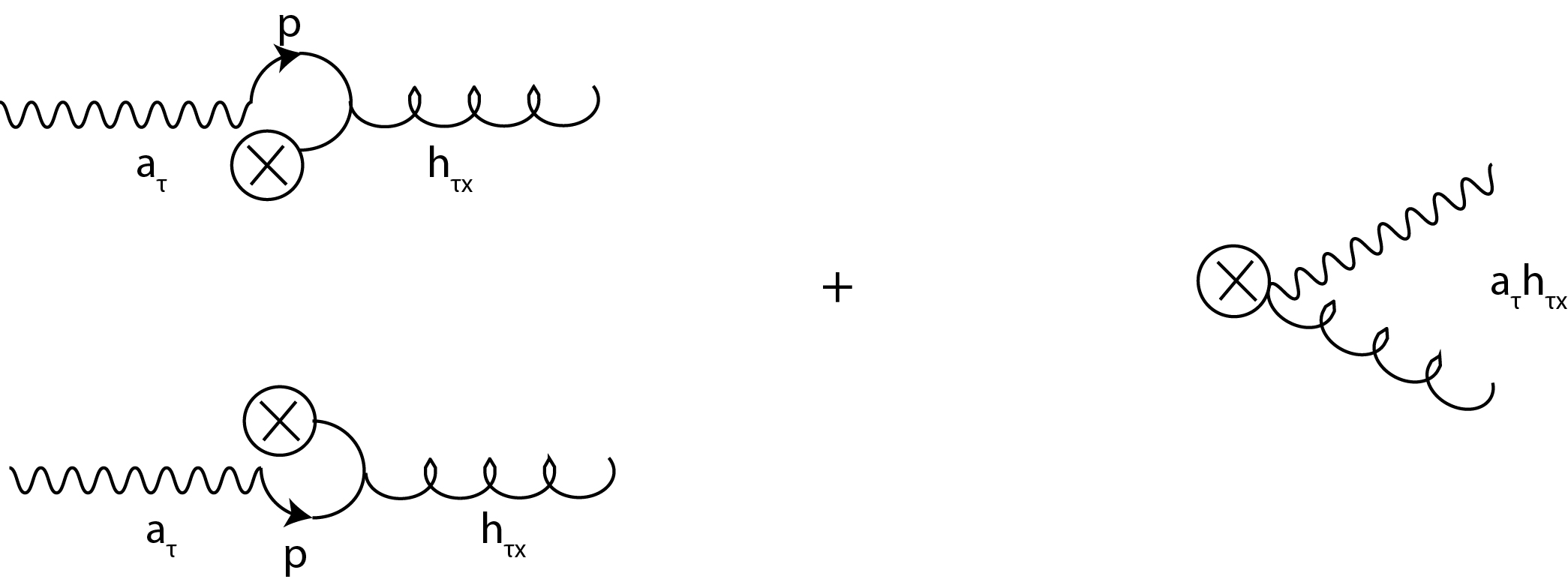} 
\caption{Diagrams contributing to  $\zeta_{\rm{Chiralboson}}^{(2)}$ }\label{jtcbdiag}
\end{figure} 
To obtain the possible contact terms, we expand the expression for for the $j^\tau$ component of the 
current, to the linear order in the metric. The relevant terms are given by 
\begin{equation}
 j^\tau = \frac{ie_b}{2}  ( - \partial_\tau + i \partial_x ) \phi  - \frac{ e_b h_{\tau x} }{4} ( -\partial_\tau + i \partial_x) \phi   + O(h, h^2). 
\end{equation}
Then the $\zeta^{(2)}$ is obtained by examining the following correlator
\begin{eqnarray}
 \frac{1}{\sqrt{g}}\frac{\delta}{\delta h_{\tau x}( p)}\left(\frac{1}{\sqrt{g}}\frac{\delta \log Z}{\delta A_\tau(- p)}
 \right) =  A + B , \\ \nonumber
 A = \langle j^\tau_{fl} (p) T^{\tau x}_{fl}(-p) \rangle , \qquad B = - \langle\frac{ \delta j^{\tau}( p)}{ h_{\tau x}(p) } \rangle. 
\end{eqnarray}
Let us now evaluate  the correlator $A$ by performing  appropriate Wick's contractions. 
In momentum space we have
\begin{eqnarray}
  j^{\tau}_{fl}(\omega_n, p)  &=&   \frac{i e_b}{2}  ( i \omega_n +  p) \phi(  \omega_n, p) , \\ \nonumber
T^{\tau x}_{fl} (- \omega_n, -  p ) & =&  \frac{i}{8\beta}\sum_{n'} 
\int\frac{dp'}{2\pi}  ( i  \omega_{n'}  + p) \left[( i \tilde \omega_{n'}   + i \omega_n) + (  p + p')     \right] \\ \nonumber
 & & \qquad\qquad \qquad \times \phi( \tilde\omega_{n'} , p') \phi( - (\omega_n +  \omega_{n'}) , -( p + p') ).
  \end{eqnarray}
For the present correlator, due to the structure of the Wick contraction and the momentum and
conservation, there is no sum over internal momenta. 
Evaluating the two possible  Wick contractions we obtain 
\begin{equation}
 A = \frac{e^2 \mu_-}{4\pi}\frac{ i \omega_n + p }{  p - i \omega_n }. 
\end{equation}
Here we have used the expectation value (\ref{curexp}). The diagrams contributing to $A$ are shown 
on the LHS of figure (\ref{jtcbdiag}).  
The contact term  $B$ is given by 
\begin{equation}
 B =  \frac{e_b}{4} ( i \omega_n  + p)  \langle \phi( \omega_n, p ) \rangle = \frac{e^2 \mu_-}{4\pi} , 
\end{equation}
where again we have used (\ref{curexp}).  The diagram contributing to $B$  is shown on the 
RHS of figure (\ref{jtcbdiag}).  
Note that as before we have suppressed the $2\pi\beta  \delta(0)$ terms in 
$A$ and $B$. 
Adding the two contributions we obtain 
\begin{equation}
 \langle j^\tau (p) T^{\tau x} ( -p) \rangle = \frac{p}{   p - i \omega_n }   \frac{e^2 \mu_-}{2\pi}.   
\end{equation}
This is identical to the expression obtained in the fermion language given in (\ref{jtchifi}). 
Therefore  taking the zero frequency limit first and then the 
zero momentum limit we see the contribution to the transport coefficient $\zeta^{(2)}$ from the chiral bosons
is identical to that 
for the fermions and is given by 
\begin{equation}
 \zeta^{(2)}_{ \; \rm chiral boson} = \frac{e^2\mu_-}{2\pi} .
\end{equation}
Note that the contact terms contributed equally just an in the case of the chiral Weyl fermion.

\subsection*{$\lambda^{(2)}$ from chiral bosons}

\begin{figure}[!htb]
  \centering
   \hspace*{-2cm}\includegraphics[scale=0.2]{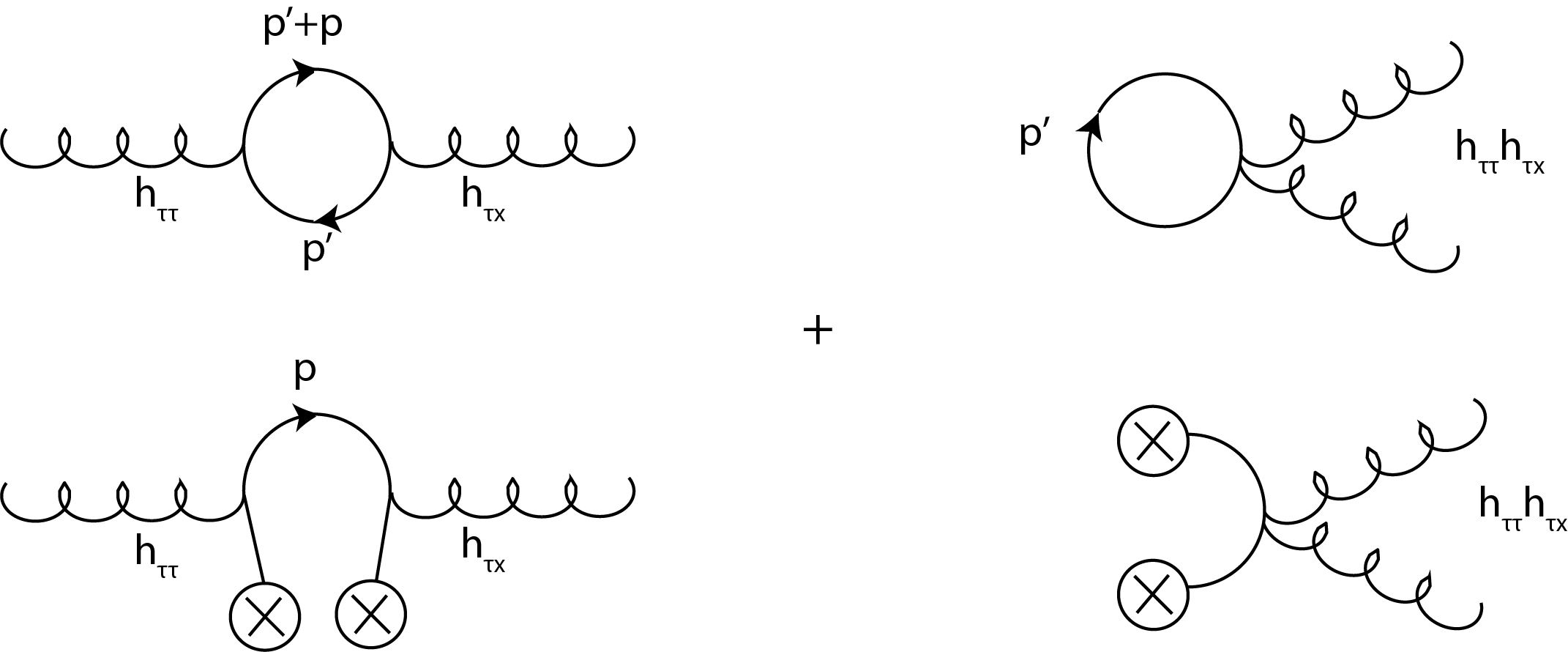} 
   \caption{Diagrams contributing to  $\lambda_{\rm{Chiralboson}}^{(2)}$ }\label{ttcbdiag}
\end{figure}   

Let us expand the $T^{\tau\tau}$ component of the stress tensor to the linear order in the metric.
The relevant terms to obtain the contribution of the  contact terms are given by 
\begin{equation}\label{chstexp}
 T^{\tau\tau} =  \frac{-1}{8} ( 1 + i h_{\tau x} ) ( - \partial_\tau + i \partial_x) \phi ( - \partial_\tau + i \partial_x ) \phi  +O(h,  h^2). 
\end{equation}
The correlator from which $\lambda^{(2)}$ is extracted is given by 
\begin{eqnarray}
2 \frac{1}{\sqrt{g}}\frac{\delta}{\delta h_{\tau x}(p)}\left(\frac{\delta \ln Z}{\delta h_{\tau\tau} (- p)} \right)
=  A + B , \\ \nonumber
A = \langle T^{\tau\tau}_{fl} (p) T^{\tau x}_{fl} ( -p)  \rangle, \qquad B =    -
\langle \frac{ \delta T^{\tau\tau}(p)  }{\delta h_{\tau x} (p)  } \rangle. 
\end{eqnarray}
In term $A$ and $B$ we need to evaluate all the contributing connected diagrams. 
The components of the flat space  stress tensor in momentum space are given by 
\begin{eqnarray}
T^{\tau\tau}_{fl} ( \omega_n , p ) &=& \frac{1}{8\beta} \sum_m \int \frac{dp'}{2\pi}
[ i \omega_m  - i \omega_n  )  + (p'- p) ][ i  \omega_m + p']  \\ \nonumber
& & \qquad\qquad\qquad \times 
\phi(\omega_m, p') \phi( \ \omega_n - \omega_m, p-p') , \\ \nonumber
T^{\tau x}_{fl} (- \omega_n, -  p ) & =&  \frac{i}{8\beta}\sum_m \int\frac{dp'}{2\pi}  ( i \omega_m  + p') \left[
( i \omega_m  + i \omega_n) + (  p + p')     \right] \\ \nonumber
 & & \qquad\qquad \qquad \times \phi( \omega_m, p') \phi( - (\omega_n +  \omega_m) , -( p + p') ).
 \end{eqnarray}
 Let us proceed to evaluate the term $A$. This quantity has two contributions due to the fact that 
 the current has  an expectation value in the thermal state. 
\begin{eqnarray}
A &=& A_1+A_2, \\ \nonumber
A_1 &=& \frac{ 2 i }{64\beta} \sum_{n'}\int \frac{dp'}{2\pi} 
( i\omega_{n'} + p')^2 ( i \omega_n' -\omega_n + p'-p)^2  S_B(\omega_{n'} , p') S_B(\omega_n - \omega_{n'}, p -p') , \\ \nonumber
A_2 &=& \frac{4 i}{ 64\beta} \sum_{n'}\int \frac{dp'}{2\pi}  \left[
( i\omega_{n'} + p')^2 ( i \omega_n' -\omega_n + p'-p)^2 S_B(\omega_{n'} , p') \right.  \\ \nonumber
& & \qquad \qquad \qquad \left. \times \langle \phi(\omega_n - \omega_n', p-p') \rangle \right]
\langle \phi( \omega_{n'} - \omega_n, p'-p \rangle . 
\end{eqnarray}
These two contributions are represented by the diagrams on the left hand side of figure \ref{ttcbdiag}. 
Note that there are 2 possible Wick contractions which contribute equally giving rise to the factor $2$ in $A_1$ and 
there are 4 possible Wick contractions which contribute equally giving rise to the factor $4$ in $A_2$. 
 Substituting the expression for the bosonic propagator $S_B$ from (\ref{propagatorcb})  in to $A_1$ we
 obtain 
 \begin{eqnarray} \nonumber
 A_1  =  \frac{i}{8\beta} \sum_{n'} \int \frac{dp'}{2\pi} 
 \frac{(  i \omega_{n'}   + p')( i  \omega_{n'}   - i \omega_n  + p' - p )}
 { ( i  \omega_m   - p')(  (  i \omega_{n'}    -  i \omega_n )  - ( p' - p) )  }. \\
 \end{eqnarray}
We can now perform the Matsubara sum and take  the zero frequency limit first  and then  zero momentum limit to obtain
\begin{eqnarray}
 \lim_{p\rightarrow 0, p_0 \rightarrow 0}  A_1 (p_0,p) 
 &=& \frac{i}{2}\int \frac{dp'}{2\pi} p'  b(p' ),\nonumber\\
&=&\frac{i\pi T^2}{12}. 
\end{eqnarray}
From the structure of the Wick contractions in $A_2$ we see that it does not involve any Matsubara sum or integral. 
We need to use (\ref{curexp}) to evaluate the expectation value of the currents and the relation between the 
charges in (\ref{eberel}). 
This leads to the following value for $A_2$
\begin{eqnarray}
 A_2  = i\frac{e^2 \mu_-}{ 4\pi}. 
\end{eqnarray} 
The contribution from the contact term $B$ is given by 
\begin{eqnarray}
 - \langle \frac{ \delta T^{\tau\tau}(p)  }{\delta h_{\tau x} (p)  } \rangle
 &=& \frac{i }{4\pi} (e^2\mu^2_- + \frac{\pi^2 T^2}{3}). 
\end{eqnarray}
This is easy to see from (\ref{chstexp}) since the contact term is  equal to $-i\langle T^{\tau\tau} \rangle$. 
The diagrams contributing to $B$ are shown on the RHS of figure (\ref{ttcbdiag}). 
Considering the contributions both from the term $A$ and $B$ we obtain 
\begin{equation}
  \lim_{p\rightarrow 0} \langle T^{\tau\tau}_{fl} ( 0  , p ) T^{\tau x}_{fl} (0  -  p ) \rangle
  = \frac{i }{2\pi} (e\mu^2_- + \frac{\pi^2 T^2}{3}). 
\end{equation}
This  is identical to that the contribution of the chiral fermions for the corresponding correlator  as seen in (\ref{euclid2dl}). 
Therefore  by the same logic we obtain the contribution of chiral bosons to $\lambda^{(2)}$ to be given by 
\begin{equation}
 \lambda^{(2)}_{{\rm chiral\, bosons}} = \frac{1 }{4\pi} (e^2\mu^2_- + \frac{\pi^2 T^2}{3}).
\end{equation}
It is indeed interesting to see how the different Feynman diagrams in the chiral boson theory 
which obeys the Bose-Einstein distribution organize themselves to give the same transport coefficients
as that of the chiral fermion theory. 

\subsection{Chiral gravitino like  system}\label{2dgravitino}

It has been suspected that fermions with spin greater than $1/2$,  
for instance gravitinos, do not obey the replacement rule \cite{Loganayagam:2012zg,Eling:2013bj}. That is the relationship
between parity odd transport and the microscopic anomaly breaks down. 
It will be instructive to see how this explicitly occurs in two dimensions, by performing a perturbative evaluation 
of the respective Kubo formula. 
 There  are no physical propagating gravitinos in two dimensions. However the 
following system mimics the features we require. 
Consider the Lagrangian 
\begin{eqnarray}
 S_E =  S_E^{g1 }   + S_E^{g2},  \nonumber \\ \label{grav2d}
 S_E^{\rm{gr}} =  \int d\tau dx \sqrt{g} e^\mu_a \psi^\dagger_\rho \gamma^a D_\mu P_- \psi^\rho, 
\end{eqnarray}
where 
\begin{equation} \label{covder}
D_\mu \psi^\rho =
  \partial_\mu \psi^\rho  + \frac{1}{2} \omega_{\mu c d} \sigma^{cd} \psi^\rho  + 
  \Gamma_{\mu\sigma}^\rho \psi^\sigma.
  \end{equation}
  This Lagrangian  in (\ref{grav2d})  is 
 the gauged fixed  Lagrangian of the gravitino in higher dimensions. The action $S_E^{g2}$ 
 is the action of the corresponding ghosts. 
 The contribution of the all the ghosts to the gravitational anomaly
  can be accounted by  subtracting the contribution of a chiral spinor \cite{AlvarezGaume:1983ig}. 
   In two dimensions 
 the above Lagrangian would not have any physical propagating degrees of freedom.
 We will see that in fact the action $S_E^{g1}$  has central charge $c=-22$.  However this 
 gravitino like system will provide a simple illustrative example of how the replacement 
 rule can break down \footnote{We thank R. Loganayagam and E. Witten for discussions
 which clarified our understanding of this point. }. 
 
 To proceed let us evaluate the central charge of this system.  There are additional terms
 in the stress tensor due to the presence of the Christoffel symbol in the covariant derivative
 given in (\ref{covder}). 
 The linearization of  the Christoffel symbol  results in 
 \begin{eqnarray}
\Gamma^\mu _{\rho \alpha} = \frac{1}{2} \eta^{\mu \sigma} (\partial_\rho h_{\sigma \alpha }+\partial_{\alpha }h_{\sigma \rho}-\partial_\sigma h_{\alpha \rho}) + O(h^3). 
\end{eqnarray} 
The stress tensor of the system is given by 
 \begin{eqnarray} \label{classicalst}
 T_{\mu\nu}  &=& T_{\mu\nu}^{g1} + T_{\mu\nu}^{g2} , \\ \nonumber
 T_{\mu\nu}^{g1}  &=&   \frac{1}{2}\psi^\dagger_\rho (x)(\gamma_\mu \partial_\nu + \gamma_\nu \partial_\mu)P_- \psi^\rho (x)\nonumber\\
T_{\mu\nu}^{g2}&=& -\frac{1}{2}(\partial_\sigma(\psi^{\dagger,\sigma} (x)\gamma_\mu P_{-}\psi_{\nu }(x)
- \partial_\sigma (\psi^\dagger_\mu (x) \gamma_\nu P_{-} \psi^\sigma(x)
+ ( \mu \leftrightarrow \nu ) ).
 \end{eqnarray}
 Note that the total derivative term in the stress tensor arises from the contribution due to the 
 linearization of the Christoffel symbol. 
 While evaluating the stress tensor, we have used the equations of motion. 
  To make it convenient to determine the central charge we 
   go over to holomorphic coordinates defined by 
   \begin{equation}\label{holanti}
   z = -i \tau + x , \qquad\qquad  \bar z = i\tau + x.
   \end{equation}
   Then the only non-trivial component of the stress tensor is given by 
   \begin{equation}\label{holstress}
   T_{zz}(z) = \frac{1}{2} \left( \hat\psi_\mu ^* \partial_z \hat \psi^\mu  -  \partial_z\hat\psi_\mu^* \psi^\mu
   \right)
    - i \epsilon^{\mu\nu} \partial_z ( \hat \psi_\mu^*  \hat \psi_\nu ). 
 \end{equation}
 Here $\hat \psi_\mu$ is the chiral component of the gravitino, $\epsilon$ is the anti-symmetric tensor
 defined with $\epsilon^{\tau x} = 1$. 
 We can now easily evaluate the central charge with the following OPE
 \begin{equation}
 \hat \psi_\mu^* (z) \hat \psi^\nu ( w) = \frac{ \delta_\mu^\nu }{z- w}. 
 \end{equation}
On performing this, it is easy to see that there are additional contributions for the 
term proportional to the total derivative in (\ref{holstress}). The central charge is given by 
\begin{equation}
c_R^{g1}  =  2 - 24 = - 22.
\end{equation}
The $2$ results from the first term in (\ref{holstress}), while the $-24$ results from the total derivative. 
Now to obtain the complete contribution, we need to take into the contribution from the 
ghosts, which is equivalent to subtracting the contribution from a complex chiral fermion.  
Therefore we obtain
\begin{equation}
c_R = c_R^{g1}  + c_R^{g2} = -22 -1 = -23.
\end{equation}
The reason that  this central charge is negative is due to the fact that the 
gravitino like system in 2 dimensions is not physical. 
The contribution of this gravitino like system to the gravitational anomaly is given by 
\begin{equation}\label{cggrav}
c_g^{\; {\rm gravitino} } = - \frac{1}{96\pi} c_R = \frac{23}{96\pi}. 
\end{equation}
Here we have used the relationship between the central charge and the coefficient of gravitational 
anomaly form \cite{Jensen:2012kj}. 
This contribution to the  gravitational anomaly of the gravitino like system coincides with what has been considered as
 the contribution of the `gravitino'   in \cite{Loganayagam:2012zg}.

 \subsection*{$\lambda^{(2)}$ from gravitino like system}
 
 We now evaluate the contribution to the transport coefficient $\lambda^{(2)}$ from the 
 gravitino like system. 
 The components of the classical stress tensor for this system in momentum space
 can be read out from (\ref{classicalst}), they are given by 
 \begin{eqnarray}
T^{\tau \tau }_{fl} (p)&=& T^{\tau\tau}_{fl (1)}  + T^{\tau\tau}_{fl(2)} , \\ 
T^{\tau\tau}_{fl (1) } (p) &=& \int \frac{d^2 p'}{(2\pi)^2} (\psi^\dagger_{\mu} (p'-p)\gamma^\tau (ip_\tau )P_-\psi^{\mu }(p')),\nonumber\\
T^{\tau\tau}_{fl (2)}(p) &= &ip_\sigma \int \frac{d^2 p'}{(2\pi)^2} 
\left[ (\psi^{\dagger \sigma}(p'-p)\gamma^\tau P_-\psi^\tau (p'))-(\psi^{\dagger \tau} (p'-p)\gamma^\tau P_- \psi^\sigma(p')) \right]. \nonumber
\end{eqnarray}
\begin{eqnarray}
T^{\tau x}_{fl}(-p)&=&T^{\tau x}_{fl (1)}  + T^{\tau x }_{fl (2)} , \\ 
 T^{\tau x }_{fl (1)} (-p) &=& 
 \int \frac{d^2p'}{(2\pi)^2 } \frac{1}{2}({\psi}^\dagger_\mu (p'+p)(\gamma^\tau ip'_{x}+
\gamma^x ip_\tau )P_-\psi^\mu (p'))),\nonumber\\
T^{\tau x }_{fl (2)} (-p)  &= & 
-\frac{1}{2}(ip_\sigma) \int \frac{d^2 p'}{(2\pi)^2}
\left[ (\psi^{\dagger \sigma}(p'+p)\gamma^\tau P_-\psi^x(p') + 
\psi^{\dagger \sigma}(p'+p)\gamma^x P_-\psi^\tau (p') ) \right. \nonumber\\
&&- \left. (\psi^{\dagger\tau}(p'+p)\gamma^x P_-\psi^\sigma(p') + \psi^{\dagger x} (p'+p)\gamma^\tau P_-\psi^\sigma (p')) \right]. \nonumber
\end{eqnarray}
Here the terms in the stress tensor, which
arises from the total derivative, have been separated and their contribution is labeled by the 
subscript $(2)$.  The two point function of the classical part  stress tensor is given by 
\begin{equation}
\langle T^{\tau\tau}_{fl} (p) T^{\tau x}_{fl} (-p)    \rangle  
=  \langle T^{\tau\tau}_{fl (1) } (p)  T^{\tau x}_{fl (1) }  (-p)  \rangle 
+ \langle T^{\tau\tau}_{fl (2) } (p)  T^{\tau x}_{fl (2) }  (-p)  \rangle. 
\end{equation}
It can be easily seen from the structure of the Wick contractions, all the remaining  cross terms vanish. 
Now the contribution from {\small $\langle T^{\tau\tau}_{fl (1) } (p)  T^{\tau x}_{fl (1) }  (-p)  \rangle  $ }
is twice that of a single chiral fermion. 
We will show that the contribution from {\small $\langle T^{\tau\tau}_{fl (2) } (p)  T^{\tau x}_{fl (2) }  (-p)  \rangle$ }
vanishes. 
We  evaluate  this correlator by performing  Wick contractions using the propagator 
\begin{equation}
\langle \psi_\mu ( \omega_m, p   ) \psi^\dagger_\nu( \omega_{m'}, p' ) =  
2\pi \beta \delta( p- p') \eta_{\mu\nu}\delta_{m, m'}
\left( \begin{array}{cc}
0 & \frac{-i}{i\omega_m-p}\\
\frac{-i}{i\omega_m +p } & 0  
\end{array} \right). 
\end{equation}
Note that we have considered the gravitino like theory at finite temperature but not at finite chemical potential. 
This is sufficient to pick out the pure gravitational contribution to the transport coefficient $\lambda^{(2)}$. 
After performing the Matsubara sums involved we obtain 
\begin{eqnarray}\label{ttgravi}
\langle T^{\tau\tau}_{fl (2) } (\omega_m , p )  T^{\tau x}_{fl (2) }  (-\omega_,  - p )  \rangle &=&
i p\frac{ i \omega_m + p}{ i\omega_m - p}   \int_{-\infty}^\infty \frac{dp' }{2\pi} \left( f( p) - f( p' + p)  \right) , \\ \nonumber
&=& -i\frac{p^2}{2\pi} \frac{ i \omega_m + p}{ i\omega_m - p}.
\end{eqnarray}
Therefore we see that in the zero momentum limit the contribution from the Christoffel  symbol vanishes in the 
zero momentum limit. 
\begin{equation}
\lim_{p\rightarrow 0, \omega_m \rightarrow 0} 
\langle T^{\tau\tau}_{fl (2) } (\omega_m , p )  T^{\tau x}_{fl (2) }  (-\omega_,  - p )  \rangle =0.
\end{equation}
For later analysis it is important to note that 
even if we did not evaluate the integral in (\ref{ttgravi}),  as long as one
is interested in the finite piece of the integral, taking the zero frequency limit and then the zero momentum 
limit ensures that the this contribution vanishes. 
Thus we conclude that the contribution to the transport coefficient $\lambda^{(2)}$
of the total derivative term in the classical  stress tensor 
(\ref{classicalst}), which arises due to the presence of the Christoffel symbol, vanishes. 
Let us now examine possible contact terms which arise from expanding the Christoffel symbol. 
On performing the expansion of the vierbein, metric as done in (\ref{mexp}), (\ref{vexp1}) and 
(\ref{vexp2}) as well as the term involving the Christoffel symbol in the 
action (\ref{grav2d}), the relevant contact terms arise from 
\begin{eqnarray} 
S^{(1)}_{\rm{Christoffell}}  =  
\int d\tau  dx  \left( h_{\tau x} \partial_x h_{\tau \tau} (\psi^\dagger_x \gamma^\tau \psi_x) + 
 \frac{h_{\tau x} \partial_\tau h_{\tau \tau}}{2} (\psi^\dagger_\tau \gamma^x \psi_\tau) 
 -\frac{h_{\tau x} \partial_x h_{\tau \tau}}{2} (\psi^\dagger_\tau \gamma^\tau \psi_\tau)  \right). \nonumber\\
\end{eqnarray}
Note that the origin of all these terms is from the expansion of the Christoffel symbol and therefore 
they involve a derivative on the metric. This is unlike the other terms in  the second line of $S^{(1)}$ given in (\ref{lexpansion1}),
which involve a derivative on the fermion.  The presence of the derivative on the metric  leads to 
 a momentum factor outside the Fermi-Dirac integral after performing the Wick contraction 
and the Matsubara sum. 
Therefore again, on taking the zero frequency and zero momentum limit, the contributions
of these terms vanish. 

The above analysis shows that the terms arising from the Christoffel symbol in the action 
for the gravitino like system in (\ref{grav2d}), do not contribute to the transport coefficient $\lambda^{(2)}$. 
Thus the contribution of this system to this transport coefficient is identical to two chiral Weyl  
fermions. We then have to subtract the ghost contribution  which is equal to  a single chiral Weyl fermion. 
The net result is that  $\lambda^{(2)}$ for a gravitino like system is given by 
\begin{equation}
\lambda^{(2)} = \frac{\pi T^2}{12}.
\end{equation}
Therefore  we have
\begin{equation}
\tilde c_{2d}^{\; {\rm gravitino} }  = \frac{\pi }{12}.
\end{equation}
Comparing the value of $c_g^{\; {\rm gravitino} }$ for the system in (\ref{cggrav}), we see that 
this system  does not satisfy the replacement rule obtained by 
 consistency of the Euclidean vacuum since 
$\tilde c_{2d}^{\; \rm{gravitino}} \neq - 8\pi^2 c_{g}^{\; \rm{gravitino}}$. 

At this point it is important to make the following observations.  The  terms proportional to the 
Christoffel symbol  contribute to the anomaly coefficient. However for the transport coefficient 
the contributions of this vanished on taking the zero momentum limit. 
It is for this reason that for the purpose of evaluating the transport coefficient the system along with
the ghosts behaved like a single chiral Weyl fermion. 
Though the system we analyzed in two dimensions  possibly does not arise  naturally in any physical 
system since the central charge of this system is negative,  it illustrates the mechanism 
of how for higher spin chiral fermions the transport coefficient is not constrained by 
the microscopic anomaly. 
We will return to this phenomenon again in 6d dimensions, where chiral gravitinos 
are physical and they contribute to the pure gravitational anomaly.

\section{Anomalies and transport in \texorpdfstring{$d=4$}{Lg}} \label{an4d}

Let us begin by  parametrizing  the constitutive for the stress tensor and the charge current. 
We will restrict our attention to terms occurring in first order in the derivative expansion in the anomaly frame. 
In $d=4$ the constitutive relations are given by 
\begin{eqnarray}
T^{\mu \nu }&=&
\left(\epsilon+P\right)u^{\mu }u^{\nu }-P \eta^{\mu \nu }+
\lambda^{(4)}_1 (u^\mu \frac{1}{2}\epsilon^{\nu \rho \alpha \beta } u_\rho F_{\alpha \beta} 
+ u^\nu \frac{1}{2}\epsilon^{\mu \rho \alpha \beta } u_\rho F_{\alpha \beta} ),\nonumber\\
&&+\lambda^{(4)}_2 (u^\mu \epsilon^{\nu \rho \alpha \beta } u_\rho \partial_\alpha u_\beta +
u^\nu\epsilon^{\mu \rho \alpha \beta } u_\rho \partial_\alpha u_\beta ),\nonumber\\
j^{\mu }&=&nu^{\mu }+\zeta^{(4)}_1 \frac{1}{2}\epsilon^{\mu \rho \alpha \beta } u_\rho F_{\alpha \beta} 
+ \zeta^{(4)}_2 \epsilon^{\mu \rho \alpha \beta } u_\rho \partial_\alpha u_\beta. 
\end{eqnarray}
where, $\epsilon,p,n,\mu , \lambda_1,\lambda_2, \zeta_1,\zeta_2$ refer to the energy density, 
pressure, charge density, chemical potential and the four anomalous transport coefficients respectively. 
The metric, we work in, has the  mostly negative signature $g_{\mu \nu }=\left(1,-1,-1,-1\right)$. 
Let us derive the Kubo formulae for the parity odd transport coefficients, by considering perturbations 
of the fluid from the rest frame. 
Let the velocity of fluid be given by  $u^\mu =\left(1,0,v^y,0\right)$ with $|v^y|<<1$, we also have 
such that $u^\mu u_\mu =1+ O(v^2) $. 
The gauge field, velocity and metric perturbations are given by,    
\begin{eqnarray}
 A_{\mu}=(0,0,a_y,0,) & & \qquad \qquad   u_{\mu }=\left(0,0,-v^y+h_{ty},0\right), \nonumber\\
 g_{\mu \nu}=\eta_{\mu \nu} + h_{\mu \nu}, &  & \qquad \qquad h_{ty}=h_{ty}\neq 0 .                         
\end{eqnarray}       
All perturbations are assumed to depend on only the $z$ direction and we can work with the Fourier modes in this direction. 
Expanding the constitutive relation to the linear order and interpreting it as an Ward identity, just as it was done in 
 the case of $d=2$, we obtain the following  Kubo formula for the anomalous transport coefficients. 
\begin{eqnarray} \label{kb4d}
\lambda^{(4)}_1 &=& \lim_{p_z \rightarrow 0, p_0 \rightarrow 0} \frac{1}{ip_z} \langle T^{tx}(p_0, p_z )j^y(-p_0, -p_z) \rangle_R ,
\\ \nonumber
\lambda^{(4)}_2 &=& 
\lim_{p_z\rightarrow 0, p_0\rightarrow 0} \frac{1}{ip_z} \langle T^{tx}(p_0, p_z )T^{ty}(-p_0, -p_z) \rangle_R  \\ \nonumber
\zeta^{(4)}_1 &=&  \lim_{p_z \rightarrow 0, p_0 \rightarrow 0} \frac{1}{ ip_z} \langle j^{x}(p_0, p_z )j^y(-p_0, -p_z ) \rangle_R, 
\\ \nonumber 
\zeta^{(4)}_2 &=&  
\lim_{p_z \rightarrow 0, p_0 \rightarrow 0} \frac{1}{ip_z} \langle j^{x}(p_0, p_z )T^{ty}(-p_0, -p_z)\rangle_R.  
\end{eqnarray}
The expectation values in all these correlators refer to real time retarded correlators at finite temperature and chemical 
potential.  As discussed earlier, we have to take the zero frequency limit first and then the zero momentum limit. 
All of these correlators involve division by a power of momentum, since these transport coefficients 
occur at the first order in  the derivative expansion. Finally note that the transport coefficients $\lambda^{(4)}_1$ and $\zeta^{(4)}_2$
are identical and therefore it is sufficient to evaluate only 3 of the Kubo formulae in (\ref{kb4d}). 
As done for the case of $d=2$ to obtain the  Minkowski real time retarded correlators, 
we will evaluate the corresponding  correlator in the Euclidean theory and then analytically continue to obtain 
the transport coefficients.

Before proceeding, let us recall  previous evaluations of these transport coefficients in the literature. 
For a theory of Weyl  fermions, the coefficient $\zeta_1^{(4)}$ was first evaluated in \cite{Kharzeev:2009pj}, the coefficient 
$\zeta_1^{(4)}$ as well as $\zeta_2^{(4)}$ was evaluated in \cite{Landsteiner:2012kd}. 
As we will see, the coefficient $\lambda_1^{(4)}$  is sensitive to the mixed  gravitational anomaly in the theory and this 
was obtained in \cite{Landsteiner:2011cp}. The evaluation of the coefficient $\lambda_2^{(4)}$  was begun in 
\cite{Landsteiner:2012kd}, and all the
diagrams  including the contact terms 
which contribute to this coefficient was finally done in \cite{Yee:2014dxa}.  The fact that the  mixed anomaly contributes to 
$\lambda_1^{(4)}$, was mentioned in \cite{Landsteiner:2012kd}. However the complete evaluation of the contribution of 
the mixed anomaly coefficient including the contact terms has not been done. 
The work in \cite{Yee:2014dxa} was interested only in the contribution of the global  gauge anomaly. 
We will  complete this small gap in literature, 
 we will also  develop a simple method of evaluating the 
resulting angular integrals.  Our calculation will also serve as a cross check   of the methods in \cite{Yee:2014dxa}
which are different from that in this paper. 
We  also show that we can take the zero frequency and 
zero momentum limit first and then perform the resulting angular integrals. 
This considerably  simplifies the evaluation of these integrals.   In all the previous work, the angular 
integrals were first performed, in fact \cite{Yee:2014dxa} provides  a formula which was guessed for 
angular integrals which occur in arbitrary dimensions.

\subsection{Chiral Fermions} \label{Chiral Fermions in d=4}
  
 Let us proceed to evaluate the corresponding correlators first, in the Euclidean theory of 
 free chiral Weyl fermions in 
 $d=4$.  Following the set up in section (\ref{an2d}),  the expression for the 
 Euclidean $\langle j^\mu j^\nu \rangle_E$ is given by 
 \begin{eqnarray}\label{contact4da}
\langle j^\mu j^{\nu} \rangle_{E,cl} &=& - \langle j^\mu \rangle_E \langle j^\nu \rangle_E + 
\frac{1}{{ \cal Z}_E^{(0)}} 
\int {\cal D}{\psi^\dagger}{ \cal D} \psi 
e^{  S_{E}^{(0)}}\frac{\delta S_E}{\delta a_\mu} \frac{\delta S_{E} } {\delta a_\nu}  \\ \nonumber
& & + \frac{1}{ { \cal Z}_E^{(0)} }  \int {\cal D}{\psi^\dagger}{ \cal D} \psi e^{ S_{E}^{(0)} } 
\frac{\delta ^2 S_E}{ \delta a_\mu \delta a_{\nu} },
\end{eqnarray}
 while the two point functions   $\langle j^\mu j^\nu \rangle_E$ and $ \langle j^\mu T^{\nu\rho} \rangle^e$ 
  are given in (\ref{jt1}) and (\ref{tt1}) respectively. 
  For convenience,  let  us define the following coefficients, which are obtained from taking limits of the 
 Euclidean correlators 
 \begin{eqnarray} \label{tr4de}
  \tilde  \zeta^{(4)}_1  &=&    \lim_{p_z \rightarrow 0,} \frac{1}{ ip_z} \langle j^{x}(0 , p_z )j^y(0 , -p_z ) \rangle_E,
  \\ \nonumber
  \tilde \zeta^{(4)}_2 &=&   \lim_{p_z \rightarrow 0} \frac{1}{ip_z} \langle j^{x}(0 , p_z )T^{\tau y}(0 , -p_z)\rangle_E,
  \\ \nonumber
  \tilde \lambda^{(4)}_2 &=&  \lim_{p_z \rightarrow 0}  \frac{1}{ip_z} \langle T^{\tau x}(0 , p_z )T^{\tau y} (0 , -p_z) \rangle_E.
  \end{eqnarray}
  To obtain the transport coefficients from the above definitions, we use the relation between Minkowski 
  real time retarded  correlators and the  Euclidean correlators which are given by 
  \begin{eqnarray}\label{EM4d}
\langle j^x (\omega_n ,p)j^y (-\omega_n ,-p) \rangle_E &=&
\langle j^x (p_0,p)j^y (-p_0,-p) \rangle_R |_{i\omega \rightarrow p_0 +i\epsilon},\nonumber\\ 
\langle j^x (\omega_n ,p)T^{\tau y} (-\omega_n ,-p) \rangle_E &=&
i\langle j^x (p_0,p)T^{ty} (-p_0,-p) \rangle_R |_{i\omega \rightarrow p_0 +i\epsilon},\nonumber\\
\langle T^{\tau x} (\omega_n ,p)T^{\tau y} (-\omega_n ,-p) \rangle_E &=&
-\langle T^{tx} (p_0,p)T^{ty} (-p_0,-p) \rangle_R |_{i\omega \rightarrow p_0 +i\epsilon}.\nonumber\\
\end{eqnarray}      
where, $\omega_n = 2\pi n T$, are the integer quantized Matsubara frequencies. 
Using (\ref{EM4d})  and the definitions of the transport coefficients in (\ref{kb4d}) and (\ref{tr4de}) we obtain 
 \begin{equation} \label{relem3e}
  \tilde \zeta^{(4)}_1 = \zeta^{(4)}_1, \
  \qquad \tilde \zeta^{(4)}_2 = i \zeta^{(4)}_2, \qquad \tilde \lambda^{(4)}_2 = - \lambda^{(4)}_2.
 \end{equation}
              
For Weyl fermions in $d=4$, the partition function ${\cal Z}_E$ is given by 
\begin{equation}
 {\cal Z}_E = \int {\cal D }\psi_E^\dagger {\cal D} \psi_E \exp(S_E),
\end{equation}  
where,
\begin{eqnarray} \label{action4d}
S_E = \int d\tau d^3x  \sqrt{g} e^{\mu}_a \psi^\dagger \gamma^a  D_\mu  P_{-}  \psi.
\end{eqnarray}  
The covariant derivative is defined as 
\begin{eqnarray}
D_\rho \psi =\partial_\rho \psi +\frac{1}{2}\omega_{\rho ab} \sigma^{ab} \psi + ieA_\rho \psi,
\end{eqnarray}                    
where, $\omega_{\rho ab}$ is the spin connection and $\sigma^{ab}=\frac{1}{4}[\gamma_a,\gamma_b]$. 
Unlike $d=2$, the contributions from the term proportional to the spin connection does not vanish for $d>2$. 
We work in Euclidean space where   the gamma matrices satisfy,
\begin{eqnarray} \label{antigamma}
\{ \gamma^a, \gamma^b\} = 2 \eta^{ab}_E, \qquad \eta^{ab}_E = - \delta^{ab},
\end{eqnarray} 
and \begin{equation}
     P_\pm = \frac{1}{2} ( 1\pm \gamma_5). 
    \end{equation}
Note raising and lowering of flat space indices are done by $\eta^{ab}_E$.  We now perturb the action in 
 \eqref{action4d} from flat space and constant chemical potential, by considering the expansion  
 with the following choice of metric and gauge field perturbations,
\begin{eqnarray}\label{4dmetgapert}
g_{\mu\nu}= - \delta_{\mu\nu} +h_{\mu\nu}, \qquad  A_\mu =A^{(0)}_\mu + a_\mu , 
\end{eqnarray}
where $A_{\tau}^{(0}$ refers to  the background chemical potential. 
For this metric perturbation following \cite{AlvarezGaume:1983ig},  
we can work with the following vierbein to the linear order in fluctuations
\begin{eqnarray}\label{4dvierp}
e_{a \mu}  =  - \delta_{a\mu}  + \frac{ h_{a\mu} }{2}  + O(h^2).  
\end{eqnarray}
Essentially we are working in a gauge which ensures that the vierbein is symmetric. 
Expanding the action to the quadratic order in the fluctuations we obtain
\begin{eqnarray}\label{action4dexp}
S_E  &=& S_E^{(0)} + S_E^{(1)} + S_E^{(2)} , \\ \nonumber
S^{(0)}_E &=&\int d^4x \psi_E^\dagger  (\gamma^\tau D_\tau +\gamma^i \partial_i )P_- \psi_E,
\nonumber\\
S^{(1)}_E &=& -\int d^4x \frac{h_{\mu \nu }}{4}\psi_E^\dagger
  (\gamma^\mu \partial^\nu + \gamma^\nu \partial^\mu)P_- \psi_E 
+ie  \int d^4x \psi_E^\dagger \gamma^\mu A_{\mu } P_-\psi_E,  \nonumber\\
S_E^{(2)}&=&  \int d^4x \frac{h_{\lambda \alpha }\partial_\mu h_{\nu \alpha}}{16}\psi_E^\dagger
 \Gamma^{\mu \lambda \nu }P_- \psi_E + O(h^3).
\end{eqnarray}
Here  coefficient of the term linear in the metric  is the  stress tensor in flat space 
and the coefficient of the term linear in the gauge field  is the charge current. 
The second order terms in the the metric, given in the last term of $(\ref{action4dexp})$, arises 
 from expanding  of the spin connection, the repeated indices 
 in the subscript refer to summation.  This term contributes to the mixed anomaly coefficient
 \cite{AlvarezGaume:1983ig}. We will see that that this term  gives rise to a contact term 
 and it is important in the evaluation of the coefficient $\lambda^{(4)}_2$. 
 $\Gamma^{\mu \nu \alpha}$ stands for totally anti-symmetric combination of gamma matrices which 
 is given by 
 \begin{equation}
 \Gamma^{\mu \nu \alpha} = \frac{1}{3!} \left( \gamma^\mu \gamma^\nu \gamma^\alpha  - 
 \gamma^\nu\gamma^\mu \gamma^\alpha + ( \rm{cyclic} ) \;\;  \right).
 \end{equation}
 Using the action (\ref{action4dexp}), we can obtain the stress tensor and charge current which are given by 
\begin{equation}\label{stress4d}
T^{\mu \nu }_{fl}(x)=\frac{1}{2}\psi_E^\dagger(\gamma^\mu \partial^\nu + \gamma^\nu \partial^\mu)P_- \psi_E,
\end{equation}
\begin{equation}\label{current4d}
j^{\mu }_{fl}= -ie \psi_E^\dagger \gamma^\mu P_-\psi_E.
\end{equation}
We now take the background chemical potential to be  
\begin{equation}
 A_\tau^{(0)} = i \mu + i \mu_c\gamma_5.
\end{equation}
The propagator derived from the action   \eqref{action4dexp} is given by 
\begin{eqnarray}\label{method2propagator}
S_E\left(q\right)&=&\frac{-i}{\gamma^\tau \left(w_m-ie\mu -ie\mu _c \gamma_5\right)+\gamma^i q_i},\nonumber\\
&=&\frac{-i}{2} \sum _{s,t=\pm} \Delta_t(i{\omega}_s,q) P_s\gamma_{\mu } \hat{q}^{\mu }_t,
\end{eqnarray}                  
where 
\begin{eqnarray}
P_{\pm}&=&\frac{(1\pm \gamma_5)}{2}, \qquad \hat{q}^{\mu }_{\pm}=(-i,\pm \hat{q}) , \qquad E_q = |q| \\  \nonumber
\Delta_{\pm}(q_0,q)&=& \frac{1}{q_0 \mp E_q}, \qquad 
\omega_s = \omega_m -ie\mu - sie\mu_c, \qquad \mu_\pm = \mu \pm \mu_c.
\end{eqnarray}

\subsection*{Evaluation of $\zeta^{(4)}_1$}

\begin{figure}[!htb]\label{jj4diag}
  \centering
   \hspace*{-2cm}\includegraphics[scale=0.2]{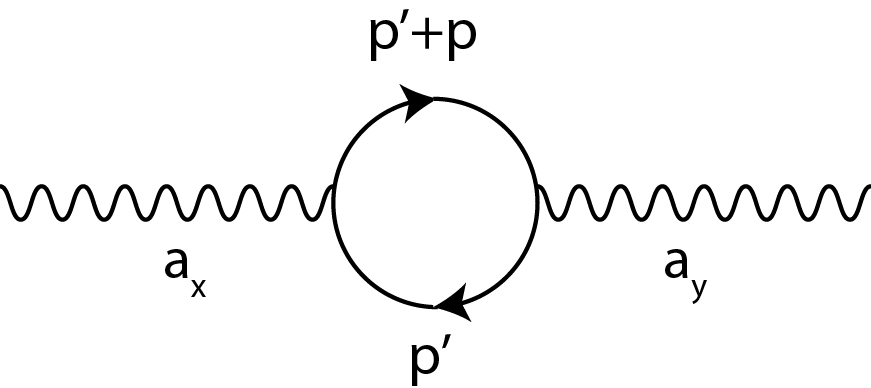} 
   \caption{Diagram  contributing to  $\zeta^{(4)}_1$ }
\end{figure}                               
                              
From the Kubo formula in \eqref{tr4de}, the transport coefficient $\tilde \zeta^{(4)}_1$ is given by 
\begin{eqnarray}
\zeta^{(4),E}_1 = \lim_{  p_z\rightarrow 0}\frac{1}{ ip_z} \langle j^{x}(0, p_z )j^y(0, -p_z) \rangle.\nonumber\\
\end{eqnarray}               
For all calculations of correlators in the rest of the paper, we will set the frequency to zero at the beginning.  
The correlator is defined using  \eqref{contact4da} and from the expansion of the action given in  
\eqref{action4dexp}, we see that there are no contact terms that  contribute to 
this correlator.  Converting the charge current given in  (\eqref{current4d}) to momentum space we obtain 
 \begin{eqnarray}
{j^\mu }_{fl}( 0, p )&=&-ie\frac{1}{\beta}\sum_{m}
\int \frac{d^3 p'  }{(2\pi)^3}\psi_E^\dagger(\omega_m ,  p'-p)\gamma^\mu P_- \psi_E(\omega_m, p').\nonumber\\
 \end{eqnarray} 
After performing  Wick contractions the correlator of interest is given by 
\begin{eqnarray} \label{defjj}
I(p ) &=&\frac{1}{ip_z}\langle j^x(0, p )j^y(0, -p)\rangle, \nonumber\\
&=&   \frac{e^2}{ i p_z\beta}\sum_{m} \int \frac{d^3p_1}{(2\pi)^3} {\rm Tr}[\gamma^x S_E(\omega_m, p' +p)\gamma^y S_E(\omega_m , p ) P_+].\nonumber\\
\end{eqnarray}
Using standard methods to  sum over the Matsubara  frequencies  we obtain
\begin{eqnarray}\label{mat4d}
& & \sum_{ m} \frac{1}{\beta}  
 \frac{1}{( i \omega_m + e\mu_-  -tE_p) ) ( i \omega_m + e\mu_- + i \omega_n - u E_{p+q}) }  
\\ \nonumber
 & &\qquad\qquad =  \frac{  tf ( E_p- te\mu_-)  - uf( E_{p+q}  - ue\mu_- - i \omega_n )}
 { i \omega_n +tE_p -uE_{p+q}}, \\ \nonumber
& & \sum_{m} \frac{1}{\beta}  
\frac{i \omega_m + e\mu_- }{ {( i \omega_m + e\mu_-  -tE_p) ) ( i \omega_m + e\mu_- + i \omega_n - u E_{p+q}) } } 
\\ \nonumber
 & & \qquad\qquad=  \frac{  E_p f ( E_p- te\mu_-)  - E_{p+q} f( E_{p+q}  - ue\mu_- - i \omega_n )}
 { i \omega_n +tE_p -uE_{p+q}} , \\ \nonumber
 & & \sum_{m} \frac{1}{\beta}  
\frac{(i \omega_m + e\mu_- )^2}{ {( i \omega_m + e\mu_-  -tE_p) ) ( i \omega_m + e\mu_- + i \omega_n - u E_{p+q}) } } 
\\ \nonumber
 & & \qquad\qquad=  \frac{  tE_p^2 f ( E_p- te\mu_-)  - uE_{p+q}^2 f( E_{p+q}  - ue\mu_- - i \omega_n )}
 { i \omega_n +tE_p -uE_{p+q}} . \\ \nonumber
\end{eqnarray}
While performing all these sums, we have ignored terms which are independent of temperature 
and chemical potential.
Using the results for the sums as well as performing the trace in (\ref{defjj}), we obtain
\begin{eqnarray}
I(p_z) &=& \sum_{t,u=\pm}\frac{e^2}{2}\int \frac{d^3p'}{(2\pi)^3} \frac{t}{E_{p'+p}} \frac{t f(E_{p'+p}-te\mu _-)-uf(E_{p'}-ue\mu _-)}{tE_{p'+p}-uE_{p'}}. \nonumber\\
\end{eqnarray}                 
After changing the variable of integration in the first term of the above equation, we obtain  
\begin{eqnarray}
I (p_z)&=& \sum_{t,u=\pm}\frac{e^2}{2}\int \frac{d^3p'}{(2\pi)^3} (\frac{f(E_{p'}-te\mu _-)}
{E_{p'}(tE_{p'}-uE_{p'+p})} +\frac{tu f(E_{p'}-ue\mu _-)}{E_{p'+p}(uE_{p'}-tE_{p'+p})}).\nonumber\\ 
\end{eqnarray}
Summing over $u$ in the first term and over $t$ in the second term, we get,
\begin{eqnarray}
I (p_z)&=& 2e^2\int \frac{d^3p_1}{(2\pi)^3}
\frac{tf(E_{p'}-te\mu _-)}{(E_{p'}^2-E_{p'+p}^2)},\nonumber\\
&=&-\frac{2e^2 }{(2\pi)^2} \sum_{t=\pm}\int dp' p^{\prime 2}
 \sin \theta tf(E_{p'}-te\mu _-) \frac{1}{(p^2+2p'p_z\cos \theta)}.
 \end{eqnarray}
 At this stage we need to perform the angular integral.  This integral can of course be easily 
 performed, however we will  show that we can take the $p_z\rightarrow 0$  limit 
 first and then perform the integral. Though the advantage of this procedure is limited in 
 $d=4$, we will see that for $d=6$ this procedure simplifies the resulting integrals considerably. 
 Expanding  in $p_z$ we obtain 
 \begin{equation}
 I(p_z) 
=-\frac{2e^2 }{(2\pi)^2} \sum_{t=\pm}\int dp' p^{\prime 2}
 \sin \theta tf(E_{p'}-te\mu _-)\left (\frac{1}{2p' p_z \cos \theta}- \frac{1}{4 p^{\prime 2}  \cos ^2 \theta} +
 \,  O(p_z^2) \right ).\nonumber\\
\end{equation}
After a change of variables $z= \cos\theta$, the required integrals are of the form
\begin{equation}
J_1 =\int_0^\pi  \frac{\sin\theta d\theta}{\cos\theta} = 
 \int_{-1}^{1} \frac{dz}{z},  \qquad \qquad 
 J_2 = \int_0^\pi \frac{\sin\theta d\theta}{\cos^2\theta} = \int_{-1}^1 \frac{dz}{z^2}.
\end{equation}
The integrals are all on the real line, we make these integrals well defined by shifting the 
integrand to $z\rightarrow z+ i\epsilon$. Then we obtain 
\begin{equation}
J_1 = 0, \qquad J_2 = -2.
\end{equation} 
Such a prescription for doing the resulting integrals renders each term in the expansion in $p_z$ 
finite. In fact  it is easy to check that performing the angular integral first  and then 
expanding in $p_z$, gives identical result to expanding in $p_z$ first and then doing the
integrals term by term using the $i\epsilon$ prescription. 
From these results for the integral, we obtain
\begin{eqnarray}
\tilde \zeta_1^{(4)} = \lim_{ p_z\rightarrow 0}\frac{1}{ip_z}\langle j^x(p)j^y(-p)\rangle &=&- \frac{e^2}{4\pi^2} \sum_{t=\pm} \int dp_1 t f(E_{p_1}-te\mu_-),\nonumber\\
&=& -\frac{e^3}{4\pi^2}\mu_-.
\end{eqnarray}
Finally  using  ( \ref{relem3e})  to obtain the 
transport coefficient $\zeta^{(4)}_1$,   we get
\begin{eqnarray}
\zeta^{(4),R}_1= \tilde \zeta^{(4)}_1  = 
- \frac{e^3}{4\pi^2} \mu_-.
\end{eqnarray}

 Let us now relate this  parity odd  transport coefficient with the coefficient of the 
 coefficient of the $U(1)$ anomaly. 
 The anomalous conservation equation for the $U(1)$ current is given by,
\begin{eqnarray}\label{gaugeanom4d}
\nabla_\mu j^\mu &=& \frac{1}{4} \epsilon^{\mu \nu \rho \lambda} [3c_A F_{\mu \nu } F_{\rho \lambda}+ c_m R^\alpha_{\beta \mu \nu}R^{\beta}_{\alpha \rho \lambda}].
\end{eqnarray}
The gauge and mixed anomaly coefficients $c_A$ and $c_m$ for a single Weyl fermion are 
given by \footnote{For example,  see (2.39) of \cite{Jensen:2013rga}.}
\begin{eqnarray}\label{gaugecoeff4d}
c_A = \frac{e^3}{24\pi^2}, \qquad c_m = \frac{e}{192\pi^2}.
\end{eqnarray}
Comparing the coefficient of the anomaly and the transport coefficient, we see that 
chiral fermions obey the relation
\begin{eqnarray}
\zeta^{(4)}_1 &=& -6 c_A \mu_-. 
\end{eqnarray}
This relation obeys the replacement rule relating the gauge anomaly to the transport 
coefficient $\zeta^{(4)}_1$ found  in  \cite{Jensen:2013rga}. Alternatively, the coefficients $c_A$ and $c_m$ can be read out from the anomaly polynomial in four dimensions ,
\begin{eqnarray}\label{anompoly4d}
{\cal P}_{d=3+1} (\hat{F},\hat{R}) &=& c_A \hat{F}\wedge \hat{F}\wedge \hat{F} +c_m \hat{F} \wedge {\rm tr}(\hat{R} \wedge \hat{R}).
\end{eqnarray}

\subsection*{Evaluation of $\zeta^{(4)}_2$}
\begin{figure}[!htb]\label{jt4diag}
  \centering
   \hspace*{-2cm}\includegraphics[scale=0.2]{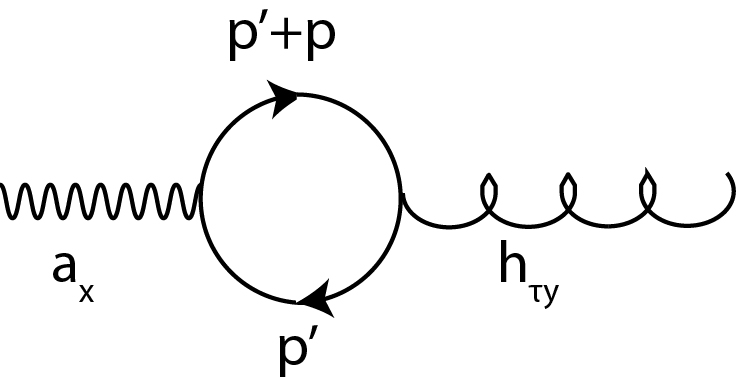} 
   \caption{Diagram  contributing to $\zeta^{(4)}_2$}
\end{figure}

We first evaluate  the  relevant two point function in the Euclidean theory, this is given by
\begin{eqnarray}
\tilde\zeta^{(4)}_2 = \lim_{ p_z\rightarrow 0}\frac{1}{ip_z} 
\langle j^{x}(0, p_z)T^{\tau y}(0, -p_z)\rangle.\nonumber\\
\end{eqnarray} 
This correlator is defined by (\ref{jt1}) and from the expansion of the action in 
(\ref{action4dexp}), we see that it does not involve any contact terms. 
The expression for the current in momentum space is given by \eqref{current4d}) . 
Converting the stress tensor in \eqref{stress4d})  to momentum space we obtain
\begin{eqnarray}
T^{\tau y}_{fl}(-p)= 
\frac{1}{\beta}\sum_{\omega_m}\frac{1}{2}\int \frac{d^3 p'}{(2\pi)^3} \psi_E^\dagger(p'+p)(\gamma^\tau ip'_{y}+\gamma^y(i\omega_m + e\mu _-))P_-\psi_E(p').\nonumber\\
\end{eqnarray}
After performing the Wick contractions we get 
\begin{eqnarray}
& &  \frac{1}{ip_z} \langle j^{x}(0, p_z)T^{\tau y}(0, -p_z)\rangle  =  \\ \nonumber
  & & \quad 
  \frac{e}{2\beta p_z} \sum_{m} \int \frac{d^3p'}{(2\pi)^3} {\rm Tr}[\gamma^x S_E(p'+p_z )(\gamma^\tau ip'{y}+\gamma^y (i\omega_m + e\mu _-))S_E(p') P_+].
\end{eqnarray} 
For convenience we break up the terms resulting from the 
 the Wick contractions  into two parts, $I_1$ and  $I_2$, which are given by
\begin{eqnarray}
I_1(  p_z) &=&  \frac{e}{i 8\beta p_z} \sum_{ m} \int \frac{d^3p' }{(2\pi)^3} 
{\rm Tr}[\gamma^x \gamma^\alpha \gamma^\tau \gamma^\beta P_+]p'_{y}(\hat p'+\hat p)_{t,\alpha} (p')_{u,\beta}\nonumber\\
&& \Delta_t(i\omega_m  ,p' +p)\Delta_u(i\omega_m,p ' ),\nonumber\\ 
&=&\frac{ie}{4(2\pi)^3} \sum_{t=\pm} \int \frac{\sin^3 \theta}{\cos^2 \theta} \cos^2 \phi d\theta d\phi
\int dp'  p'  tf(p' -te\mu_-) + O(p_z ), \nonumber\\
I_2(   p_z) &=&-  \frac{e}{ 8\beta p_z } \sum_{ m} \int \frac{d^3p' }{(2\pi)^3}
{\rm Tr}[\gamma^x \gamma^\alpha \gamma^y \gamma^\beta P_+](i\omega_m+e\mu _-)(\hat p' +\hat p)_{t,\alpha} \hat p'_{u,\beta}\nonumber\\
&& \Delta_t(i\omega_m   ,p' +p)\Delta_u(i\omega_m,p' ),\nonumber\\
&=& \frac{ie}{4(2\pi)^3} \sum_{t=\pm} \int \frac{\sin \theta}{\cos^2 \theta}
d\theta d\phi \int dp'  p'  tf(p' -te\mu_-) + O(p_z ). \nonumber\\
\end{eqnarray}
Here, in the second and the last line of the above equation, we have performed the 
Matsubara sums,  expanded the integrand in the external momenta $p_z$ and finally performed 
the angular integrals using the $i\epsilon$ prescription defined earlier. 
Note that the angular integrals are considerable simpler than the ones done first in 
\cite{Landsteiner:2011cp} to evaluate this correlator. 
Using  these results, we obtain 
\begin{eqnarray}
\tilde \zeta^{(4)}_2 & =&  
 \lim_{ p_z\rightarrow 0}\frac{1}{ip_z} 
\langle j^{x}(0, p_z)T^{\tau y}(0, -p_z)\rangle= 
\lim_{p_z\rightarrow 0} ( I_1(p_z) +I_2(p_z) ),  \\ \nonumber
&=& -\frac{i e}{4\pi^2} 
 \left(  \int_0^{\infty} dp'  p'  ( f(p'  -e\mu _-) + f( p' + e\mu_-) \right), \nonumber\\
&=&- \frac{ie}{8\pi^2} (e^2 \mu _-^2 + \frac{\pi^2 T^2}{3}). 
\end{eqnarray} 

We can now relate the above Euclidean correlator, to the desired  transport coefficient using (\ref{relem3e}). This results in 
the following expression 
\begin{eqnarray}\label{4dcorrelator2}
\zeta^{(4)}_2&=&\frac{-e}{8\pi^2} (e^2 \mu _-^2 + \frac{\pi^2 T^2}{3}). 
\end{eqnarray}

Finally let us relate the two terms, that occur in the  transport coefficient $\tilde\zeta^{(4)}$, to the two anomaly 
coefficients $c_A$ and $c_m$, defined in (\ref{gaugeanom4d}) and (\ref{gaugecoeff4d}). 
First parametrize  $\zeta^{(4)}$ following  \cite{Jensen:2013rga} as 
\begin{eqnarray}
\zeta^{(4)}_2=-3c_A \mu_-^2+\tilde{c}_{4d} T^2.
\end{eqnarray}
Comparing with the results from Kubo formulae \eqref{4dcorrelator2}, we obtain
\begin{eqnarray} \label{zeta4dans}
c_A=\frac{e^3}{24 \pi^2}, \qquad\qquad  \tilde{c}_{4d}= -\frac{e}{24}.
\end{eqnarray}
From \eqref{gaugeanom4d}, \eqref{gaugecoeff4d} and \eqref{anompoly4d}, we see that value of $c_A$ obtained using the Kubo formula 
for $\zeta^{(4)}_1$ agree. This is a consistency check on our calculations. 
Finally we obtain the relation of the 
\begin{eqnarray}
\tilde{c}_{4d}= -8\pi^2 c_m.
\end{eqnarray}
This relation was obtained using  consistency of the  Euclidean vacuum in 
in \citep{Jensen:2013rga}.  This direct perturbative calculation provides a check on this  argument. 
Note that the transport coefficient $\zeta^{(4)}_2$ 
has been evaluated earlier by \cite{Landsteiner:2011cp},  here we have demonstrated that 
the angular integrals simplify considerably on taking the zero momentum limit first. In fact 
doing this leads directly  to the moments of the Fermi-Dirac distribution.

\subsection*{Evaluation of $\lambda^{(4)}_2$}
\begin{figure}[!htb]\label{tt4diag}
  \centering
   \hspace*{-2cm}\includegraphics[scale=0.2]{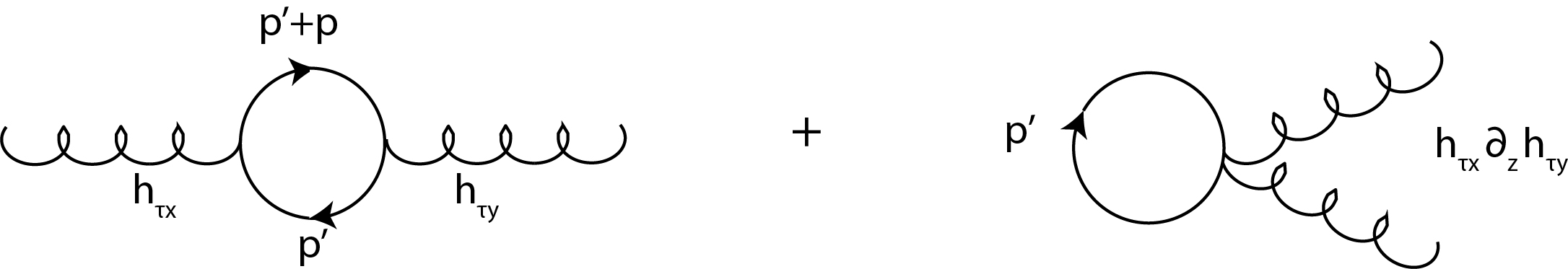} 
   \caption{Contributions to $\lambda^{(4)}_2$  }
\end{figure}

 The last transport coefficient of interest in  4 dimensions is $\lambda^{(4)}_2$.   The corresponding 
coefficient in the Euclidean theory is given by the following  two point function 
\begin{eqnarray}
\tilde\lambda^{(4)}_2 &=&\lim_{p_z\rightarrow 0} \frac{1}{ip_z} \langle T^{\tau x}(0, p_z)T^{\tau y}(0, p_z )\rangle\nonumber\\
\end{eqnarray}    
From (\ref{tt1}) and the perturbative expansion of the action in (\ref{action4dexp}), we see that there is  a contact term due to 
 $S^{(2)}$  from the action.  Therefore we obtain 
\begin{eqnarray} \label{lambcont}
\tilde\lambda^{(4)}_2  &=&  A + B,  \\ \nonumber
A &=& \lim_{ p\rightarrow 0, p_0\rightarrow 0}\frac{1}{ip_z} \langle T^{\tau x}_{fl}(0, p_z) T^{\tau y}_{fl}(0, -p_z )\rangle_E, \\
\nonumber
B  &=& \lim_{ p_z\rightarrow 0}\frac{1}{ip_z}  ( \frac{-ip_z }{8 \beta } ) 
 \sum_{ m} \langle \int \frac{d^3p' }{(2\pi)^3} \psi_E^\dagger(\omega_m , p') \Gamma^{zxy} P_- \psi_E (\omega_m, p' ) 
 \rangle. 
\end{eqnarray}                      
The expression for the stress tensors in momentum space can be read out from  the following expressions 
\begin{eqnarray}
T^{\tau x}_{fl}(p) &=& \frac{1}{2 \beta }\sum_{m}\int \frac{d^3p' }{(2\pi)^3} \psi_E^\dagger(p' -p) 
[\gamma^\tau ip'_{x} + \gamma^x ( i\omega + e\mu_-)  ] P_-\psi_E(p'),\nonumber\\
T^{\tau y}_{fl}(-p) &=& \frac{1}{2\beta } \sum_{m}
\int \frac{d^3p' }{(2\pi)^3} \psi_E^\dagger(p' +p) [\gamma^\tau ip'_{y} + \gamma^y ( i\omega_m  + e\mu_- ) ]P_-\psi_E(p'). 
\nonumber\\ 
\end{eqnarray}   
This contribution from the contact term $B$  in (\ref{lambcont})  
was absent in the calculations of \cite{Landsteiner:2012kd}  and was considered in 
 the recent work of \cite{Yee:2014dxa}.  We will see that only after this contact term is considered, we will obtain the result 
 consistent with the replacement rule. 
Performing the Wick contractions and the Matsubara sums, the term $A$ reduces to                                             
\begin{eqnarray}\label{4dA}
A &=&  \lim_{ p_z\rightarrow 0} \frac{\langle T^{\tau x}(0, p_z )T^{\tau y}(0, - p_z)\rangle}{ip_z},  \\ \nonumber
&=& \frac{1}{16 \pi^2} \left( (e\mu _{-})^3+ \pi^2 T^2\mu _{-} \right).
\end{eqnarray}                                       
We now need to compute the contribution due to the contact term.  This is given by 
\begin{eqnarray}\label{4dB}
B  &=& \frac{i}{\beta}\sum_{m, t=\pm} \frac{1}{8} \int \frac{d^3p'}{(2\pi)^3}  
{\rm Tr}[\gamma^\alpha \gamma^z \gamma^x \gamma^y P_-] \left( 
\frac{1}{2} \hat{p'}_{t,\alpha} \Delta_t(i\omega_m, p') \right), \nonumber\\
&=&  \frac{1}{8} \sum_{t=\pm} \int \frac{d^3 p' }{(2\pi)^3} t f(E_{p' }-te\mu _-),\nonumber\\
&=&\frac{1}{16\pi^2}\int dp' p'^2  \left(  f(E_{p_1}-e\mu _-) - f(E_{p_1}+e\mu _-)  \right), \nonumber\\
&=&\frac{1}{48\pi^2} ((e\mu _{-})^3+ \pi^2 T^2\mu _{-}).\nonumber\\ 
\end{eqnarray}                  
Combining the results of term $A$ and $B$, from (\ref{4dA}) and (\ref{4dB}), we get          
\begin{eqnarray}
\tilde \lambda^{(4)}_2
&=&\frac{1}{12 \pi^2} ((e\mu _{-})^3+ e\pi^2 T^2\mu _{-}).
\end{eqnarray}
We can now use the relation (\ref{relem3e}) to obtain the transport result of interest
\begin{eqnarray}\label{4dcorrelator3}
\lambda^{(4)}_2 &=& - \tilde \lambda^{(4)}_2,\nonumber\\
&=&\frac{-1}{12 \pi^2} ((e\mu _{-})^3+ e\pi^2 T^2\mu _{-}). 
\end{eqnarray}

Let us now verify that this result is consistent with the replacement rule. 
 Following \cite{Jensen:2013rga}, we  parametrize   transport coefficient as 
\begin{eqnarray}
 \lambda^{(4)}_2 &=& 2(-c_A \mu_-^3 + \tilde{c}_{4d} T^2 \mu_-).
\end{eqnarray}
Comparing this parametrization  with \eqref{4dcorrelator3}, we  obtain 
\begin{eqnarray}
c_A = \frac{e^3}{24\pi^2}, \qquad\qquad   \tilde{c}_{4d} =-  \frac{e}{24}. \nonumber\\
\end{eqnarray}
From \eqref{gaugecoeff4d}, we see that the value of $c_A$ is that of a single Weyl fermion.  Furthermore $\tilde c_{4d}$
is identical to that obtained evaluating $\zeta^{(4)}_1$ in (\ref{zeta4dans}) and it also satisfies the relation
\begin{equation}
\tilde{c}_{4d}= -8\pi^2 c_m.
\end{equation} 
This relation was found by the argument involving the consistency of the Euclidean 
vacuum in \cite{Jensen:2013rga}. 
It is important to note that the contribution of the contact term $B$ was important to arrive at these results.

The transport coefficients involving chiral fermions in  $d=4$ are summarized in the table \eqref{table:2}.
    \begin{table}
\centering
 \begin{tabular}{|c|c|c|}
 \hline
 Dimension & Correlator  & Value \\ [1.0ex] 
 \hline\hline
 
 d=4 & $\zeta^{(4)}_1$ &  $\frac{-e^3}{4\pi^2}\mu _-$ \\ [1ex]
 \hline
 & $\zeta^{(4)}_2$ &  $\frac{-e}{8\pi^2} (e^2 \mu _-^2 + \frac{\pi^2 T^2}{3}) $ \\ [1ex]
 \hline
& $\lambda^{(4)}_2$ & $-\frac{1}{12 \pi^2} ((e\mu _{-})^3+ e\pi^2 T^2\mu _{-})$ \\ [1ex]
 \hline 
\end{tabular}
\caption{Transport coefficients In 4 Dimensions}
\label{table:2}
\end{table}      
We note that in \cite{Loganayagam:2012zg},  the contribution of chiral gravitinos to these transport coefficients was discussed. 
However gravitinos cannot be charged in flat space without violating the
gauge symmetry, $\psi_\mu \rightarrow \psi_\mu + \partial_\mu \epsilon$.   Therefore they do not admit any $U(1)$ charge 
current which implies there is no contribution to these transport coefficients from chiral gravitinos \footnote{We thank
Zohar Komargodski for emphasizing this to us. }.

\section{Anomalies and transport in d=6}\label{an6d}

In this section we evaluate all the parity odd transport coefficients occurring in $d=6$, at the leading order in 
derivatives, using the same methods developed in the previous section. 
This analysis will serve as a check of the computation in \cite{Yee:2014dxa}, of transport coefficients 
determined only  by the charge current gauge anomaly. This was  done using different approach
than the one adopted in this paper. 
Our  analysis will also  use   the $i\epsilon$ prescription, developed in the previous section, to perform the 
angular integrals.  Therefore  the results of the $d=6$ calculations also test this prescription. 
Finally we also determine the 
contributions to the transport coefficients from the mixed as well as the pure gravitational anomaly, 
which was not done in \cite{Yee:2014dxa}.  Our results are in agreement with the replacement rule 
for the mixed and pure gravitational anomaly, found by \cite{Jensen:2012kj, Jensen:2013rga} for chiral fermions. 
But for the case of chiral gravitinos which contribute to the pure gravitational anomaly, we see 
that the their contribution to the transport coefficients does not obey the replacement rule. 
We precisely identify the terms which contribute to the anomaly polynomial but not to the
transport coefficient.

We begin with parametrization of the constitutive relations in $d=6$, where we focus on the 
the  leading parity odd transport coefficients.  Following  the parametrization given in \cite{Yee:2014dxa},  the charge 
current and the stress tensor is given by 
\begin{eqnarray} \label{6dconsti}
j^{\mu }&=& n u^{\mu }  + 
\frac{\zeta^{(6)}_1}{3} \epsilon^{\mu \nu \alpha \beta \gamma \delta} u_{\nu }F_{\alpha \beta}F_{\gamma \delta}\nonumber\\
&&+\frac{\zeta^{(6)}_2}{3} \epsilon^{\mu \nu \alpha \beta \gamma \delta} u_{\nu }\partial_\alpha u_ \beta F_{\gamma \delta}
+\frac{\zeta^{(6)}_3}{3} \epsilon^{\mu \nu \alpha \beta \gamma \delta} 
u_{\nu }\partial_{\alpha} u_{\beta} \partial_{\gamma} u_{\delta},\nonumber\\
T^{\mu \nu }&=&(\epsilon+p)u^{\mu }u^{\nu }-p g^{\mu \nu }+ u^{\mu }(\frac{\lambda^{(6)}_1}{3} \epsilon^{\nu \rho \alpha \beta \gamma \delta} u_{\rho }F_{\alpha \beta}F_{\gamma \delta}\nonumber\\
&&+\frac{\lambda^{(6)}_2}{3} \epsilon^{\nu \rho \alpha \beta \gamma \delta} u_{\rho }\partial_\alpha u_ \beta F_{\gamma \delta}+\frac{\lambda^{(6)}_3}{3} \epsilon^{\nu \rho \alpha \beta \gamma \delta} u_{\rho }\partial_{\alpha} u_{\beta} \partial_{\gamma} u_{\delta})+(\mu \rightarrow \nu ),\nonumber\\    
\end{eqnarray} 
where, $\epsilon,p,n,\mu $ are the thermodynamic variables denoting energy density, pressure and charge density. 
The six anomalous transport coefficients are denoted by  
$\lambda^{(6)}_1,\lambda^{(6)}_2, \lambda^{(6)}_3, \zeta^{(6)}_1,\zeta^{(6)}_2, \zeta^{(6)}_3$. 
 The Minkowski metric, we work in, has the signature $( \eta_{\mu \nu }) ={\rm diag}\, \left(1,-1,-1,-1,-1,-1\right)$.
 Let us also label the directions as follows, the position vector or momentum vector  has 
 components in directions labeled by $(t, a, z, x, y, b)$.  We choose the  orientation in $d=6$  so that  $\epsilon^{tazxyb} = 1$. 
 To arrive at the Kubo formula, we assume that the fluid is perturbed from rest and its 
 velocity field is given by  $u^\mu =\left(1,0,v^z,v^x,0,0)\right)$ . The  velocity field 
 $v^z( t, b) $ depends only on time $t$ and  coordinate $b$, while the velocity field $v^x(t, z)$ depends only on time 
 and the coordinate $z$. 
 We then consider the  metric perturbation $h_{tz}(t, b)$ and $h_{tx}(t, y)$, while the rest of the fluctuations 
 are set to zero. Similarly we consider 
 gauge fluctuations $a_z(t, b)$ and $a_x(t, y)$. We summarize the perturbations  and the coordinates
 they depend on in the 
 following equations
\begin{eqnarray} \label{perturb6d}
u^\mu &=& ( 1,\,  0,\,  v^z( t, b),\,  v^x (t, z),\,  0, \, 0), \\ \nonumber
  u_{\mu }&=& (1,\, 0,\, -v^z(t, b) +h_{tz}(t, b) ,\, -v^x(t, z) +h_{tx}(t, z) ,\, 0,\, 0),\\ \nonumber
   A_{\mu} &=& ( \mu, \, 0,\, a_z(t, b), \, a_x (t, y),\, 0,\, 0), \\ \nonumber
 g_{\mu \nu}&=&\eta_{\mu \nu} + h_{\mu \nu}, \qquad {\rm with}\;   h_{tz}(t, b) ,h_{tx}(t, y) \neq 0.                         
\end{eqnarray}       
 With this choice of the polarizations and perturbations, 
it can be easily seen that the  contributions proportional to the anomalous transport coefficients in (\ref{6dconsti}), appear 
at second order in fluctuations and derivatives. 
Therefore the Kubo formulae for these transport coefficients are given by three point functions. 
Expanding the constitutive relations  to second order, we obtain  following expressions relating 
the three point functions to the transport coefficients
\begin{eqnarray} \label{6dkubo}
\frac{8}{3} \zeta^{(6)}_1&=&- \lim_{k_y , p_b \rightarrow 0}\frac{\langle j^a(p+k)j^x(-k)j^z(-p)\rangle}{ik_y ip_b}, 
\\ \nonumber
\frac{2}{3} \zeta^{(6)}_2&=&- \lim_{k_y , p_b \rightarrow 0}
-\frac{\langle j^a(p+k) j^{x}(-k) T^{tz}(-p)\rangle}{ik_y ip_b},\nonumber\\
\frac{2}{3} \zeta^{(6)}_3&=& - \lim_{k_y , p_b \rightarrow 0}
-\frac{\langle j^a(k+p) T^{tx}(-k) T^{tz}(-p)\rangle}{ik_y ip_b},\nonumber\\ \nonumber
\; & \; \\ \nonumber
\frac{8}{3} \lambda^{(6)}_1 &=&-\lim_{k_y , p_b \rightarrow 0}\frac{\langle j^{a}(p+k)j^x(-k)T^{tz}(-p)\rangle}{ik_y ip_b}, 
\\ \nonumber
\frac{2}{3} \lambda^{(6)}_2 &=& -\lim_{k_y , p_b \rightarrow 0} \frac{\langle j^{a}(p+k)T^{tx}(-k)T^{tz}(-p)\rangle}{ik_y ip_b},
\nonumber\\
\frac{2}{3} \lambda^{(6)}_3 &=&-\lim_{k_y , p_b \rightarrow 0}\frac{\langle T^{ta}(p+k) T^{tx}(-k)T^{tz}(-p)\rangle}{ik_y ip_b}.
\nonumber
\end{eqnarray}
To un-clutter our notations, it is understood that the external frequencies are set to zero from the beginning 
and that the only non-zero component of the  momentum vectors  $p, k$ are in the $b$ and the $y$ directions
respectively. From (\ref{6dkubo}), we arrive at the following relations between the transport coefficients
\begin{eqnarray}
4 \lambda^{(6)}_1 =\zeta^{(6)}_2, \qquad \qquad 
\lambda^{(6)}_2 = \zeta^{(6)}_3. 
\end{eqnarray}
Therefore it is sufficient to focus on evaluating the coefficients $\zeta^{(6)}_1, \zeta^{(6)}_2, \zeta^{(6)}_3$ and 
$\lambda^{(6)}_3$. 
The strategy for evaluating the correlators will be same, we will first evaluate them in the Euclidean theory
and 
then analytically continue them to the Lorentzian theory. 
Let the value of the corresponding correlators in the Euclidean theory by $\tilde \zeta^{(6)}_1, \tilde \zeta^{(6)}_2, \tilde \zeta^{(6)}_3$
and $\tilde \lambda^{(6)}_3$. Then going carrying out the analytic continuation leads to the following relations
\begin{eqnarray}\label{6danlcont}
 \tilde \zeta^{(6)}_1 = \zeta^{(6)}_1, &\quad& \tilde\zeta^{(6)}_2 = i  \zeta^{(6)}_2, \\ \nonumber
 \tilde \zeta^{(6)}_3 = - \zeta^{(6)}_3, &\qquad& \tilde \lambda^{(6)}_3 = -i \lambda^{(6)}_3.
\end{eqnarray}

We will now proceed to evaluation of the contribution of these correlators, first for chiral fermions and then the contribution of the
chiral gravitinos to $\lambda^{(6)}_3$. 

\subsection{Chiral Fermions }

For Weyl fermions in $d=6$, the partition function is given by 
\begin{eqnarray}
{\cal Z} &=&\int {\cal D }\psi^\dagger {\cal D} \psi \exp(S_E),
\end{eqnarray}    
where $\psi$ is a  Euclidean Dirac spinor in $d=6$ and the Euclidean action is given by 
\begin{eqnarray} \label{action6d}
S_E = \int d\tau d^5x  \sqrt{g} e^{\mu}_a \psi^\dagger \gamma^a  D_\mu  P_{-}  \psi.
\end{eqnarray}  
The covariant derivative is defined as 
\begin{eqnarray}
D_\rho \psi =\partial_\rho \psi +\frac{1}{2}\omega_{\rho ab} \sigma^{ab} \psi + ieA_\rho \psi,
\end{eqnarray}                    
where, $\omega_{\rho ab}$ is the spin connection and $\sigma^{ab}=\frac{1}{4}[\gamma_a,\gamma_b]$. 
The gamma matrices follow the same conventions as before. 
The conventions for the metric, vierbein and their perturbations  remain same as that in 
$d=4$ which is given in equations (\ref{antigamma}), 
(\ref{4dmetgapert})  and (\ref{4dvierp}).  The chiral projection operator is defined by 
\begin{equation}
P_{\pm} = \frac{1}{2} ( 1 \pm \gamma^7), 
\qquad\qquad \gamma^7 = i \gamma^\tau\gamma^a\gamma^z\gamma^x\gamma^y \gamma^b.
\end{equation}
To evaluate the Feynman diagrams, we will need the following trace
\begin{equation}
{\rm Tr} ( \gamma^p\gamma^q\gamma^r\gamma^s\gamma^t\gamma^u\gamma^7) = 
-8 \epsilon^{pqrstu}, \qquad \qquad \epsilon^{\tau azxyb} = i.
\end{equation}
Wick contractions are performed by the propagator given in 
(\ref{method2propagator}). 
Expanding the action in in these fluctuations, we obtain
\begin{eqnarray}\label{action6dexp}
S^{(0)}_E &=&\int d\tau d^5x   \psi^\dagger  (\gamma^\tau D_\tau +\gamma^i \partial_i )P_- \psi,\nonumber\\
S^{(1)}_E &=& -\int d\tau  d^5x   \frac{h_{\mu \nu }}{4}
\psi^\dagger   (\gamma^\mu \partial^\nu + \gamma^\nu \partial^\mu)P_- \psi \nonumber\\
&&+ie\int d\tau  d^5x \psi^\dagger \gamma^\mu A_{\mu }\psi + O(ha) +  O(h^2, a) + O(h^3),  \nonumber\\
S_E^{(2)}&=& \int d\tau  d^5x 
\frac{h_{\lambda \alpha }\partial_\mu h_{\nu \alpha}}{16}\psi^\dagger   \Gamma^{\mu \lambda \nu }P_- \psi + O(h^3).
\end{eqnarray}
At first glance one would have expected that we must expand the action to at least 
$O(h^3), O(h^2 a)$ to obtain all contributions to the correlators of interest. 
This is because now we are evaluating  three point functions. 
However terms involving $ O(h, a) , O(h^3), O(h^2, a)$  in the expansion $S_E^{(1)}$  of  \eqref{action6dexp},  do not have sufficient 
number of $\Gamma$ matrices to ensure a non-zero result for the correlators of interest. 
We need  $6$  $\Gamma$ matrices to ensure non-zero result in $d=6$. 
There  is another reason the $O(h A)$ terms do not contribute. 
From the structure of the  perturbations  in (\ref{perturb6d}) of the metric and the gauge field, 
it is easy to see there are no terms of the type $O(h_{\tau x} a_z ) , O( h_{\tau z} a_x)$ and therefore
there is no contributions to the correlators of interest.
Finally the reason $O(h^3)$ term is not considered in $S_E^{(2)}$ of  \eqref{action6dexp} is again due to the fact that such 
terms do not have sufficient number of gamma matrices to saturate the correlator of interest. 
Therefore the terms in the expansion (\ref{action6dexp}) is sufficient 
for evaluating all the transport coefficients of interest.

\subsection*{Evaluation of $\zeta^{(6)}_1$}

\begin{figure}[!htb]
  \centering
   \hspace*{-2cm}\includegraphics[scale=0.2]{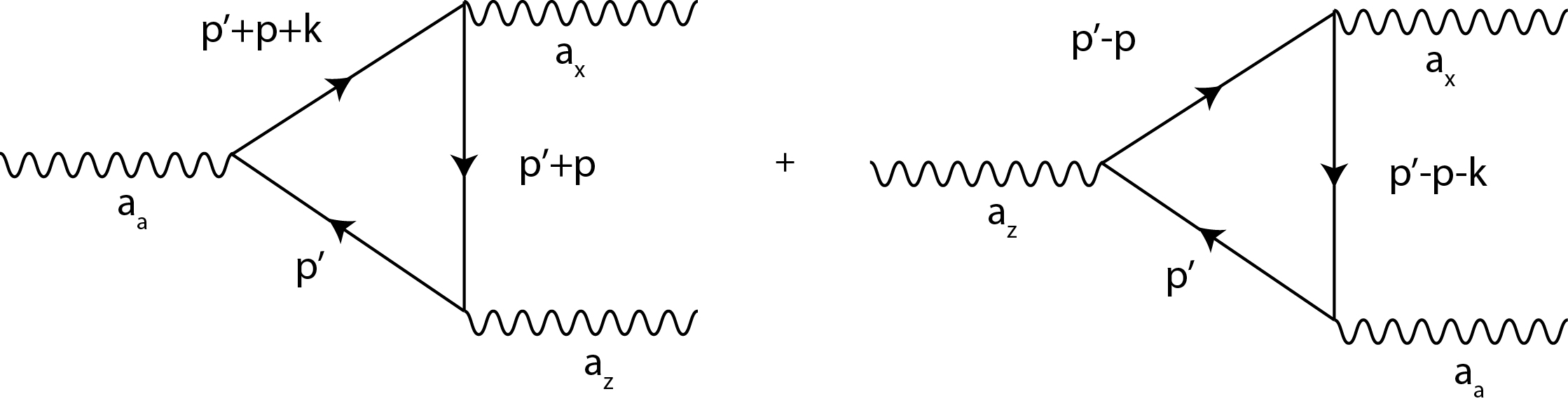} 
   \caption{Diagrams  contributing to  $\zeta^{(6)}_1$ }
   \label{jjjd}
\end{figure}

We begin by evaluating the coefficient $\zeta^{(6)}_1$, the corresponding  correlator is given by 
 \begin{eqnarray}\label{contact6d1}
\langle j^\mu j^{\nu}j^{\rho} \rangle &=&   
-\frac{1}{{ \cal Z}_E^{(0)}}
 \int {\cal D}{\psi^\dagger}{ \cal D} \psi e^{  S_{E}^{(0)}}\left(\frac{\delta S_E}{ \sqrt{g} \delta A_\mu}
\frac{\delta S_{E} } { \sqrt{g} \delta A_\nu}\frac{\delta S_{E} } {\sqrt{g}  \delta A_\rho} \right.  \\ \nonumber
&&+ \frac{\delta ^2 S_E}{  \sqrt{g} \delta A_\mu  \sqrt{g} \delta A_{\nu} }
 \frac{\delta S_{E} } {\sqrt{g} \delta  A_\rho}
+\frac{\delta ^2 S_E}{ \sqrt{g} \delta A_\mu \sqrt{g} \delta A_{\rho} }
 \frac{\delta S_{E} } {\sqrt{g} \delta A_\nu} \\ \nonumber
& & \left. +\frac{\delta S_E}{ \sqrt{g} \delta A_\mu }\frac{\delta^2 S_{E} }
   {\sqrt{g} \delta A_\nu \sqrt{g} \delta A_\rho}
+ \frac{\delta^3 S_{E} } {\sqrt{g} \delta A_\mu \sqrt{g} \delta A_\nu \sqrt{g} \delta A_\rho}\right).\nonumber
\end{eqnarray}
Here it is understood that only the connected correlators should be evaluated while performing Wick contractions. 
The correlator of  interest is given by 
\begin{eqnarray}\label{jjj}
\tilde \zeta^{(6)}_{1}&=& -\lim_{p_b,k _y\rightarrow 0}  \frac{3}{8}\frac{\langle j^a(k+p)j^x(-k)j^z(-p)\rangle }{ip_b ik_y}.            
\end{eqnarray}  
As we have mentioned earlier, we choose the kinematic configurations in which 
the momentum vectors $p, k $  have components only in the $b, y$ directions. 
The frequency dependence of the external momenta has been set to zero from the 
start. 
First note that   from the expansion of the action in (\ref{action6dexp}),  
the gauge fluctuations appear linearly and therefore there are no contributions 
from terms involving 2 derivatives of the action in (\ref{contact6d1}). 
Therefore the correlator in (\ref{jjj}) is obtained by performing Wick contractions 
of the flat space charge  current which is   given by 
\begin{equation}\label{6dcurr} 
j^\mu_{fl} (p) = -i e\frac{1}{\beta} \sum_{m} \int \frac{d^5 p'}{(2\pi)^5} 
\psi^\dagger(\omega_m, p'-p) \gamma^\mu P_-
\psi(\omega_m, p').
\end{equation}
After performing the two possible Wick contractions, we obtain
\begin{eqnarray}\label{jjjw}
& & \langle j^a(p+k) j^x(-k) j^z(-p)\rangle = 
-\frac{i e^3}{\beta}\sum_{ m}\int \frac{d^5 p_1}{(2\pi)^5} \\ \nonumber
& & \qquad\qquad\qquad \qquad\qquad \times \left( 
{\rm Tr}[\gamma^a S (\omega_m ,  p' +  k +  p)\gamma^x S( \omega_m,  p' +  p)\gamma^z S( \omega_m, p )P_+] \right. \\
\nonumber
 &&\qquad\qquad\qquad \qquad\qquad \left. + {\rm Tr}[\gamma^z S(\omega_m,   p'-  p )\gamma^x 
S(\omega_m ,  p' -  p-k )\gamma^a S_E( \omega_m, p' )P_+] \right). 
\end{eqnarray}        
The two Wick contractions are summarized in the figure  \ref{jjjd}. 
 The details of carrying out the Matsubara sums and then the angular integrals using the $i\epsilon$ prescription is
 given in the appendix \ref{Appendix A1}. 
 The end result for the Euclidean coefficient is 
\begin{eqnarray}
\tilde\zeta^{(6)}_{1}&=& \frac{3e^4}{64\pi^3} \int_0^\infty ( f( p - e\mu_-) - f( p+ e\mu_-)  ) , \\ \nonumber
 &=& \frac{3e^4}{64\pi^3} \mu_- . 
\end{eqnarray}  
 Using the relations for analytic continuation  given in  (\ref{6danlcont}), we obtain  
 \begin{eqnarray}
 \zeta^{(6)}_{1}&=& \frac{3e^4}{64\pi^3} \mu_-.  
 \end{eqnarray}
 
 Let us now compare these coefficients to that in the anomaly polynomial in $d=6$. 
 The anomaly polynomial for a single Weyl fermions in $d=6$ is given by \cite{Loganayagam:2012zg}
\begin{eqnarray}\label{anompoly6d}
{\cal P}_{d=5+1} ({\hat F} ,{ \hat R})&=& c_\alpha \hat F^4  + c_\beta  F^2 {\rm Tr} (\hat{R}^2) +
c_\gamma ({\rm Tr}(\hat{R}^2))^2 +c_\delta (\frac{1}{4}{\rm Tr }(\hat R^4)-\frac{1}{8}{\rm Tr}(\hat R^2)^2),\nonumber\\
\end{eqnarray}
where 
\begin{eqnarray}
c_\alpha =-\frac{e^4}{192\pi^3}, & \qquad&  c_\beta =-\frac{e^2}{768\pi^3}, \\
c_\gamma =-\frac{7}{184320 \pi^3}, & \qquad&  c_\delta =-\frac{1}{11520 \pi^3}. \nonumber
\end{eqnarray}
Thus the transport coefficient satisfies the relation
\begin{eqnarray}
\zeta^{(6)}_{1}= -9c_\alpha \mu_-.
\end{eqnarray}
This relation  is the expected relation between the gauge anomaly coefficient and the transport coefficient of 
interest, predicted by general considerations of hydrodynamics in \cite{Banerjee:2012cr,Jensen:2013rga}.

\subsection*{Evaluation of $\zeta^{(6)}_2$}

\begin{figure}[!htb]
  \centering
   \hspace*{-2cm}\includegraphics[scale=0.2]{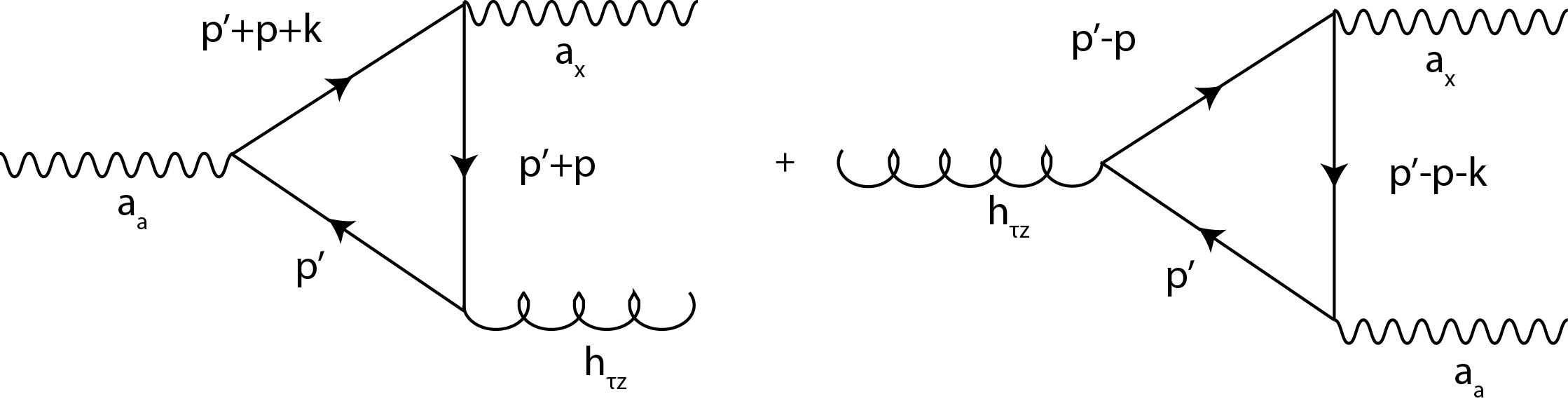} 
   \caption{ Contributions to  $\zeta^{(6)}_2$ }\label{jjtdiag}
\end{figure}

The correlator of interest for the transport coefficient $\zeta^{(6)}_2$ is given by 
\begin{eqnarray}
 \langle j^\mu j^{\nu}T^{\rho\sigma} \rangle  &=& -\frac{2}{{ \cal Z}_E^{(0)}} \int {\cal D}{\psi^\dagger}{ \cal D} \psi e^{  S_{E}^{(0)}} \times\\ \nonumber  
 &&\left(\frac{\delta S_E}
 {\delta \sqrt{g}  A_\mu} \frac{\delta S_{E} } {\sqrt{g} \delta A_\nu}
 \frac{\delta S_{E} } {\sqrt{g} \delta g_{\rho\sigma}} + \frac{\delta ^2 S_E}{ \sqrt{g} \delta A_\mu \sqrt{g}  \delta A_{\nu} }\frac{\delta S_{E} } 
 {\sqrt{g} \delta  g_{\rho\sigma}} \right.\nonumber\\
 &&+ \frac{\delta ^2 S_E}{ \sqrt{g} \delta A_\mu \sqrt{g} \delta  g_{\rho\sigma} }\frac{\delta S_{E} } 
 {\sqrt{g} \delta A_\nu}+ \frac{\delta S_E}{ \sqrt{g} \delta A_\mu }\frac{\delta^2 S_{E} } {\sqrt{g} \delta A_\nu\delta  \sqrt{g}  g_{\rho\sigma}}\nonumber\\
 &&\left. + \frac{\delta^3 S_{E} } {\sqrt{g} \delta a_\mu \sqrt{g} \delta A_\nu \sqrt{g} \delta g_{\rho\sigma}}\right).\nonumber
 \end{eqnarray}
Here again it is understood that only connected Wick contractions need to be evaluated. 
After evaluating the correlator in Euclidean space, we extract the following coefficient
\begin{equation}
 \label{jjt}
\tilde \zeta^{(6)}_{2}= -\lim_{p_b,k_y \rightarrow 0}
 \frac{3}{2}\frac{\langle j^a(k+p)j^x(-k)T^{\tau z}(-p)\rangle}{ip_b ik_y}.                    
\end{equation}
From the expansion of the action in (\ref{action6dexp}), we see that none of the contact terms contribute to this correlator and 
therefore we can evaluate the correlator in (\ref{jjt}), by performing Wick contractions  on the 
flat space charge current given in (\ref{6dcurr}) as well as the stress tensor given by 
\begin{eqnarray}
T^{\tau z}_{fl}(-p)&=&\frac{1}{\beta}\sum_{ m} 
\int \frac{d^5 p' }{(2\pi)^5}\psi^\dagger(p'+p)(\gamma^\tau ip'_{z}+\gamma^z ( i\omega_m + \mu_-) )P_-\psi(p').\nonumber\\
\end{eqnarray}
The two possible Wick contractions are summarized in the figure \ref{jjtdiag}. 
The details of the Wick contractions, Matsubara sums and angular integrals resulting in the calculation 
are provided in \ref{Appendix A2}. The end result of the corresponding Euclidean transport  coefficient is given by 
\begin{eqnarray}
\tilde \zeta^{(6)}_{2} &=& i \frac{3e^2}{16\pi^3} \int_0^\infty p ( f( p - e\mu_-)  - f( p+ e\mu_-) , \\ \nonumber
&=&i\frac{3e^2}{32\pi^3}(e^2\mu _{-}^2+\frac{\pi^2 T^2}{3}).
\end{eqnarray} 
Again using the relations in (\ref{6danlcont}), we obtain the transport coefficient of interest to be 
\begin{eqnarray} \label{6dfinalr}
\zeta^{(6)}_{2}&=&\frac{3e^2}{32\pi^3}(e^2\mu _{-}^2+\frac{\pi^2 T^2}{3}).
\end{eqnarray}

Let us compare this coefficient with the coefficients in the anomaly polynomial. 
For this, following \cite{Jensen:2013rga},  we first parametrize the transport coefficient as 
\begin{eqnarray}
\zeta^{(6)}_{2}&=& 3(-6c_\alpha \mu_-^3 + \tilde{c}^{6d}_m T^2),
\end{eqnarray}
where $c_\alpha$ for Weyl fermions is read out from   \eqref{anompoly6d}. Comparing (\ref{6dfinalr}),  we see that 
$\tilde{c}^{6d}_m$ is related to the mixed gravitational anomaly coefficient $c_\beta$ by
\begin{eqnarray}
\tilde{c}^{6d}_m &= -8\pi^2 c_\beta.
\end{eqnarray}
In \cite{Jensen:2013rga}, such a relation was obtained using the argument 
involving the   consistency of the Euclidean  vacuum. 
Here we have explicitly  demonstrated this relation using the perturbative computation.

\subsection*{Evaluation of  $\zeta^{(6)}_3$}

\begin{figure}[!htb]
  \centering
   \hspace*{-2cm}\includegraphics[scale=0.2]{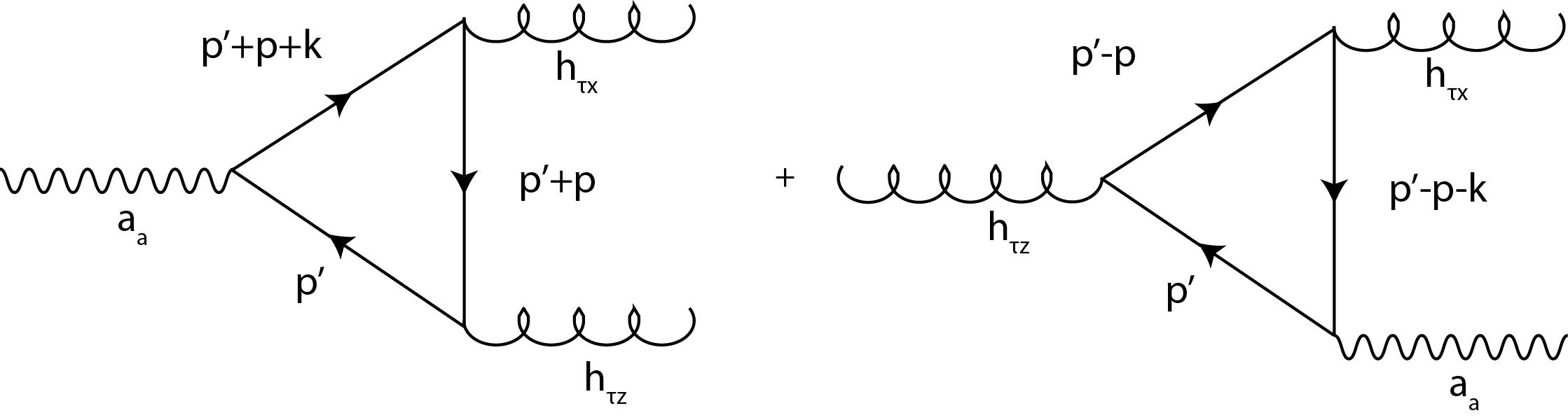} 
   \caption{ Contributions to $\zeta^{(6)}_3$ from Wick contractions of flat currents.}
    \label{jttdiag}
\end{figure}

\begin{figure}[!htb]
  \centering
   \hspace*{-2cm}\includegraphics[scale=0.2]{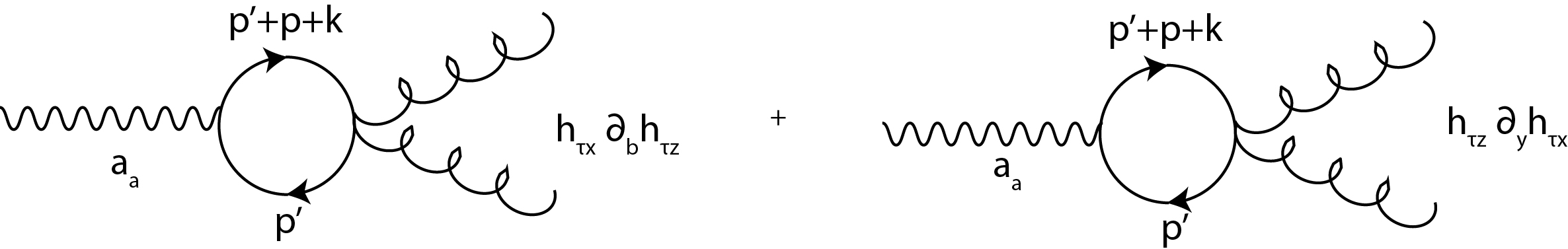}
   \caption{ Contributions to $\zeta^{(6)}_3$ from contact terms.}
    \label{jttdiag1}
\end{figure}

Now consider the correlator
\begin{eqnarray}
  \langle j^\mu T^{\nu\beta}T^{\rho\sigma} \rangle &=& -  \frac{4}{{ \cal Z}_E^{(0)}} \int {\cal D}{\psi^\dagger}{ \cal D}
 \psi e^{  S_{E}^{(0)}} \times\\ \nonumber
 & &\left( \frac{\delta S_E}{\sqrt{g} \delta A_\mu} 
 \frac{\delta S_{E} } {\sqrt{g} \delta g_{\nu\beta}}\frac{\delta S_{E} }
 {\sqrt{g} \delta g_{\rho\sigma}} +  \frac{\delta ^2 S_E}{ \sqrt{g} \delta A_\mu \sqrt{g} \delta  g_{\nu\beta} }
 \frac{\delta S_{E} } {\sqrt{g} \delta  g_{\rho\sigma}}\right.\nonumber\\
 & &+ \frac{\delta ^2 S_E}{ \sqrt{g}  \delta A_\mu \sqrt{g} \delta  g_{\rho\sigma} }
 \frac{\delta S_{E} } {\sqrt{g} \delta  g_{\nu\beta}}+  \frac{\delta S_E}{ \sqrt{g} \delta A_\mu }\frac{\delta^2 S_{E} } 
 {\sqrt{g} \delta  g_{\nu\beta} \sqrt{g} \delta g_{\rho\sigma}}\nonumber\\
 & &\left. + \frac{\delta^3 S_{E} } {\sqrt{g} \delta A_\mu  \sqrt{g}\delta g_{\nu\beta} \sqrt{g} \delta  g_{\rho\sigma}}\right). \nonumber
 \end{eqnarray}
 The corresponding transport coefficient in the Euclidean theory is given by 
\begin{eqnarray}\label{jtt}
\tilde \zeta^{(6)}_{3}&=&-\lim_{p_b,k_y \rightarrow 0} 
 \frac{3}{2}\frac{\langle j^a(k+p)T^{\tau x}(-k)T^{\tau z}(-p)\rangle}{ip_b ik_y}.\nonumber\\                     
\end{eqnarray}
Now from examining the expansion of the action in (\ref{action6dexp}), we see that the only non-zero 
contributions to the correlator are given by 
\begin{eqnarray}
  \langle j^\mu T^{\nu\beta}T^{\rho\sigma} \rangle &=& = A + B,  \\ \nonumber
 A & = &-  \frac{4}{{ \cal Z}_E^{(0)}} \int {\cal D}{\psi^\dagger}{ \cal D}
 \psi e^{  S_{E}^{(0)}}\frac{\delta S_E}{\sqrt{g} \delta A_\mu} 
 \frac{\delta S_{E} } {\sqrt{g} \delta g_{\nu\beta}}\frac{\delta S_{E} }
 {\sqrt{g} \delta g_{\rho\sigma} }, \\ \nonumber
 B &=& -  \frac{4}{ { \cal Z}_E^{(0)} }  \int {\cal D}{\psi^\dagger}{ \cal D} \psi e^{ S_{E}^{(0)} } 
 \frac{\delta S_E}{ \sqrt{g} \delta A_\mu }\frac{\delta^2 S_{E} } 
 {\sqrt{g} \delta  g_{\nu\beta} \sqrt{g} \delta g_{\rho\sigma}}.\nonumber
\end{eqnarray}
Therefore we have a contribution from the Wick contractions of the flat space currents 
denoted by $A$, 
together  with a contact term which involves Wick contraction with a flat space current denoted by $B$. 
These are summarized in  figure \ref{jttdiag}.  and figure \ref{jttdiag1}.  respectively. 
We have evaluated these contributions in detail in appendix \ref{Appendix A3}. 
The end result for the Euclidean correlator is given by 

\begin{eqnarray}
\tilde \zeta^{(6)}_{3}                   
&=&\frac{-3e}{16\pi^3}\int_0^\infty dp p^2 \left( f(p_1-e\mu _{-}) - f(p+ e\mu_-) \right) ,\nonumber\\
&=&\frac{-e}{16\pi^3}[(e\mu _{-})^3+ e\pi^2 T^2\mu _{-}].
\end{eqnarray}
Again using the relations in (\ref{6danlcont}) to go from the Euclidean correlator 
to the   transport coefficient  of interest in the Lorentzian theory, we obtain 
\begin{eqnarray}\label{6drjtt}
\zeta^{(6),R}_{3}           
&=&\frac{e}{16\pi^3}[(e\mu _{-})^3+ e\pi^2 T^2\mu _{-}]. 
\end{eqnarray}

Let us compare this transport coefficient to what is expected from the replacement rule.
Following \cite{Jensen:2013rga}, we parametrize the coefficient as 
\begin{eqnarray}
\zeta^{(6),R}_{3}&=&6(-2c_\alpha \mu_-^3 + \tilde{c}^{6d}_m T^2 \mu_- ).
\end{eqnarray}
Comparing the final result of the perturbative evaluation in (\ref{6drjtt}) and the above equation 
we see that indeed the results for the coefficient  $c_\alpha$  agrees with the 
gauge anomaly of a single Weyl fermion given in (\ref{anompoly6d}). 
Also we have the relation
\begin{equation}
\tilde c^{6d}_m  =   -8\pi^2c_\beta,
\end{equation}
which is again consistent with that obtained by the argument involving the consistency of the  
Euclidean 
vacuum.

\subsection*{Evaluation of $\lambda^{(6)}_3$} \label{6dfermig}

The last transport coefficient in $d=6$ is given by the following correlator 
\begin{eqnarray} 
  \langle T^{\mu\alpha} T^{\nu\beta}T^{\rho\sigma} \rangle &=& 
 -\frac{8}{{ \cal Z}_E^{(0)}} \int {\cal D}{\psi^\dagger} {\cal D}{ \psi} \exp( S_E^{(0)} )  \times  
 \\ \nonumber
  & & 
\left(  \frac{\delta S_E}{\sqrt{g} \delta g_{\mu\alpha}} 
  \frac{\delta S_{E} } {\sqrt{g} \delta g_{\nu\beta}}\frac{\delta S_{E} } {\sqrt{g} 
  \delta g_{\rho\sigma}} + 
\frac{\delta ^2 S_E}{ \sqrt{g} \delta g_{\mu\alpha} \sqrt{g} \delta  g_{\nu\beta} }
\frac{\delta S_{E} } {\sqrt{g} \delta  _{\rho\sigma}} \right. \nonumber\\
 && + 
 \frac{\delta ^2 S_E}{ \sqrt{g} \delta g_{\mu\alpha} \sqrt{g} \delta  g_{\rho\sigma} }\frac{\delta S_{E} } 
 {\sqrt{g} \delta  g_{\nu\beta}}+
 \frac{\delta S_E}{ \sqrt{g} \delta g_{\mu\alpha} }\frac{\delta^2 S_{E} } 
 {\sqrt{g} \delta  g_{\nu\beta} \sqrt{g} \delta  _{\rho\sigma}}\nonumber\\
 &&\left. 
 + 
 \frac{\delta^3 S_{E} } {\sqrt{g} \delta g_{\mu\alpha}  \sqrt{g} \delta  g_{\nu\beta} \sqrt{g} \delta  g_{\rho\sigma}} \right).
 \nonumber
 \end{eqnarray} 
The coefficient in the Euclidean theory is given by 
\begin{eqnarray}\label{ttt}
\tilde \lambda^{(6)}_{3}&=& -\frac{3}{2} \lim_{p_b,k_y \rightarrow 0}
\frac{-\langle T^{\tau a}(k+p)T^{\tau x}(-k)T^{\tau z}(-p)\rangle}{ip_b ik_y}                   .
\end{eqnarray}  
From the expansion of the action in (\ref{action6dexp} ), 
we see that only non-zero contributions to this correlator arise from  
\begin{eqnarray} \nonumber
 \langle T^{\mu\alpha} T^{\nu\beta}T^{\rho\sigma} \rangle &=&   A + B , \\ \nonumber
 A &=& - \frac{8}{{ \cal Z}_E^{(0)}} \int {\cal D}{\psi^\dagger} {\cal D}{ \psi} \exp( S_E^{(0)} ) 
\frac{\delta S_E}{\sqrt{g} \delta g_{\mu\alpha}} 
  \frac{\delta S_{E} } {\sqrt{g} \delta g_{\nu\beta}}\frac{\delta S_{E} } {\sqrt{g} 
  \delta g_{\rho\sigma}}, \\ \label{tttc}
  B &=& -\frac{8}{{ \cal Z}_E^{(0)}} \int {\cal D}{\psi^\dagger} {\cal D}{ \psi} \exp( S_E^{(0)} )    
  \left( 
  \frac{\delta ^2 S_E}{ \sqrt{g} \delta g_{\mu\alpha} \sqrt{g} \delta  g_{\nu\beta} }
\frac{\delta S_{E} } {\sqrt{g} \delta  _{\rho\sigma}}  \right.  \\ \nonumber
& &  \left.   + 
 \frac{\delta ^2 S_E}{ \sqrt{g} \delta g_{\mu\alpha} \sqrt{g} \delta  g_{\rho\sigma} }\frac{\delta S_{E} } 
 {\sqrt{g} \delta  g_{\nu\beta}}+
 \frac{\delta S_E}{ \sqrt{g} \delta g_{\mu\alpha} }\frac{\delta^2 S_{E} } 
 {\sqrt{g} \delta  g_{\nu\beta} \sqrt{g} \delta  _{\rho\sigma}} \right). 
 \end{eqnarray}
 Term $A$ involves Wick contractions of the flat space stress tensor while 
 $B$ is a contact term.  From the explicit evaluation of term B 
 in appendix \ref{Appendix A4}, it is seen that all terms in $B$ contribute equally. 

\begin{figure}[!htb]
  \centering
   \hspace*{-2cm}\includegraphics[scale=0.2]{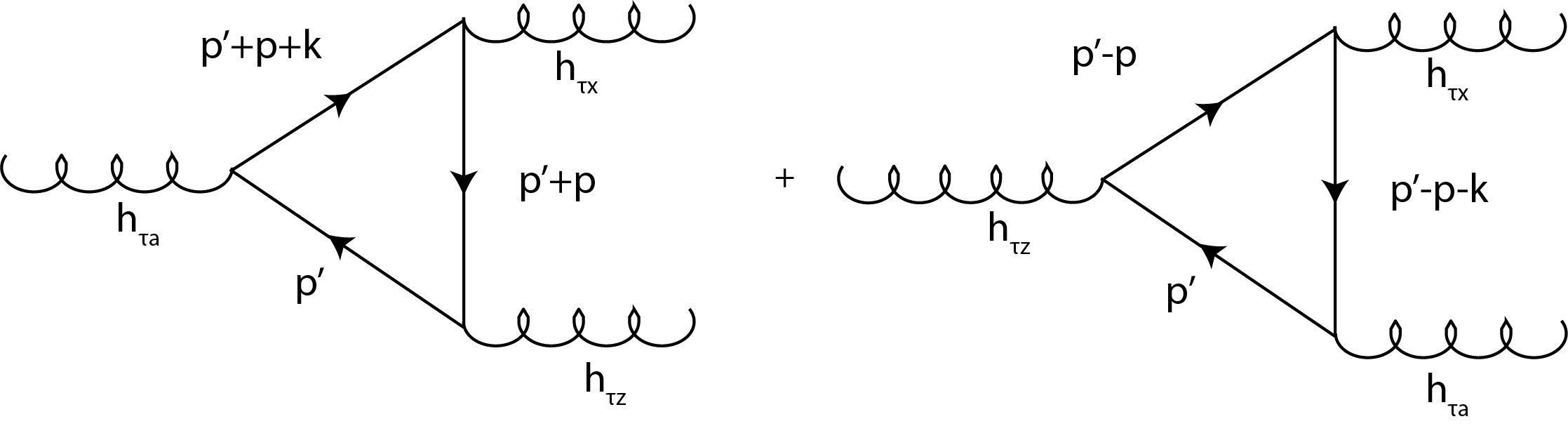} 
   \caption{Contributions to $\lambda^{(6)}_3$ from Wick contractions of 
    the stress tensor} \label{tttdiag}
\end{figure}

\begin{figure}[!htb]
  \centering
   \hspace*{-2cm}\includegraphics[scale=0.2]{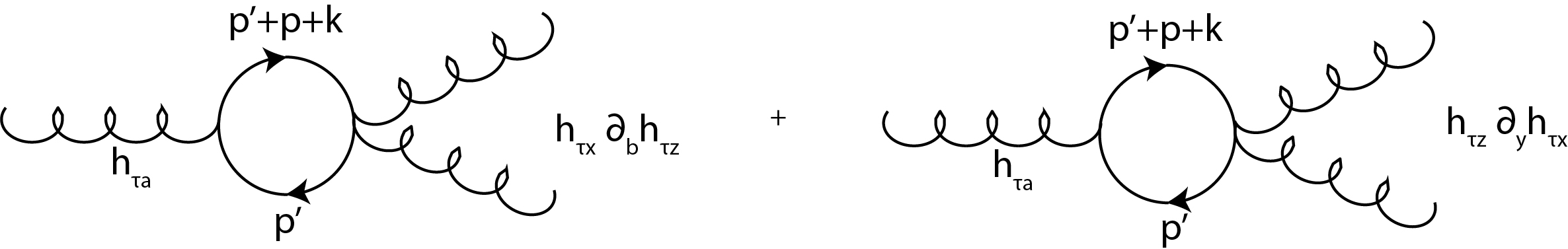} 
   \caption{Contributions to $\lambda^{(6)}_3$ from contact terms } \label{jttcontact}
\end{figure}

The contributions to $A$ and one of the terms in $B$ are summarized   in 
 figure \ref{tttdiag} and \ref{jttcontact} respectively. 
The detailed evaluation of all the diagrams yields the following result
 \begin{eqnarray}
 \tilde 
 \lambda^{(6)}_{3}&=&\frac{-3i}{16\pi^3}\int_0^\infty dp  p_1^3  ( f(E_{p_1}-e\mu_{-}) 
 - f(E_{p_1}+ e\mu_{-})  ),  \\ \nonumber
 &=& \frac{-3i}{64 \pi^3}(\frac{7\pi^4 T^4}{15}+2 e^2 \pi^2 T^2\mu _{-}^2 + e^4\mu _{-}^4 ).
\end{eqnarray}              
Writing down the transport coefficient in the Lorentzian theory using the relations
in (\ref{6danlcont}) gives
\begin{eqnarray} \label{tttrf}
 \lambda^{(6)}_{3}&=&
 \frac{3}{64 \pi^3}(\frac{7\pi^4 T^4}{15}+2 e^2 \pi^2 T^2 \mu _{-}^2 + e^4\mu _{-}^4 ).
\end{eqnarray}
The importance of this transport coefficient  is that it captures one of the  pure 
gravitational anomaly $d=6$. Again following \cite{Jensen:2013rga}, we parametrize this 
coefficient as
\begin{eqnarray}
\lambda^{(6)}_{3}&=&9(-c_\alpha \mu_-^4+ \tilde{c}_m^{6d}T^2\mu_-^2 + \tilde{c}_g^{6d} T^4).
\end{eqnarray}
Comparing the result of the perturbative evaluation in (\ref{tttrf}), we see that the values of the 
gauge anomaly $c_\alpha$ and the  coefficient $\tilde c_m^{6d}$  is related to the 
mixed anomaly, as seen in the earlier correlators, while 
 $\tilde{c}_g^{6d}$ is given by
\begin{eqnarray}
\tilde{c}_g^{6d}=\frac{7\pi}{2880},
\end{eqnarray}
Comparing with the coefficients of the pure gravitational  anomaly of a Weyl fermion 
$c_\gamma$ given in 
\eqref{anompoly6d}, we obtain
\begin{eqnarray}
\tilde{c}_g^{6d}= - (8\pi^2)^2 c_\gamma.
\end{eqnarray}
This relation is what is expected from the replacement rule and the 
consistency of the Euclidean vacuum \footnote{We have not checked 
 if the precise coefficient of $-(8\pi^2)^2$ agrees with the normalizations of  \cite{Jensen:2013rga}.}. 

  \subsection{Chiral Gravitinos }

  Chiral gravitinos cannot be consistently coupled to a charge current. 
  Therefore they contribute only to the 
  pure gravitational anomaly and 
  only to the transport coefficient 
  $\lambda_3^{(6)}$ . 
In this section, we show that analogous to the coefficient $\lambda^{(2)}$ 
  for the chiral gravitino like theory  $d=2$ in section \ref{2dgravitino} , 
 the  $\lambda_3^{(6)}$ for chiral gravitinos in $d=6$  do 
  not obey the relation  $\tilde c_g^{6d} = -(8\pi^2)^2 c_\gamma$. 
  We will identify the terms which do not contribute to the transport coefficient but contribute to the 
  anomaly polynomial. The gauge fixed gravitino Lagrangian is given by 
  \begin{eqnarray}
   S_E = S_E^{{\rm gr} } + S_E^{{\rm gh} },    \\ \nonumber
   S_E^{{\rm gr} } = \int d\tau dx^5 \sqrt{g} e^\mu_a \psi_\rho^\dagger D_\mu P_- \psi^\rho,
  \end{eqnarray}
  where the covariant derivative is given by 
  \begin{equation}\label{6dcov}
   D_\mu \psi^\rho = \partial_\mu \psi^\rho + \frac{1}{2} \omega_{\mu cd} \sigma^{cd} \psi^\rho +
   \Gamma^\rho_{\mu\sigma} \psi^\sigma.
  \end{equation}
The action $S_E^{{\rm gh}}$  is the action of the corresponding ghosts. The contribution of all the ghosts can be accounted 
by subtracting the contribution of a chiral spinor. 
The action $S_E^{{\rm gr} }$ is similar to the action of $6$ chiral fermions except that the 
covariant derivative in (\ref{6dcov}) contains the Christoffel symbol in addition to the spin connection. 
The linearization of the Christoffel symbol term results in the following interaction term in addition to the 
terms of the type present in (\ref{action6dexp}). 
\begin{eqnarray}
 S^{(2)}_{{\rm Christoffel} } =\frac{1}{2} \int d\tau d^5x  \sqrt{g} e^\mu_a \psi_\rho^\dagger \gamma^a  
 \eta^{\rho\alpha} ( \partial_\mu h_{\alpha\sigma} + \partial_\sigma h_{\alpha\mu} - \partial_\alpha h_{\mu \sigma}  + O(h^2) ) \psi^\sigma . \nonumber \\
\end{eqnarray}
The linear terms in the metric modify the stress tensor. 
There are higher order terms in the expansion of the metric and one would have expected possible contributions
to contact terms from them. However such terms do not have sufficient number of gamma matrices to contribute 
to the correlator involving 3 stress tensors. We need 6 gamma matrices, so that  the correlator is non-zero. Any  possible 
contact term from higher order terms in the above action have lesser number of gamma matrices. 
Therefore we conclude  that 
the only difference  of the chiral gravitino, compared to that of the chiral fermion, is that
its flat space tensor acquires additional term, due to the linearization of the Christoffel symbol. 
The stress tensor  is given by 
\begin{eqnarray} \label{classicalst6d}
 T^{\mu\nu}  &=& T^{\mu\nu}_{g1} + T^{\mu\nu}_{g2} , \\ \nonumber
 T^{\mu\nu}_{g1}  &=&   \frac{1}{2}\psi^\dagger_\rho (x)(\gamma^\mu \partial^\nu +
  \gamma^\nu \partial^\mu)P_- \psi^\rho (x)\nonumber\\
T^{\mu\nu}_{g2}&=& -\frac{1}{2}(\partial_\sigma(\psi^{\dagger,\sigma} (x)\gamma^\mu P_{-}\psi^{\nu }(x)
- \partial_\sigma (\psi^{\mu \dagger} (x) \gamma^\nu P_{-} \psi^\sigma(x)
+ ( \mu \leftrightarrow \nu ) ).
 \end{eqnarray}
Note that the contribution $T_{\mu\nu}^{g2}$ is due to the linearization of the Christoffel symbol. 
In momentum space the stress tensor is given by 
\begin{eqnarray} \label{6dgrstress}
 T_{fl}^{\tau\mu}(k) &=&  T_{g1}^{\tau s} (k)  + T_{g1}^{\tau s}(k) , \\ \nonumber
 T_{g1}^{\tau s} (k) &=& \frac{1}{2\beta} \sum_m \int \frac{d^5p'}{(2\pi)^5} 
 ( \psi^\dagger ( \omega_m, p' -k) ( \gamma^\tau i p'_s  + \gamma^x i \omega_m ) P_- \psi^\mu (\omega_m, p') , \\ \nonumber
 T_{g2}^{\tau s}( k) & & = i \frac{ k_\sigma }{2\beta}  \sum_m 
 \frac{d^5p'}{(2\pi)^5}  \left( 
 \psi^{\sigma \dagger} (\omega_m,  p' -p) \gamma^\tau P_- \psi^s(\omega_m,  p')  - 
 \psi^{\tau^\dagger} ( \omega_m,  p' -p) \gamma^z
 P_- \psi^\sigma( \omega_m, p')  \right.  \\ \nonumber
 & & \left.  \qquad\qquad\qquad + ( \tau \leftrightarrow s)  \right). 
\end{eqnarray}
We can now evaluate the correlator 
\begin{equation}
\lambda^{(6)}_3 = 
 -\frac{3}{2} \lim_{p_b,k_y \rightarrow 0}
\frac{-\langle T^{\tau a}(k+p)T^{\tau x}(-k)T^{\tau z}(-p)\rangle}{ip_b ik_y}.   
\end{equation}
Apart from the contact terms arising from the linearization of the spin connection, 
we have to evaluate the following contribution to the  correlator
\begin{equation}
A = \langle T^{\tau a}_{cl} (k+p)T^{\tau x}_{cl} (-k)T^{\tau z}_{cl}(-p)\rangle.
\end{equation}
Substituting the expression for $T^{\tau s}_{cl}$ from (\ref{6dgrstress}), 
it can seen from the structure of the Wick  contractions, that the only non-zero terms 
are 
\begin{eqnarray}
A &=& A_1 + A_2  \\ \nonumber
A_1 &=&  \langle T^{\tau a}_{g1} (k+p)T^{\tau x}_{g1} (-k)T^{\tau z}_{g1}(-p)\rangle
\\ \nonumber
A_2 &=&  \langle T^{\tau a}_{g2} (k+p)T^{\tau x}_{g2} (-k)T^{\tau z}_{g1}(-p)\rangle
+ \langle T^{\tau a}_{g2} (k+p)T^{\tau x}_{g1} (-k)T^{\tau z}_{g2}(-p)\rangle \\ \nonumber
& & + \langle T^{\tau a}_{g2} (k+p)T^{\tau x}_{g2} (-k)T^{\tau z}_{g2}(-p)\rangle.
\end{eqnarray}
After performing Wick contractions, the sum of the terms in $A_2$ organize themselves to 
\begin{eqnarray}
A_2 &=& -\frac{1}{4\beta} \sum_m \int \frac{d^5 p'}{(2\pi)^5}  \times  \\ \nonumber
& &  \left(  p_b^2 {\rm Tr} \left[
\gamma^a S ( \omega_m,  p' +  p+ k) (\gamma^\tau p'_{x} + \gamma^x i\omega_m )S( 
\omega_m, p' +  p)\gamma^z 
S( \omega_m, p') P_+  \right] \right.  \\ \nonumber
& & + p_b^2 {\rm Tr} \left[    
\gamma^z S( \omega_m, 
 p' -p) (\gamma^\tau  p'_{x} + \gamma^x i\omega_m )
 S( \omega_m, p' -  p -k)\gamma^a S(\omega_m, p') P_+ \right] \\ \nonumber
 & & + k_y^2 {\rm Tr}  \left[
\gamma^a S ( \omega_m,  p' +  p+ k)  \gamma^x S( 
\omega_m, p' +  p) (\gamma^\tau p'_{z} + \gamma^z i\omega_m )
S( \omega_m, p') P_+  \right] \\ \nonumber
& &  \left.  + k_y^2 {\rm Tr} 
 \left[(\gamma^\tau p'_{z} + \gamma^z i\omega_m )
 S ( \omega_m,  p' -  p)  \gamma^x S( 
\omega_m, p' -  p-k )\gamma^a 
S( \omega_m, p') P_+  \right] \right). 
 \end{eqnarray}
 After inserting the propagator and performing the Matsubara sums, it can be seen that each of the terms in $A_2$  is proportional to $p_b k_y$.  Therefore, the leading contribution to the transport coefficient
 from $A_2$ is of the form
 \begin{equation}
 \lim_{p_b, k_y \rightarrow 0} \frac{A_2}{i p_b i k_y}  = O( p_b^2) + O( k_y^2). 
 \end{equation}
The entire contribution from $A_2$ to the transport coefficient vanishes in the zero momentum 
limit. 
This is the same mechanism by which, the contribution  from terms arising from the 
linearization of the Christoffel symbol to $\lambda^{(2)}$ in the gravitino like theory in $d=2$,
vanishes. 
The contribution from $A_1$ is 6 times that of a chiral fermion. 
The contribution from the  contact terms arising from the linearization of the 
spin connection is also 6 times that of a chiral fermion. 
To these contributions we must subtract the contribution of  a single chiral fermion 
to account for the ghosts.  Using the results of the the section \ref{6dfermig}
equation (\ref{tttrf}),  we conclude that the 
contribution of a chiral gravitino to  $\lambda^{(6)}_3$ is given by 
\begin{equation} \label{6dfinr1}
\lambda^{(6)}_{3, {\rm gravitinos} } = 5 \times \frac{3}{64 \pi^3}  \frac{ 7 \pi ^4 T^4}{15}
= 9 \tilde c^{6d}_{g, \rm gravitinos} .
\end{equation}
Here the last equality arises from the parametrization following from  the replacement rule. 

Now let us compare the transport coefficient to the coefficient of the pure gravitational 
anomaly. 
The terms arising from the linearization of the Christoffel symbol do give a contribution 
to the anomaly polynomial of the gravitino \cite{AlvarezGaume:1983ig} \footnote{See equation (35) of \cite{AlvarezGaume:1983ig}}.  In fact such terms give rise to the 
interaction of the magnetic moment of the gravitino with the gravitational field. 
These terms modify the gravitational anomaly coefficient $c_\gamma$ for gravitinos which 
is given by 
\begin{equation}\label{6dfinr2}
c_{\gamma, {\rm gravitinos}} = - 275\times \frac{1}{184320 \pi^3}.
\end{equation}
Comparing (\ref{6dfinr1}) and (\ref{6dfinr2}), we see that gravitinos do not obey the 
relation $\tilde c_g^{6d} = - ( 8\pi^2)^2 c_\gamma$, which 
is argued from the consistency of the Euclidean vacuum in \cite{Jensen:2013rga}. 
Here we have identified the terms which contribute to the anomaly polynomial but
not to the transport coefficient. 
From our analysis in $d=2, d=6$ we see that gravitinos  contribute  to correlators
involving only the stress tensor, since they are not charged. Their contribution is 
essentially  $d-1$ times
that of a chiral fermion to the transport coefficient.  This is the same
conclusion arrived at, by analysis of the helicity of the thermal state, in 
\cite{Loganayagam:2012zg}. 

\newpage
\begin{table}
\centering
 \begin{tabular}{|c|c||c|c|}
 \hline
 Dimension & Correlator & Species & Value \\ [1.0ex] 
 \hline\hline

 d=6 & $\zeta^{(6)}_1$ & Fermions & $\frac{3e^4}{64\pi^3} \mu _-$ \\ [1ex]
 \hline
  & $\zeta^{(6)}_2$ & Fermions & $\frac{3e^2}{32\pi^3}(e^2(\mu _{-})^2+\frac{\pi^2 T^2}{3})
$ \\ [1ex]
 \hline
   & $\zeta^{(6)}_3$ & Fermions & $\frac{e}{16\pi^3}[(e\mu _{-})^3+ e\pi^2 T^2\mu _{-}] 
$ \\ [1ex]
 \hline
 & $\lambda^{(6)}_3$ & Fermions & $\frac{3}{64 \pi^3}(\frac{7\pi^4 T^4}{15}+2 e^2 \pi^2 T^2 (\mu _{-})^2 + e^4\mu _{-}^4 ) 
$ \\ [1ex]
 \hline
 & $\lambda^{(6)}_3$ & Gravitinos & $\frac{15}{64 \pi^3}(\frac{7\pi^4 T^4}{15}) 
$ \\[1ex]
 \hline
\end{tabular}
\caption{Transport coefficients In d=6}
\label{table:1}
\end{table}

\section{Hydrodynamic modes in  \texorpdfstring{$d=2$}{} Weyl gas}
\label{dispersion}

A charged $1+1$ dimensional relativistic fluid admits $2$  hydrodynamic modes, 
the sound mode and the charge diffusion mode. 
The leading dispersion relation for these modes, without anomalous terms in the constitutive 
relations for the stress tensor and the charge current, was obtained in \cite{David:2010qc}. 
In this section we would like to show that due to  the presence of anomalous terms, 
proportional to $\zeta^{(2)}$ and $\lambda^{(2)}$ in the constitutive relation, the 
charge diffusion mode acquires a non-zero speed. 
We then apply the discussion to the case of ideal  $d=2$ Weyl gas, held at a chiral chemical 
potential $\mu_-$. As seen in section \ref{an2d}, such a gas has a non-zero value for 
$\zeta^{(2)}$ and $\lambda^{(2)}$. 
The speed of sound for this gas remains the same as the speed of light, however the 
speed of the charge diffusion mode is non-zero and is equal half the speed of light.

Consider a relativistic fluid in $d=2$ with anomalous transport $\zeta^{(2)}$ and $\lambda^{(2)}$ 
non-zero.  The constitutive relations to the first derivative in velocity in the Lorentz frame is given by 
\begin{eqnarray} \label{dispcons}
T^{\mu \nu }=\left(\epsilon+P\right)u^{\mu }u^{\nu }-P \eta^{\mu \nu } -\xi(u^{\mu }u^{\nu }-\eta^{\mu \nu })\partial_\lambda u^\lambda,\nonumber\\
j^{\mu }=\rho u^{\mu }-\sigma T(\eta^{\mu \nu }-u^{\mu }u^{\nu })\partial_{\nu }(\frac{\mu }{T}) +\zeta_v \epsilon^{\mu \nu }u_{\nu }                .   
\end{eqnarray}
The  signature  of the metric is given by $\eta_{\mu\nu} = {\rm diag} (1,-1)$. 
$u^\mu $ is the velocity field
with $u^{\mu }u_{\mu }=1$, $\xi$ is the bulk viscosity, $\sigma$ the conductivity  and  
$\zeta_v$ is the anomalous transport coefficient. 
$\epsilon,P,\rho,\mu $ refer to the energy density, pressure, charge density and chemical potential respectively. 
$\zeta_v$ in the Lorentz frame is related to the transport coefficients in the anomaly frame by 
\begin{equation}\label{anflf}
 \zeta_v = \zeta^{(2)} - \frac{\rho}{\epsilon + P } \lambda^{(2)}. 
\end{equation}
 To study linearized hydrodynamic modes, consider small fluctuations from the rest frame with velocity 
 field $(1, v^z)$ ( $|v_z|<<1$). 
 We also have fluctuations in the energy density, pressure which 
 are related to fluctuations in the stress tensor  and charge current, using the constitutive relations in 
 (\ref{dispcons}). These are given by 
\begin{eqnarray}\label{fluctu}
\delta T^{00} &=& \delta \epsilon , \qquad
\delta T^{0z} = (\epsilon + P)\delta u^z, \qquad
\delta T^{zz} = \delta P - \frac{\xi \partial_z \delta T^{0z}}{\epsilon+P},\nonumber\\
\delta j^0 &=& \delta \rho - \zeta_v \delta u^z, \qquad 
\delta j^z = \rho \delta u^z - \sigma T \partial_z (\delta \bar{\mu }) - \delta \zeta_v . 
\end{eqnarray}      
It is convenient to use
 the energy density $\epsilon$ and charge density $\rho$ as independent 
variables.  We can write the variation in  chemical potential and  pressure as 
\begin{equation}
\delta \bar{\mu }= \partial_\epsilon \bar{\mu }\delta T^{00}+ \partial_\rho \bar{\mu }\delta \rho, \qquad 
 \delta P = \partial_\epsilon P \delta T^{00}+ \partial_\rho P \delta \rho. \nonumber\\
\end{equation}
where $\bar \mu = \mu/T$. 
The hydrodynamic modes are eigen modes of the linearized equations of hydrodynamics, which
are given by 
\begin{equation}
 \partial_\mu \delta T^{\mu \nu} =0, \qquad \partial_\mu \delta j^\mu =0.
\end{equation}
Here we set the background gauge field strength ,including its  
fluctuations, to zero. 
Writing out these equations in Fourier space 
\footnote{For eg. Fourier mode of  $\delta T^{zz} =  e^{-i\omega t + i k z} \delta T^{zz}(\omega, k)$.}, 
eliminating $\delta T^{zz}$, and  $\delta u^z$ using
(\ref{fluctu}),  we obtain  the following set of equations  from the conservation equation for the 
stress tensor
\begin{eqnarray}\label{fluceom}
 & & - i \omega  \delta T^{00} + i k \delta T^{0z} =0,  \\ \nonumber
 && -i\omega \delta T^{0z} + ik(\partial_\epsilon P \delta T^{00}+ 
 \partial_\rho P \delta \rho)+k^2 \xi \frac{\delta T^{0z}}{\epsilon+P}
= 0.
\end{eqnarray}
Now using the first equation in (\ref{fluceom}), to eliminate $\delta T^{0z}$, we obtain 
\begin{eqnarray}\label{a3}
 \delta \rho( \partial_\rho P ik ) +\delta T^{00}\left(\frac{-i\omega^2}{k}+
 \frac{\omega k \xi}{\epsilon+P} + ik\partial_\epsilon P\right)&=&0.
\end{eqnarray}  
A similar analysis of the conservation equation for the current yields 
\begin{eqnarray}\label{a4} 
& &\delta \rho (-i\omega + k^2 \sigma T \partial_\rho \bar{\mu }- ik \partial_\rho \zeta_v) \\ \nonumber
&& \qquad \qquad
+\delta T^{00} \left(k^2\sigma T \partial_\epsilon \bar{\mu } + 
\frac{i\omega \rho}{ \epsilon+P} - ik\partial_\epsilon \zeta_v + 
\frac{i\omega^2 \zeta_v}{k(\epsilon+P)} \right)=0.
\end{eqnarray}
From the homogeneous equations (\ref{a3}) and (\ref{a4}),  we see that to obtain 
a non-trivial solution for the fluctuations $\delta\rho, \delta T^{00}$ we must have the condition
\begin{eqnarray}\label{a5}
&  & (\partial_\rho P ik)\left(k^2\sigma T \partial_\epsilon \bar{\mu } + \frac{i\omega \rho}
{ \epsilon+P} - ik\partial_\epsilon \zeta_v + \frac{i\omega^2 \zeta_v}{k(\epsilon+P)}\right)\nonumber\\
& & \quad -\left(\frac{-i\omega^2}{k}+
\frac{\omega k \xi}{\epsilon+P} + ik\partial_\epsilon P\right)
\left(-i\omega + k^2 \sigma T \partial_\rho \bar{\mu }- ik \partial_\rho \zeta_v\right)=0.
\end{eqnarray}  
To obtain solutions which satisfy the condition in (\ref{a5}), 
we assume the dispersion relation of the form
\begin{eqnarray}\label{ansatz}
\omega &=& v k-iD k^2 + O(k^3). 
\end{eqnarray} 
Substituting this anstaz into (\ref{a5}), we see that the  velocity $v$  is determined by 
the following cubic equation 
 \begin{eqnarray} \label{a6}
v^3+v^2\left(-\frac{\partial_\rho P \zeta_v}{\epsilon+P}+
\partial_\rho \zeta_v\right) - 
v\left(\frac{\rho\partial_\rho P}{\epsilon+P} +
\partial_\epsilon P\right)- 
\partial_\epsilon P \partial_\rho \zeta_v + \partial_\rho P \partial_\epsilon \zeta_v =0.  
\end{eqnarray}
This equation is obtained by setting the coefficient of the  $O(k^2)$ term to zero on substituting 
the anstaz (\ref{ansatz}) in (\ref{a5}). 
As a simple check of this equation, consider setting the 
anomalous transport coefficients $\zeta_v =0$, then 
the equation reduces to 
\begin{equation}
 v^2 - \left( \frac{\rho\partial_\rho P}{\epsilon+P}+ \partial_\epsilon P \right) v =  0.
\end{equation}
The three solutions to this equation 
are 
\begin{equation}
 v= \pm \left( \frac{\rho\partial_\rho P}{\epsilon+P}+ \partial_\epsilon P \right)^{\frac{1}{2}} ,\quad  v=0.
\end{equation}
The first two roots correspond to the speed of sound and the last one is the 
speed of the charge diffusion mode. 
Now one can show for $\zeta_v \neq 0$ but small,  there are corrections to 
both speed of sound and the speed of the charge diffusion mode that depend on $\zeta_v$. 
On further analysis, one can show that the the dissipative constant $D$ for both the sound and the charge diffusion mode 
receives corrections on turning on $\zeta_v$. 

We will now  show for the ideal Weyl gas the three roots of (\ref{a6}) are $v = \pm 1, v =-1/2$. 
The thermodynamics of the ideal Weyl gas is given by \cite{Loganayagam:2012pz}

\begin{eqnarray} \label{thermweyl}
\epsilon &=& \frac{e^2 \mu _-^2}{4\pi}+\frac{T^2 \pi}{12 } =P,\nonumber\\
\rho &=& \frac{e^2\mu _-}{2\pi}.
\end{eqnarray}
Using (\ref{anflf}) and the values of $\zeta^{(2)}$ and $\lambda^{(2)}$ we obtain 
the transport coefficient  to be given by 
\begin{eqnarray}\label{anftran}
\zeta_v &=& \frac{e^2\mu_ -}{2\pi} -  \frac{1}{4\pi}\frac{\rho}{P+\epsilon} 
 \left(e^2 \mu _-^2 + \frac{\pi^2 T^2}{3}\right ),\nonumber\\
&=& \frac{\rho}{2}.
\end{eqnarray}
To obtain the last  equality  we have used the expression for energy density and $\epsilon =P$. 
Since the ideal Weyl gas is conformal and there are no interactions the bulk viscosity $\xi$ as well as the conductivity 
$\sigma$ vanishes. 
Now from the  relations in  (\ref{thermweyl})  and (\ref{anftran}) we see that 
\begin{eqnarray}
\partial_\rho P=0, \quad \partial_\epsilon P=1, \qquad\qquad  
\partial_\rho \zeta_v=\frac{1}{2}, \quad \partial_\epsilon \zeta_v= 0. 
\end{eqnarray}                
Substituting these values in equation (\ref{a6}) for the velocity of the modes we obtain 
\begin{eqnarray}
v^3+\frac{1}{2} v^2-v - \frac{1}{2} &=& 0.
\end{eqnarray}
Therefore the  3 roots are  
\begin{eqnarray}
v &=& \pm 1, \,  -\frac{1}{2}.
\end{eqnarray}
As we discussed earlier, the first two roots correspond to the sound mode which moves with the 
velocity of light. The last root corresponds to the speed of the charge diffusion mode. 
Note that there is a unique sign for this speed, this is because of the definite chirality of the 
Weyl fermions. If the we had considered Weyl fermions with opposite chirality the 
sign would have been positive. 

It will be interesting to explore this  phenomenon further to see if it is  observable in
realistic low dimensional quantum systems
like quantum wires. 
It will be also  interesting to study the effect of the anomalous transport coefficients for 
other relativistic systems in $1+1$ dimensions. One such example is the charged D1-brane
which was studied in \cite{David:2010qc}.

\section{Conclusions}

We have evaluated all  the leading parity odd transport coefficients for chiral fermions in $d=2, 4, 6$ as well as chiral bosons 
in $d=2$ at one loop. The analysis was done using finite temperature field theory. 
We have seen that  chiral fermions as well as chiral bosons in $d=2$ obey the replacement rule of
\cite{Jensen:2013rga} which relates the coefficient of anomalies including both mixed and pure gravitational anomalies 
to the respective transport coefficient. 
As far as we are aware the calculations in $d=2$ is the first direct perturbative evaluation of
transport coefficients using the Kubo formula.
The calculations in $d=4, 6$ serve as an alternative approach as well as  a check of the calculations  done in    
\cite{Yee:2014dxa} which 
involved Schwinger-Keldysh propagators. We have also kept track of the contributions 
from the gravitational anomalies which was ignored in \cite{Yee:2014dxa}. 
For a chiral gravitino like theory in $d=2$ as well as chiral gravitinos 
in $d=6$ the relationship between the pure gravitational anomaly and the corresponding 
transport coefficient found using the argument involving the   consistency of the
Euclidean vacuum  breaks down. 
We have identified the mechanism of how this occurs in the perturbative analysis. 
We also developed a simple method to perform the angular integrals that occur in the 
transport coefficients which involves taking the zero momentum limit first and the use of 
an $i\epsilon$ prescription. 
Finally we have shown that the charge diffusion mode of the ideal Weyl gas in $d=2$ at finite chemical potential 
acquires a speed which is $1/2$ the speed of light due to the presence of 
anomalous transport coefficients. 

There are several directions in which the work done in this paper can be extended. 
It will be interesting to extend the analysis done in this paper to all  even dimensions. A more challenging 
task would be to evaluate the contribution of self dual tensors in $d=4k +2$ for $k>0$ to the 
transport coefficient 
Though the coupling of these tensors to the charge current is not explicitly known, one can evaluate 
contributions related to the pure gravitational anomaly of these tensors. 
Direct holographic checks of these transport coefficients in $AdS_3, AdS_7$ backgrounds 
similar to the ones done in $AdS_5$  in \cite{Landsteiner:2011iq}
is another 
direction  direction of interest.  
Finally it will be interesting to study to what extent  these parity odd  transport coefficients 
are one loop exact analogous to what was done for the coefficient $\zeta^{(4)}_2$ in $d=4$ by 
\cite{Golkar:2012kb}.
 
\acknowledgments

We thank  Sayantani Bhattacharyya, 
Zohar Komargodski, R. Loganayagam, Diptiman Sen, Aninda Sinha and Amos Yarom
for discussions  on   various issues  
related to  this work. S.D.C.  thanks Edward Witten for the discussion during the poster session of 
Strings 2015 which  helped us to clarify our understanding of the gravitino like theory in $d=2$. 
We also thank Kristan Jensen for useful correspondence.

\appendix

\section{Moments of statistical distributions}\label{moments}

To evaluate the various transport coefficients of interest in this paper we require several  moments of the 
Fermi-Dirac distribution. Let $f(p)$ be the Fermi-Dirac distribution which is given by
\begin{equation}
 f(p) = \frac{1}{e^{ \beta p} +1}
\end{equation}
The results for the following moments are used at several instances in the paper
\begin{eqnarray} \label{int1}
\int_0^\infty dp [f(p-e\mu)-f(p+e\mu)]&=& e\mu \\ \label{int2}
\int_0^\infty dp \, p\, [f(p-e\mu)+f(p+e\mu)]&=&
 \frac{1}{2}(e^2\mu^2+\frac{\pi^2T^2}{3})\\ \label{int3}
\int_0^\infty dp \, p^2\, [f(p-e\mu)-f(p+e\mu)]&=& 
\frac{1}{3}(e^3\mu^3 + e\pi^2 T^2 \mu) \label{int4} \\
\int_0^\infty dp \, p^3 \, [f(p-e\mu)+f(p+e\mu)]&=& \frac{1}{4}(\frac{7\pi^4 T^4}{15}+
2 e^2 \pi^2 T^2 \mu ^2 + e^4\mu ^4 ) 
\end{eqnarray}
These integrals are derived in \cite{Loganayagam:2012pz}. 
Let  $b(p)$ by the Bose-Einstein distribution which is given by 
\begin{equation}
 b(p) = \frac{1}{ e^{\beta p } -1}
\end{equation}
Then the  following integral is required in the analysis of chiral bosons. 
\begin{eqnarray}\label{int5}
\int_{0}^\infty dp \, p \, b(p) &=&
 \frac{\pi^2T^2}{6}
\end{eqnarray}

\section{Evaluation of transport coefficients in \texorpdfstring{$d=6$}{Lg}}\label{an6da}

In this appendix we detail the calculations involved in obtaining the leading 
parity odd transport coefficients in $d=6$. 

\subsection{Contributions to \texorpdfstring{$\zeta^{(6)}_1$}{Lg}}
\label{Appendix A1}   

Before we begin, let us recall the kinematic set up of the correlator considered in (\ref{jjj}). 
This kinematic configuration will be the same for all the other correlators considered in $d=6$. 
There are two external momenta $p, k$. We set the frequency components of both these external 
momenta to zero. The momentum vector $p$ has a non-zero component only along the $b$ direction while 
momentum vector $k$ has component only along the $y$ direction. 
While performing the integration over the spatial directions of the internal momentum 
labelled as $p'$ it is convenient to 
parametrize its components  in terms of angular variables as 
\begin{eqnarray}\label{19}
p'_b = |p'| \cos\phi_1,  \qquad p'_y = |p'| \sin\phi_1 \cos\phi_2, \quad p'_x = |p'| \sin\phi_1\sin\phi_2\cos\phi_3, \\ \nonumber 
p'_z = |p'|\sin\phi_1\sin\phi_2\sin\phi_3\cos\phi_4, \qquad 
p'_a = |p'|\sin\phi_1\sin\phi_2\sin\phi_3\sin\phi_4.
\end{eqnarray}
We choose orientation of the axis so that 
\begin{equation}
 {\rm Tr} ( \gamma^{p}\gamma^q\gamma^r\gamma^s\gamma^t\gamma^u\gamma_c ) = - 8 \epsilon^{pqrstu}, 
 \qquad\qquad \epsilon^{\tau azxyb } = i .
\end{equation}
With this kinematic configurations, let us write down the expression for the various energies which will occur in the 
expression for the propagators. 
\begin{eqnarray}\label{energies}
E_{p'+p}&=&( {|p'|}^2+{|p|}^2+2|p||p'| \cos \phi_1)^{\frac{1}{2}},  \\ \nonumber
E_{p'+k}&=& ( {|p'|}^2+{|k|}^2+2|k||p'|\sin \phi_1 \cos \phi_2 )^{\frac{1}{2} } ,\\  \nonumber
E_{p'+p+k} &=&(
{|p'|}^2+{|k|}^2+{|p|}^2+2|k||p'|\sin \phi_1 \cos \phi_2 + 2|p||p'|\cos \phi_1 )^{\frac{1}{2}}.
\end{eqnarray}
For convenience  in notation we will also define 
\begin{equation}
 i\omega_- = i\omega_m + e(\mu - \mu_c), \qquad\qquad  \omega_m = (2m + 1) \pi T . 
\end{equation}
Now substituting the expression for the propagators from (\ref{method2propagator}), 
into (\ref{jjjw}), we obtain
\begin{eqnarray} \label{jjjwa}
& & \langle j^a(p+k) j^x(-k) j^z(-p)\rangle =  \frac{e^3}{2^3 \beta} 
\sum_{m} \int \frac{d^5 p'}{(2\pi)^5}  \times  \\ \nonumber
&&\left( \sum_{t,v,u=\pm} {\rm Tr}
[\gamma^a \gamma^{\alpha} \gamma^x \gamma^\beta \gamma^z \gamma^\nu P_+]  
\Delta_t(i\omega_-  ,p'+k+p) (\hat{p'}+\hat{k}+\hat{p})_{(t,\alpha)} \times  \right. \\ \nonumber
& & \qquad \qquad  \left. 
\Delta_u(i\omega_- ,p' +p)(\hat{p'}+\hat{p})_{(\beta ,u)} 
\Delta_v(i\omega_-,p')\hat{p'}_{(\nu ,v)}  \right)  + 
\\ \nonumber
& & ( {\rm Exchange \,  term}). 
\end{eqnarray}
The exchange term is the second term in equation (\ref{jjjw}), arising due to the second Wick contraction. 
We can obtain the exchange term from the term written down by the performing the 
following operation on the first term written in (\ref{jjjwa}). 
\begin{eqnarray}
  a\rightarrow z,  &\qquad& z \rightarrow a, \\ \nonumber
 (p' + p +k) \rightarrow ( p' - p), &\qquad&  ( p' + p) \rightarrow ( p'-p-k).
\end{eqnarray}
Performing the trace over the $\gamma$ matrices, we obtain
\begin{eqnarray}
& & {\rm Tr} 
[\gamma^a \gamma^{\alpha}\gamma^x \gamma^\beta \gamma^z \gamma^{\nu }P_{+}]
(\hat p' +\hat k+\hat p)_{(t,\alpha)}(\hat p'+\hat p)_{(u,\beta)} (\hat p')_{(v,\nu )} \\ \nonumber
&=& \qquad \left( \frac{4 t\, u\, v \, p'_{a}(p'+p)_z (p'-q)_x }{E_{p'}E_{(p'+p)}E_{(p'-q)}}+
\textrm{permutations in} \, (a, z, x)  \right) \nonumber\\
&& + \left( \frac{4t\, u( k_{y}p'_{b}+p_bk_y) }{E_{p'+p+k}E_{p'+p}}-\frac{4u\, v \, p'_{y}\, p_{b} }
{E_{p'}E_{p'+p}}
-\frac{4t\, v( k_{y}p'_{b}-p_bp'_{y} ) }{E_{p'+p+k}E_{p'}} \right).\nonumber
\end{eqnarray}
The next step is to perform the following four integrals
\begin{eqnarray}
\frac{-A_1}{ip_b ik_y}&=&\frac{e^3}{p_bk_y 2^3 \beta} \sum_{m}
 \int \frac{d^5 p'}{(2\pi)^5}\sum_{t,v,u=\pm}\left
 (\frac{4p'_{a}(p'+p)_z (p'+p+k)_x}{E_{p'}E_{(p'+p)}E_{(p'-q)}}+
\textrm{permutations}\right) \times  \nonumber\\
&&\qquad\qquad\qquad\qquad\qquad
\Delta_t(i\omega_-  ,p'+k+p) \Delta_u(i\omega_- ,p' +p) \Delta_v(i\omega_-,p'), \nonumber\\ 
\frac{-A_2}{ip_b ik_y}&=&\frac{e^3}{p_bk_y 2^3 \beta} \sum_{m} \int \frac{d^5 p'}{(2\pi)^5}
\sum_{t,v,u=\pm}\frac{4t\, u\, \, p_b\, k_y}{E_{p'+p+k}E_{p'+p}} \times \nonumber\\
&& \qquad\qquad\qquad\qquad\qquad
\Delta_t(i\omega_- ,p'+k+p)  \Delta_u(i\omega_- ,p' +p) \Delta_v(i\omega_-,p'), \nonumber\\ 
\frac{-A_3}{ip_b ik_y}&=&\frac{e^3}{p_b k_y 2^3 \beta} \sum_{ m} \int \frac{d^5 p'}
{(2\pi)^5}\sum_{t,v,u=\pm}\left
(\frac{4\, t\, u\, k_{y}\, p'_{b} }{E_{p'+p+k}E_{p'+p}}-\frac{4\, t\, v\,  k_{y}\, p'_{b} }
{E_{p'+p+k}E_{p'}}
\right) \times \nonumber\\
&& \qquad\qquad\qquad\qquad\qquad
\Delta_t(i\omega_-   ,p'+k+p)  \Delta_u(i\omega_- ,p' +p) \Delta_v(i\omega_-,p'), \nonumber\\ 
\frac{-A_4}{ip_b ik_y}&=&\frac{e^3}{p_bk_y 2^3 \beta} \sum_{m}
 \int \frac{d^5 p'}{(2\pi)^5}\sum_{t,v,u=\pm}\left
 (-\frac{4\, u\, v\, p'_{y}\, p_{b}}{E_{p'}E_{p'+p}}+
 \frac{4\, t\, v\, p_b\, p' _{y} }{E_{p'+p+k}E_{p'}} \right)\nonumber\\
&&\qquad\qquad\qquad\qquad\qquad
\Delta_t(i\omega_-  ,p' +k+p)]  \Delta_u(i\omega_- ,p'  +p) \Delta_v(i\omega_-,p'). \nonumber\\ 
\end{eqnarray}
Here we have divided by the external momenta and multiplied by a negative sign 
in each term so that we obtain terms directly related to the transport coefficient of 
interest. 
Each of the terms involve a sum over Fermionic Matsubara frequencies. They are 
evaluated using the following expression
\begin{eqnarray}\label{ms6d1}
&&\frac{1}{\beta}\sum_{ m}\Delta_t(i\omega_-,p'+p+k)\Delta_u(i\omega_-,p'+p)
\Delta_v(i\omega_-,p')\\ \nonumber
&&=\frac{tf(E_{p'+p+k}-te\mu _{-})}{(tE_{p'+p+k}-vE_{p'})(tE_{p'+p+k}-uE_{p'+p})}+
\frac{uf(E_{p'+p}-ue{\mu }_{-})}{(uE_{p'+p}-tE_{p'+p+k})(uE_{p'+p}-vE_{p'})}\nonumber\\
&&\qquad \qquad+\frac{vf(E_{p'}-ve{\mu }_{-})}{(vE_{p'}-tE_{p'+p+k})(vE_{p'}-uE_{p'+p})}\nonumber\\
&\equiv& {\cal M}. 
\end{eqnarray}
In arriving at the above sum, we have neglected terms independent of the 
temperature and chemical potential.  An important observation from (\ref{energies}),  is that all terms in the 
Matsubara sum is independent of the angles $\phi_4$ due to our choice of directions of the external momenta. 
 Let us first evaluate $A_1$. The numerator is proportional to $p'_{a}p'_{x}p'_{z}$.
 After substituting for these components from (\ref{19}), we that the numerator is 
  proportional to $\sin \phi_4 \cos \phi_4$. Then on performing the 
  angular integral over $\phi_4$, this term vanishes since each term in the Matsubara sum in (\ref{ms6d1}) is 
  independent of $\phi_4$. 

The  term $A_2$ term is given by 
\begin{eqnarray}
\frac{-L_2}{ip_b ik_y}&=&\frac{e^3}{p_bk_y 2^3 \beta} \sum_{\omega_m}
\int \frac{d^5 p_1}{(2\pi)^5}\sum_{t,v,u=\pm}\frac{4tu[p_bk_y]}{E_{p_1+p+k}E_{p_1+p}} \times {\cal M}. \nonumber\\
\end{eqnarray} 
We perform the sum over a pair  from $\{t, u, v \}$ in each of the 3 terms in the Matsubara sum ${\cal M}$ as follows. 
For the first term in the ${\cal M}$ we sum over the pair $u, v$, for the 2nd term the pair $t, v$ and for the
3rd term we sum over the pair $t, u$. 
We can then change variables of the $p'$ integration so that the numerator in each of the terms in 
the Matsubara sum is a function of  the integration variable $p'$.  For instance in the first term we substitute 
$ p'  + p + k = \tilde p'$, while in the second term we substitute $p'+ p = \tilde p'$. 
This ensures that the numerator of each term in ${\cal M}$ is a function of $E_{p'}$ with
$p'$ being the integration variable. 
After performing these manipulations we obtain 
\begin{eqnarray}
\frac{-A_2}{ip_b ik_y}&=&\frac{2e^3}{(2\pi)^5}\sum_{t=\pm}
\int d^5 p'  tf(E_{p'}-te\mu_-)\nonumber\\
&&\frac{1}{(E_{p'}^2-E_{p'+p}^2)(E_{p'}^2-E_{p'+k+p}^2)}+\frac{1}{(E_{p'}^2-E_{p'-k}^2)(E_{p'}^2-E_{p'+p}^2)}\nonumber\\
&&\frac{1}{(E_{p'}^2-E_{p'+k}^2)(E_{p'}^2-E_{p'+p+k}^2)}.\nonumber\\
\end{eqnarray} 
We can now take the $p, k \rightarrow 0$ limit and expand each term in the above expression as a series in $p, k$. 
The order of limit in these expansions does not matter. 
The leading singular term and finite terms are 

\begin{eqnarray}
\lim_{p_b,k_y\rightarrow 0}
\frac{-A_2}{ip_b ik_y}&=& \frac{2e^3}{(2\pi)^5} \sum_{t=\pm}\int d^5 p' \,   t\, f(E_{p'}-te\mu_-) \\ \nonumber
&&\times
\left( 
\frac{1}{8p^{\prime 4} \cos^2 \phi_1 \sin^2 \phi_1 \cos^2 \phi_2}-\frac{1}{4 p_b p_1^3 \cos^2 \phi_1 \sin \phi_1 \cos \phi_2}\right) 
\\ \nonumber
& & + O(p_b, k_y). 
\end{eqnarray}
Note that the singular term  in $p_b$  on $\phi_2 \rightarrow \pi -\phi_2$ and therefore vanishes on performing the angular integration 
over $\phi_2$ using the $i\epsilon$ prescription. 
Therefore one is left with the following expression 
\begin{eqnarray}
\lim_{p_b,k_y \rightarrow 0}\frac{-A_2}{ip_b ik_y}&=&\frac{2e^3}{8(2\pi)^5} \
\sum_{t=\pm}\int dp' d\phi_1 d\phi_2 d\phi_3 d \phi_4 \times  \\ \nonumber
& & \qquad \qquad \frac{\sin \phi_1}{\cos^2 \phi_1} \frac{\sin^2 \phi_2}{\cos^2 \phi_2} \sin \phi_3 
  t\, f(|p' |-te\mu_-).\nonumber
\end{eqnarray} 
The angular integration using the $i\epsilon$ prescription  can be performed using the following results 
\begin{eqnarray}
\int_0^\pi \frac{\sin \phi}{\cos^2 \phi} = -2, \qquad 
\int_0^\pi \frac{\sin^2 \phi}{\cos^2 \phi}=-\pi, \qquad 
\int_0^\pi \sin \phi = 2. 
\end{eqnarray}
Substituting these values for the angular integrations we obtain 
\begin{eqnarray}
\lim_{p_b,k_y\rightarrow 0}\frac{-A_2}{ip_b ik_y} &=& 
\int_0^\infty  dp'\frac{e^3}{16\pi^3} [f( p'-e\mu _{-})-f(p' +e\mu _{-})]\nonumber\\
&=&  \frac{e^4 \mu_-}{16\pi^3} .
\end{eqnarray}
Proceeding in a similar manner, we find that  when we sum over pairs in $\{ t, u, v \}$ as discussed earlier  in term 
$A_3$ it vanishes. The term $A_4$ also vanishes due to the same reason. 
Now the exchange term in (\ref{jjjwa}) also contributes equally. Therefore using the transport coefficient in 
the Euclidean theory is given by   
\begin{eqnarray}
\tilde \zeta^{(6)}_1=\frac{3e^4}{64\pi^3} \mu_-.
\end{eqnarray}
Note that  from (\ref{6dkubo}) the transport coefficient is $3/8$ times the  $\langle jjj \rangle $ correlator.

\subsection{Contributions to \texorpdfstring{$\zeta^{(6)}_2$}{Lg}}\label{Appendix A2}

After performing  Wick contractions on the correlator corresponding to this 
transport coefficient we obtain 
\begin{eqnarray}\label{jjtap}
& & \langle j^a(p+k) j^x(-k) T^{\tau z}(-p)\rangle =  i\frac{e^2}{2^4 \beta} \sum_{m} \int \frac{d^5 p'}{(2\pi)^5}\sum_{t,v,u=\pm} {\rm Tr}  \left( 
[\gamma^a\gamma^{\alpha}\gamma^x\gamma^\beta(\gamma^\tau ip'_{z}+i\omega_- \gamma^z) \gamma^{\nu }P_{+}] \times  \right. \nonumber\\
&& \qquad  \left. \Delta_t(i\omega_-,p' +p+k)(\hat p'+\hat p+\hat k)_{(\alpha,t)}\Delta_u(i\omega_-,p'+p)
(\hat p'+\hat p)_{(\beta,u) } \Delta_v(i\omega_-,p')\hat{p'}_{(\nu,v) } \right), \nonumber\\ 
&&+i\frac{e^2}{2^4 \beta} \sum_{m}  \int \frac{d^5 p'}{(2\pi)^5}\sum_{t,v,u=\pm} \left(  {\rm Tr}
 [(\gamma^\tau ip'_{z}+i\omega_- \gamma^z)\gamma^{\alpha}\gamma^x\gamma^\beta \gamma^a\gamma^{\nu }P_{+}] \times  \right. \nonumber\\
&& \qquad \left.  \Delta_t(i\omega_-,p'-p) (\hat p'-\hat p)_{(\alpha,t) } \Delta_u(i\omega_-,p'-p-k) 
(\hat p'-p-k)_{(\beta,u) }  \Delta_v(i\omega_-,p')\hat{ p'}_{(\nu,v)} \right). \nonumber\\ 
\end{eqnarray}
The second term in the curved brackets is obtained by the the exchange Wick contraction. 
Examining the terms we see that we need to evaluate the following integrals
\begin{eqnarray}
& & A_1 = i^2\frac{e^2}{2^4 \beta} \sum_{ m} \int \frac{d^5 p'}{(2\pi)^5}
\sum_{t,v,u=\pm} {\rm Tr}
[\gamma^a\gamma^{\alpha}\gamma^x\gamma^\beta\gamma^\tau \gamma^{\nu }P_{+}]
p'_{z} \times  \\
&&\ \Delta_t(i\omega_-,p'+p+k) ( \hat p'+\hat p+\hat k)_{(\alpha,t)}
\Delta_u(i\omega_-,p'+p) (\hat p'+\hat p)_{(\beta,u)}
  \Delta_v(i\omega_-,p') \hat p'_{(\nu,v)},\nonumber\\
& & A_2 = i\frac{e^2}{2^4 \beta} \sum_{ m} \int \frac{d^5 p_1}{(2\pi)^5}\sum_{t,v,u=\pm} {\rm Tr} [\gamma^a\gamma^{\alpha} \gamma^x\gamma^\beta\gamma^z \gamma^{\nu }P_{+}]i\omega_-
\times  \nonumber\\
&&\ \Delta_t(i\omega_-,p'+p+k) (\hat p'+\hat p+\hat k)_{(\alpha,t)}
\Delta_u(i\omega_-,p'+p) (\hat p'+ \hat p)_{(\beta,u) } 
 \Delta_v(i\omega_-,p') \hat p'_{(\nu,v) }.\nonumber 
\end{eqnarray}
Evaluating the trace over the $\gamma$ matrices, we end up with 
\begin{eqnarray}
A_1&=&i\frac{e^2}{2^4 \beta} \sum_{\omega_m} \int \frac{d^5 p'}{(2\pi)^5}\sum_{t,v,u=\pm}
\frac{4\, t\, u\, v\, k_y\, p_b \, p^{\prime 2}_{z} }{E_{p'}E_{p'+p+k}E_{p'+p}} \times  \\
&&  \Delta_t(i\omega_-,p'+p+k)\Delta_u(i\omega_-,p'+p)
  \Delta_v(i\omega_-,p'),\nonumber\\
A_2&=&
i\frac{e^2}{2^4 \beta} \sum_{\omega_m} \int \frac{d^5 p'}{(2\pi)^5}\sum_{t,v,u=\pm} 
\left(\frac{4tu(k_{y}p'_{b}+p_bk_y)}{E_{p'+p+k}E_{p'+p}}
-\frac{4u\, v \, p'_{y}\, p_{b} }{E_{p'}E_{p'+p}} \right. 
\nonumber\\
&& \left. -\frac{4tv(k_{y}p_{1b}-p_bp'_{y})}{E_{p'+p+k}E_{p'}} \right)
i\omega_- \Delta_t(i\omega_-,p'+p+k)\Delta_u(i\omega_-,p'+p) 
 \Delta_v(i\omega_-,p').\nonumber 
\end{eqnarray}
The sum over Matsubara frequencies in $A_1$ is performed using \eqref{ms6d1}. 
To perform the sum in $A_2$, we use the following
\begin{eqnarray}\label{ms6d2}
&&\frac{1}{\beta}\sum_{ m}i\omega_-\Delta_t(i\omega_-,p'+p+k)\Delta_u(i\omega_-,p'+p)\Delta_v(i\omega_-,p')\nonumber\\
&&=\frac{E_{p'+p+k}f(E_{p'+p+k}-te\mu _{-})}{(tE_{p'+p+k}-vE_{p'})(tE_{p'+p+k}-uE_{p'+p})}+
\frac{E_{p'+p}f(E_{p'+p}-ue{\mu }_{-})}{(uE_{p'+p}-tE_{p'+p+k})(uE_{p'+p}-vE_{p'})}\nonumber\\
&&\qquad\qquad 
+\frac{E_{p'}f(E_{p'}-ve{\mu }_{-})}{(vE_{p'}-tE_{p'+p+k})(vE_{p'}-uE_{p'+p})}.\nonumber\\
\end{eqnarray}
We can now proceed as discussed in \ref{Appendix A1} to perform the sum over $\{t, u, v\}$ and 
then change variable  of integrations. Finally after performing the angular 
integrals using the $i\epsilon$ prescription, we obtain 
\begin{eqnarray}
\lim_{p_b,k_y \rightarrow 0}\frac{-A_1}{ip_b ik_y}&=&
i\frac{e^2}{32\pi^3} \sum_{t=\pm} \int_0^\infty dp'\,  p' f( p' -te\mu _{-}) \\
\lim_{p_b,k_y\rightarrow 0} \frac{-A_2}{ip_b ik_y}&=&i\frac{e^2}{32\pi^3} \sum_{t=\pm} \int_0^\infty
 d p' \, p p' f(p' -te\mu _{-}).\nonumber 
\end{eqnarray}
The exchange term in (\ref{jjtap}) also contributes an equal  amount to the correlator. Therefore, we 
have 
\begin{eqnarray}
\lim_{p_b,k_y \rightarrow 0}
- \frac{j^a(k+p)j^x(-k)T^{tz}(-p)}{ip_b ik_y}
&=&\lim_{p_b,k_y \rightarrow 0} -2 \frac{A_1+ A_2}{ip_b ik_y},\nonumber\\
&=&i\frac{e^2}{8\pi^3} \sum_{t=\pm} \int _0^\infty dp' p' f( p' -te\mu _{-}),\nonumber\\
&=&i\frac{e^2}{16\pi^3}(e^2(\mu _{-})^2+\frac{\pi^2 T^2}{3}).
\end{eqnarray}

\subsection{Contributions to \texorpdfstring{$\zeta^{(6)}_3$}{Lg}}\label{Appendix A3}
         
  The correlator corresponding to   the transport coefficient   $\zeta^{(6)}_3$     
  also involves contact terms.  We will first evaluate the contribution from Wick contractions of 
  the flat space currents in the correlator 
   $\langle j^a_{fl}T^{tx}_{fl}T^{tz}_{fl}\rangle$. We  then proceed to evaluate the 
   contribution from the  contact term.
\begin{eqnarray}
&& \langle j^a_{fl}(p+k) T_{fl}^{\tau x}(-k) T_{fl}^{\tau z}(-p)\rangle  \\ \nonumber
&=&-\frac{e}{32 \beta}\sum_{ m} \int \frac{d^5 p'}{(2\pi)^5}\sum_{t,v,u=\pm}{\rm Tr}
[\gamma^a\gamma^\alpha(\gamma^\tau ip'_{x}+\gamma^x i\omega_-)
\gamma^\beta(\gamma^\tau ip_{'z}+\gamma^z i\omega_-)\gamma^\nu P_{+}]\nonumber\\
&& \qquad \times (\hat p'+\hat p+\hat k)_{(t,\alpha)} (\hat p' +\hat p)_{(u,\beta) }
p'_{(v,\nu )}    \Delta_t(i\omega_-,p'+k+p)
\Delta_u(i\omega_- ,p'+p)
\Delta_v(i\omega_-,p' )\nonumber \\
&&+(\textrm{Exchange term}) .  \nonumber
\end{eqnarray}         
Among  the four possible traces  it can be seen that only 3 of them contribute, they are
given by 
\begin{eqnarray}
A_1&=&-\frac{ie}{32 \beta}\sum_{ m} \int \frac{d^5 p'}{(2\pi)^5}\sum_{t,v,u=\pm}{\rm Tr}
[\gamma^a\gamma^\alpha\gamma^\tau \gamma^\beta\gamma^z\gamma^\nu P_{+}]p'_{x}
i\omega_-\nonumber\\
&&\qquad 
\times (\hat p'+\hat p+\hat k)_{(t,\alpha)} (\hat p'+\hat p)_{(u,\beta)} \hat p'_{(v,\nu)} 
 \Delta_t(i\omega_- ,p'+k+p)\Delta_u(i\omega_- ,p'+p)
\Delta_v(i\omega_-,p')\nonumber\\
&&+( \textrm{Exchange  term}) ,\nonumber\\ 
A_2&=&-\frac{ie}{32 \beta}\sum_{m} \int \frac{d^5 p'}{(2\pi)^5}\sum_{t,v,u=\pm}{\rm Tr}
[\gamma^a\gamma^\alpha\gamma^x\gamma^\beta\gamma^\tau \gamma^\nu P_{+}]p'_{z}i\omega_-\nonumber\\
&&\qquad 
\times (\hat p'+\hat p+\hat k)_{(t,\alpha)}(\hat p'+\hat p)_{(u,\beta)}\hat{p'}_{(v,\nu) } 
 \Delta_t(i\omega_- ,p'+k+p)
\Delta_u(i\omega_- ,p' +p)\Delta_v(i\omega_-,p' )\nonumber\\
&&+(\textrm{Exchange term}),\nonumber\\ 
A_3&=&\frac{-e}{32 \beta}\sum_{m} \int \frac{d^5 p'}{(2\pi)^5}\sum_{t,v,u=\pm}{\rm Tr}
[\gamma^a\gamma^\alpha\gamma^x\gamma^\beta\gamma^z\gamma^\nu P_{+}](i\omega_-)^2\nonumber\\
&&\qquad \times (\hat p'+\hat p+\hat k)_{(t,\alpha)}(\hat p'+\hat p)_{(u,\beta)}\hat p'_{(v,\nu )} 
\Delta_t(i\omega_-,p'+k+p)\Delta_u(i\omega_-,p'+p)\Delta_v(i\omega_-,p')\nonumber\\
&&+(\textrm{Exchange term}). \nonumber\\
\end{eqnarray}
Now we follow the methods adopted to evaluate 
 the previous correlators.  We first perform the traces in each of them and follow that up with the Matsubara sum and then take the limit, $p_b,k_y \rightarrow0$. 
 Lastly we perform the angular integrals with $i\epsilon$  trick.
  We find that the terms $A_1$ and $A_2$ contribute 
  equally and they are given by  
\begin{eqnarray}
A_1&=&\frac{-e}{32 \beta}\sum_{m} \int \frac{d^5 p'}{(2\pi)^5}
\sum_{t,v,u=\pm}\frac{4\, t\, u\, v\, k_y\, p_b\,  p^{\prime 2}_{x} \, i\omega_- }{E_{p'}E_{p'+p+k}E_{p'+p}} \Delta_t(p'+k+p)\Delta_u(p'+p)\Delta_v(p')
 \nonumber\\
&&+(\textrm{Exchange term}), \nonumber\\
A_2&=&\frac{-e}{32 \beta}\sum_{ m} \int \frac{d^5 p'}{(2\pi)^5}\sum_{t,v,u=\pm}
\frac{4\, t\, u\, v\, k_y\, p_b\,   p_{z}^{\prime 2}\,  i\omega_-}{E_{p'}E_{p'+p+k}E_{p'+p}} 
  \Delta_t(p'+k+p)\Delta_u(p'+p)\Delta_v(p')\nonumber\\
&&+(\textrm{Exchanged term}). \nonumber\\
\end{eqnarray} 
The Matsubara sums are performed using equation \eqref{ms6d2}. 
After  a change of variables of integration and taking the external momenta to zero, we obtain 
\begin{eqnarray}
\lim_{p_b,k_y \rightarrow 0}\frac{-A_1}{ip_b ik_y}&=&-
\frac{2 e}{64\pi^3}\sum_{t=\pm}\int_0^\infty  dp'  \, t\, p^{\prime 2}f(p'-te\mu _{-}).
\end{eqnarray}
Here again we have used the $i\epsilon$ prescription to evaluate the integrals. 
The  factor of $2$ results from taking into account the equal contribution of the 
exchange term.  
 Evaluating the $A_3$ term involves performing the following Matsubara sum
 \begin{eqnarray}\label{ms6d3}
&& \frac{1}{\beta}\sum_{m}(i\omega_-)^2\Delta_t(i\omega_-,p'+p+k)\Delta_u(i\omega_-,p'+p)
\Delta_v(i\omega_-,p_1)\nonumber\\
&&=\frac{tE_{p'+p+k}^2 f(E_{p'+p+k}-te\mu _{-})}{(tE_{p'+p+k}-vE_{p'})(tE_{p'+p+k}-
uE_{p'+p})}+\frac{uE_{p'+p}^2 f(E_{p'+p}-ue{\mu }_{-})}{(uE_{p'+p}-tE_{p'+p+k})(uE_{p'+p}-vE_{p'})}\nonumber\\
&&+\frac{vE_{p'}^2 f(E_{p'}-ve{\mu }_{-})}{(vE_{p'}-tE_{p'+p+k})(vE_{p'}-uE_{p'+p})},\nonumber\\
& & \equiv {\cal N}. 
\end{eqnarray}
Performing the trace over the $\gamma$ matrices and substituting the Matsubara sum ${\cal N}$, 
we obtain 
\begin{eqnarray}
A_3&=&\frac{e}{32 \beta}\sum_{m} \int \frac{d^5 p'}{(2\pi)^5}\sum_{t,v,u=\pm}\left( 
\frac{4\, t\, u\, (k_{y}p'_{b}+p_bk_y) }{E_{p'+p+k}E_{p'+p}}-\frac{4\, u\, v\,  p'_{y}\, p_{b} }{E_{p'}E_{p'+p}}\right. \\ \nonumber
&& \qquad\qquad\qquad\qquad
\left. -\frac{4\, t\, v\, ( k_{y}\, p'_{b}-p_b p'_{y}) }{E_{p'+p+k}E_{p'}} \right)\times {\cal N}  \\ \nonumber
& & +(  \textrm{Exchange term})  .
\end{eqnarray} 
Now summing over $\{t, u, v\}$, making appropriate shifts in the integrations and taking 
the zero momentum limit, we get 
\begin{equation}
\lim_{p_b,k_y\rightarrow 0}\frac{-A_3}{ip_b ik_y} = - \frac{2 e}{64\pi^3}\sum_{t=\pm}\int_0^\infty dp'\,  
t\, p^{\prime 2}f(p'-t\, e\, \mu _{-}).
\end{equation}
Therefore the putting terms $A_1, A_2, A_3$ together we see that the 
 contributions from the Wick contractions of the flat space currents is 
\begin{eqnarray}
\lim_{p_b,k_y \rightarrow 0}
-\frac{\langle j^a_{fl}(-p-k) T^{tx}_{fl}(k) T^{tz}_{fl}(p)\rangle}{ip_b ik_y}  = 
- \frac{3e}{32\pi^3}\sum_{t=\pm}\int_0^\infty  dp'\,  
p^{\prime 2} t\, f(p'-t\, e\, \mu _{-}) .\nonumber \\
\end{eqnarray}

We will now obtain the contribution from the contact terms. For this we first Fourier transform the 
interaction $S_E^2$ given in (\ref{action6dexp}),   this yields 
\begin{eqnarray}
\frac{\delta S_E^{(2)}} {\delta h_{x \tau} (k) \delta h_{z, \tau}(p)  } &=&
-\frac{i}{16 \beta } \sum_{ m}\int \frac{d^5 p'}{(2\pi)^5} 
\psi^\dagger ( \omega_m, p+k+p')(p_b \gamma^b \gamma^x \gamma^z + k_y \gamma^y \gamma^z \gamma^x )
P_- \psi( \omega_m , p').\nonumber\\
\end{eqnarray}
Considering the Wick contraction of the above contact term along with the flat space charge current, results in the 
following two terms 
\begin{eqnarray}
\frac{-1}{ip_b ik_y}B&=& B_1 +B_2 \\ \nonumber
B_1 &=& \frac{e  }{16  k_y} \frac{1}{\beta}\sum_{m}\int \frac{d^5 p'}{(2\pi)^5} 
{\rm Tr} [\gamma^a S(\omega_m, p+k+p')\gamma^b \gamma^x \gamma^z S (\omega_m, p')P_+],\nonumber\\
B_2 &=& \frac{e  }{16 p_b } \frac{1}{\beta}\sum_{m}\int 
\frac{d^5 p'}{(2\pi)^5}{\rm Tr} [\gamma^a S(\omega_m, p+k+p')\gamma^b 
\gamma^z \gamma^x S(\omega_m, p')P_+].\nonumber
\end{eqnarray}
We proceed to evaluate $B_1$. 
Substituting the expression for thermal propagators we obtain,
\begin{eqnarray}
\lim_{p_b,k_y\rightarrow 0}\lim_{p_0,k_0\rightarrow 0}B_1&=&
\lim_{p_b,k_y\rightarrow 0}\frac{-e}{16k_y} \frac{1}{\beta}\sum_{m}\int \frac{d^5 p'}{(2\pi)^5}
\sum_{t,u=\pm} {\rm Tr} [\gamma^a \gamma^\alpha \gamma^b \gamma^x \gamma^z \gamma^\beta P_+]\nonumber\\
&&\times \frac{1}{4} (\hat p'+\hat p+\hat k)_{(t,\alpha)}\hat{p'}_{(u,\beta)}
\Delta_t(i\omega_-, p'+p+k) \Delta_u(i\omega_-, p'),\nonumber\\ 
&=&\lim_{p_b,k_y\rightarrow 0}
\frac{e}{16}\sum_{t,u=\pm}\int \frac{d^5 p'}{(2\pi)^5} 
\frac{t}{E_{p'+p+k}} \frac{tf(E_{p'+p+k}-te\mu _-)-uf(E_{p'}-ue\mu _-)}{tE_{p'+p+k}-uE_{p'}},\nonumber\\
&=&\frac{e}{4 (2\pi)^5} \int dp' d\phi_1 d\phi_2 d\phi_3 (2\pi) 
\sum_{t=\pm} p^{\prime 4} \sin^3 \phi_1 \sin^2 \phi_2 \sin \phi_3 \nonumber\\
&& \qquad\qquad \times \left(-\frac{1}{2p_bp' \cos \phi_1}+ \frac{1}{4p^{\prime 2} \cos^2 \phi_1} \right) 
+ O( k_y/p_b, p_b). 
\end{eqnarray}
To obtain the last equality we have summed over $\{t, u \}$ and then taken the external momenta to zero. 
We have taken $k_y \rightarrow 0$ first and then taken the limit $p_b\rightarrow 0$. 
But the final result for the  non-singular term is independent of the order of limits. 
Also note that the singular term in $p_b$  is odd under $\phi_1 \rightarrow \pi -\phi_1$. 
Therefore it vanishes on performing the angular integration over $\phi_1$. 
If the order of limit is interchanged the resulting singular terms in $k_y$ also vanish 
on angular integration and the final result for the non-singular term is invariant. 
The angular integration in this term can be evaluated easily by the $i\epsilon$ prescription,
\begin{eqnarray}
\int_0^\pi \frac{\sin^3 \phi_1}{ \cos^2 \phi_1}&=& -4. \nonumber\\
\end{eqnarray}
Substituting this result into  $B_1$ we obtain 
\begin{eqnarray}
\lim_{p_b,k_y\rightarrow 0}B_1
&=&-\frac{e}{64\pi^3}\sum_{t=\pm}\int_0^\infty  dp' \, t\, p^{\prime 2}f(p' -t\, e\, \mu _{-}).
\end{eqnarray}
Going through a similar analysis on $B_2$ it can be shown that the result for $B_2$ is equal 
to that of $B_1$. 
Thus the total contribution from the contact term is given by 
\begin{eqnarray}
\lim_{p_b,k_y\rightarrow 0}B &=&-\frac{2 e }{64\pi^3} \sum_{t=\pm}
\int_0^\infty  dp'  \, t\, p^{\prime 2} f(p' -t\, e\, \mu _{-}). 
\end{eqnarray}
Therefore putting all the terms from $A$ and $B$ together we obtain the final
result for the correlator of interest to be 
\begin{eqnarray}
\frac{-1}{ip_b ik_y}\langle j^a(p+k) T^{\tau x}(-k) T^{\tau z}(-p)\rangle 
&=& \frac{-e }{8\pi^3}\sum_{t=\pm}\int_0^\infty  dp'\,  t\, p^{\prime 2}f(p_1-t\, e\, \mu _{-})\nonumber\\ 
&=&  \frac{-e}{ 24\pi^3} \left( e^3\mu_-^3  + e \pi^2 T^2 \mu_-\right). 
\end{eqnarray}

\subsection{Contributions to \texorpdfstring{$\lambda^{(6)}_3$}{Lg}} \label{Appendix A4}

There are two contributions to this correlator, one from the direct Wick contractions 
of the flat space currents and the other from the contact terms. 
Let us first evaluate the former contribution, performing Wick contractions leads to 
\begin{eqnarray}
&&\langle T_{fl}^{\tau a}(p+k) T_{fl}^{\tau x}(-k) T_{fl}^{\tau z}(-p)\rangle =\frac{-i}{\beta}\sum_{m}
\frac{1}{64}\int \frac{d^5 p'}{(2\pi)^5} \times  \\
&&\left( 
{ \rm Tr} [\gamma^\tau ip'_{a} +\gamma^a i\omega_-)\gamma^\alpha (\gamma^\tau ip'_{x}
+\gamma^xi\omega_-)\gamma^\beta(\gamma^\tau ip'_{z}+\gamma^z i\omega_-)\gamma^{\nu }P_{+}]
\right. \nonumber\\
&&\left. \Delta_{t}(i\omega_-, p'+p+k)\Delta_u (i\omega_-,p'+p)\Delta_v (i\omega_-, p')
(\hat p'+\hat p+\hat k)_{(t,\alpha)}
(\hat p'+\hat p)_{(u,\beta)}\hat p'_{(v,\nu)} \right)  \nonumber \\
& & +( \textrm{exchange term} ).\nonumber
\end{eqnarray}
Expanding the above express it can be seen that there are $4$ terms which are non-vanishing
these are given by 
\begin{eqnarray}
A_1&=&-\frac{1}{64 \beta }\sum_{m}\int
 \frac{d^5 p'}{(2\pi)^5}{\rm Tr}
  [\gamma^\tau \gamma^\alpha \gamma^x \gamma^\beta \gamma^z\gamma^{\nu }P_{+}]p'_{a}
  \omega_-^2  \times  \nonumber \\
 & &  \Delta_{t}(i\omega_-, p'+p+k)\Delta_u (i\omega_- ,p'+p)\Delta_v (i\omega_-, p')
(\hat p'+\hat p+\hat k)_{(t,\alpha)}(\hat p'+\hat p)_{(u,\beta)} \hat p'_{(v,\gamma)}\nonumber\\
& & +( \textrm{exchange term} ), \nonumber \\ 
A_2&=&-\frac{1}{64 \beta }\sum_{m}\int \frac{d^5 p'}{(2\pi)^5}{\rm Tr} 
[\gamma^a \gamma^\alpha \gamma^x \gamma^\beta \gamma^\tau \gamma^{\nu }P_{+}]
p'_{z}\omega_-^2 \times \nonumber\\
&&\Delta_{t}(i\omega_- p'+p+k)\Delta_u (i\omega_-,p'+p)\Delta_v (i\omega_-, p')
(\hat p'+\hat p+\hat k)_{(t,\alpha)}(\hat p'+\hat p)_{(u,\beta)}\hat{p'}_{(v,\nu)}\nonumber\\
& & +( \textrm{exchange term} ), \nonumber \\
A_3&=&-\frac{1}{64 \beta }\sum_{m}\int \frac{d^5 p'}{(2\pi)^5}{ \rm Tr} 
[\gamma^a \gamma^\alpha \gamma^\tau \gamma^\beta \gamma^z \gamma^{\nu }P_{+}]p'_{x}
\omega_-^2   \times \nonumber \\ 
&&\Delta_{t}(i\omega_- , p'+p+k)\Delta_u (i\omega_-,p'+p)
\Delta_v (i\omega_-, p')
(\hat p'+\hat p+\hat k)_{(t,\alpha)} (\hat p'+ \hat p)_{(u,\beta)}\hat{p'}_{(v,\nu) }\nonumber\\
& & +( \textrm{exchange term} ), \nonumber \\
A_4&=&-\frac{1}{64\beta }\sum_{m}\int \frac{d^5 p'}{(2\pi)^5}{\rm Tr} 
[\gamma^a \gamma^\alpha \gamma^x \gamma^\beta \gamma^z \gamma^{\nu }P_{+}]
\omega_-^3\nonumber\\
&&\Delta_{t}(i\omega_- , p'+p+k)\Delta_u (i\omega_-,p'+p)\Delta_v (i\omega_-, p')
(\hat p'+\hat p+\hat k)_{t,\alpha}(\hat p'+\hat p)_{u,\beta}\hat{p'}_{(v,\nu)}\nonumber\\
& & +( \textrm{exchange term} ). 
\end{eqnarray}
We find that contribution of the terms $A_1$,$A_2$ and $A_3$  are equal.  Furthermore the 
contribution of the direct Wick contraction is equal to that of the exchange term in each 
$A_1, A_2, A_3, A_4$. After performing the trace over $\gamma$ matrices, we find that 
$A_1$ is given by 
\begin{eqnarray}
A_1&=& \frac{2i }{64 \beta}\sum_{m}\int \frac{d^5 p'}{(2\pi)^5} \sum_{t,u,v=\pm} 
\frac{4\, t\, u\, v\, p^{\prime 2}_{a}}{E_{p'+p+p}E_{p'+p}E_{p'}}\omega_-^2 \times \\
&&\Delta_{t}(i\omega_-, p'+p+k)\Delta_u (i\omega_-,p'+p)\Delta_v (i\omega_-, p').\nonumber
\end{eqnarray}
Here the factor of $2$ is due to the contribution of the exchange term. 
Proceeding as in discussed in the previous examples we obtain 
\begin{eqnarray}
\lim_{p_b,k_y \rightarrow 0}
- \frac{A_1}{ip_bik_y}&=&-\frac{2 i}{128\pi^3} \sum_{t=\pm}\int_0^\infty
  dp' \, p^{\prime 3}  f( p'-t\, e\, \mu_{-}).
\end{eqnarray}
Since the contribution of $A_2$ and  $A_3$ is identical to that of $A_1$ we have
\begin{eqnarray}
\lim_{p_b,k_y\rightarrow 0} -\frac{A_1+A_2+ A_3}{ip_bik_y}&=&-\frac{6i}{128\pi^3}  \sum_{t=\pm}
\int_0^\infty  dp' \, p^{\prime 3 }f( p' -t\, e\, \mu_{-}). 
\end{eqnarray}
To evaluate the term $A_4$ 
we need to perform the matsubara sum which is given by 
\begin{eqnarray}\label{ms6d4}
 &&\frac{1}{\beta}\sum_{m}(i\omega_-)^3\Delta_t(i\omega_-,p'+p+k)
 \Delta_u(i\omega_-,p'+p)\Delta_v(i\omega_-,p')\nonumber\\
&&=\frac{E_{p'+p+k}^3 f(E_{p'+p+k}-te\mu _{-})}{(tE_{p'+p+k}-vE_{p'})(tE_{p'+p+k}-uE_{p'+p})}+
\frac{E_{p'+p}^3 f(E_{p'+p}-ue{\mu }_{-})}{(uE_{p'+p}-tE_{p'+p+k})(uE_{p'+p}-vE_{p'})}\nonumber\\
&&+\frac{E_{p'}^3 f(E_{p'}-ve{\mu }_{-})}{(vE_{p'}-tE_{p'+p+k})(vE_{p'}-uE_{p'+p})}.
\end{eqnarray}
At the end of the usual manipulations, we obtain
\begin{eqnarray}
\lim_{p_b,k_y\rightarrow 0}- \frac{A_4}{ip_bik_y}&=&-\frac{2i}{128\pi^3} \sum_{t=\pm}\int_0^\infty dp'\, 
 p^{\prime 3}  f( p'-t\, e\, \mu_{-}).
\end{eqnarray}
Therefore the contribution from the Wick contractions of the flat space current is given by 
\begin{eqnarray}
\lim_{p_b,k_y\rightarrow 0} -\frac{ \langle T^{\tau a}(p+k) T^{\tau x}(-k) T^{\tau z}(-p)\rangle}{ip_bik_y} 
&=&\frac{-8i}{128\pi^3} \sum_{t=\pm} \int dp' p^{\prime 3} f( p' -t\, e\, \mu_{-}) . \nonumber \\
\end{eqnarray}

Now lets examine the contact terms. As seen in the equation (\ref{tttc}), there are 3 contact
terms.  We will evaluate one of them, it can be seen that the contributions from
all the 3 contact terms are equal.  Performing the Wick contraction of the 
contact term with the flat space stress tensor we obtain
\begin{eqnarray}
\frac{-B_1}{ip_b i k_y} &=& -\frac{1}{32i \beta} \sum_{m}\int \frac{d^5 p'}{(2\pi)^5}
\left ( \frac{1}{k_y}{\rm Tr}[(\gamma^\tau ip'_{a}+\gamma^a i\omega_-)
S( \omega_m, p' + p + k) (\gamma^b \gamma^x \gamma^z)S( \omega_m, p')P_+] \right. \nonumber\\
&&\left. + \frac{1}{p_b}{ \rm Tr}[(\gamma^\tau ip'_{a}+\gamma^a i\omega_-)S(\omega_m,  p' + p + k) (\gamma^y \gamma^z \gamma^x)S(\omega_m, p')P_+] \right).\nonumber\\
\end{eqnarray}
We then follow the same steps as before to evaluate this integral and get
\begin{eqnarray}
\lim_{p_b,k_y \rightarrow 0}-\frac{B_1}{ip_bi k_y}&=&\frac{-8i}{3}\frac{1}{128\pi^3} \sum_{t=\pm}\int_0^\infty dp' \, p^{\prime 3} f( p' -t\, e\, \mu_{-}).\nonumber\\
\end{eqnarray}
The other two contact terms 
$B_2$ and $B_3$ also give the same amount of contribution.  Therefore the total contribution from  the contact terms is given by 
\begin{equation}
B=- \frac{8i}{128\pi^3}\sum_{t=\pm}\int_0^\infty  dp'\,  p^{\prime 3}  f( p'-t\, e\, \mu_{-}).
\end{equation}
Combing the terms $A$ and $B$ we see that the 
 value for the three point function  of interest is given by 
\begin{equation}
\lim_{p_b,k_y  \rightarrow 0} -\frac{\langle T^{\tau a}(k+p)T^{\tau x}(-k)T^{\tau z}(-p)\rangle}{ip_b ik_y}=\frac{-16i}{128\pi^3}\sum_{t=\pm}\int_0^\infty dp' p^{\prime 3}   f(p' -t\, e\,  \mu_{-}).                    
\end{equation}

\providecommand{\href}[2]{#2}\begingroup\raggedright\endgroup

\end{document}